%% file: main.tex
\documentclass[12pt]{article}
\let\oldphi\varphi \let\varphi\phi \let\phi\oldphi
\let\oldepsilon\varepsilon \let\varepsilon\epsilon \let\epsilon\oldepsilon

\usepackage{amsmath,amssymb}
\usepackage{amsfonts}
\usepackage[utf8]{inputenc}
\usepackage{graphicx}
\usepackage[T1]{fontenc}
\usepackage[english]{babel}
\usepackage{graphicx}
\usepackage{fancyhdr}
\usepackage{float}
\usepackage{afterpage}
\usepackage{placeins}
\usepackage{titlesec}
\usepackage{setspace}
\usepackage[font=small,labelfont=bf]{caption}
\usepackage{url}
\usepackage[left=2cm,right=2cm,top=2.5cm,bottom=2.5cm]{geometry}
\usepackage{hyperref}
\usepackage{comment}
\usepackage{tikz-cd}		% for arrow diagrams
\usepackage{tikz}	% for the special graphs/tables
\usetikzlibrary{decorations.pathreplacing,calc,shapes.geometric,arrows} % needed for tikz

\usepackage{tensor}	% for tensor notation
\usepackage{wasysym} % for astronomical Symbols lime solar mass: M_Sun
\usepackage[header]{appendix} % for the appendix to have numbering from A-Z
\usepackage{setspace} % include to change the line spacing
\usepackage[nottoc,numbib]{tocbibind} % to give the rbibliography numbering in the table of contents
\usepackage{xstring}
\usepackage{subfigure}
\usepackage{multirow}

\usepackage{nicematrix}

\graphicspath{
	{pics/}
}

\newcommand{\e}[1]{\cdot 10^{#1}} % Wissenschaftliche Notation für Potenzen

\tikzstyle{startstop} = [rectangle, rounded corners, 
minimum width=3cm, 
minimum height=1cm,
text centered, 
draw=black, 
fill=red!30]

\tikzstyle{io} = [trapezium, 
trapezium stretches=true, % A later addition
trapezium left angle=70, 
trapezium right angle=110, 
minimum width=3cm, 
minimum height=1cm, text centered, 
draw=black, fill=blue!30]

\tikzstyle{process} = [rectangle, rounded corners,
minimum width=3cm, 
minimum height=2cm, 
text centered, 
text width=3cm, 
draw=red!50, 
fill=orange!20]

\tikzstyle{decision} = [diamond, 
minimum width=3cm, 
minimum height=1cm, 
text centered, 
draw=black, 
fill=green!30]

\tikzstyle{arrow} = [thick,->,>=stealth]

\begin{document}
%%%%%%%%%%%%%%%%%%%%%%%%% title page:
\thispagestyle{empty}
% add a nice looking tikz image for a border around the title page:
   \begin{tikzpicture}[remember picture,overlay,inner sep=0,outer sep=0]
     \draw[blue!70!black,line width=4pt] ([xshift=-1.5cm,yshift=-2cm]current page.north east) coordinate (A)--([xshift=1.5cm,yshift=-2cm]current page.north west) coordinate(B)--([xshift=1.5cm,yshift=2cm]current page.south west) coordinate (C)--([xshift=-1.5cm,yshift=2cm]current page.south east) coordinate(D)--cycle;

     \draw ([yshift=0.5cm,xshift=-0.5cm]A)-- ([yshift=0.5cm,xshift=0.5cm]B)--
     ([yshift=-0.5cm,xshift=0.5cm]B) --([yshift=-0.5cm,xshift=-0.5cm]B)--([yshift=0.5cm,xshift=-0.5cm]C)--([yshift=0.5cm,xshift=0.5cm]C)--([yshift=-0.5cm,xshift=0.5cm]C)-- ([yshift=-0.5cm,xshift=-0.5cm]D)--([yshift=0.5cm,xshift=-0.5cm]D)--([yshift=0.5cm,xshift=0.5cm]D)--([yshift=-0.5cm,xshift=0.5cm]A)--([yshift=-0.5cm,xshift=-0.5cm]A)--([yshift=0.5cm,xshift=-0.5cm]A);

     \draw ([yshift=-0.3cm,xshift=0.3cm]A)-- ([yshift=-0.3cm,xshift=-0.3cm]B)--
     ([yshift=0.3cm,xshift=-0.3cm]B) --([yshift=0.3cm,xshift=0.3cm]B)--([yshift=-0.3cm,xshift=0.3cm]C)--([yshift=-0.3cm,xshift=-0.3cm]C)--([yshift=0.3cm,xshift=-0.3cm]C)-- ([yshift=0.3cm,xshift=0.3cm]D)--([yshift=-0.3cm,xshift=0.3cm]D)--([yshift=-0.3cm,xshift=-0.3cm]D)--([yshift=0.3cm,xshift=-0.3cm]A)--([yshift=0.3cm,xshift=0.3cm]A)--([yshift=-0.3cm,xshift=0.3cm]A);

   \end{tikzpicture}
\begin{center}
\centering
$\:$ \\
{\LARGE \textbf{ Scalar- and Vector Dark Matter \\ $\:$ \\ Admixed Neutron Stars }} \\

$\:$ \\
$\:$ \\
$\:$ \\
Dated \today \\
$\:$ \\
$\:$ \\
$\:$ \\

{\Large
Master Thesis\\
Cédric Jockel \\
}
$\:$ \\
(jockel@itp.uni-frankfurt.de)
$\:$ \\
$\:$ \\

\includegraphics[width=0.3\textwidth]{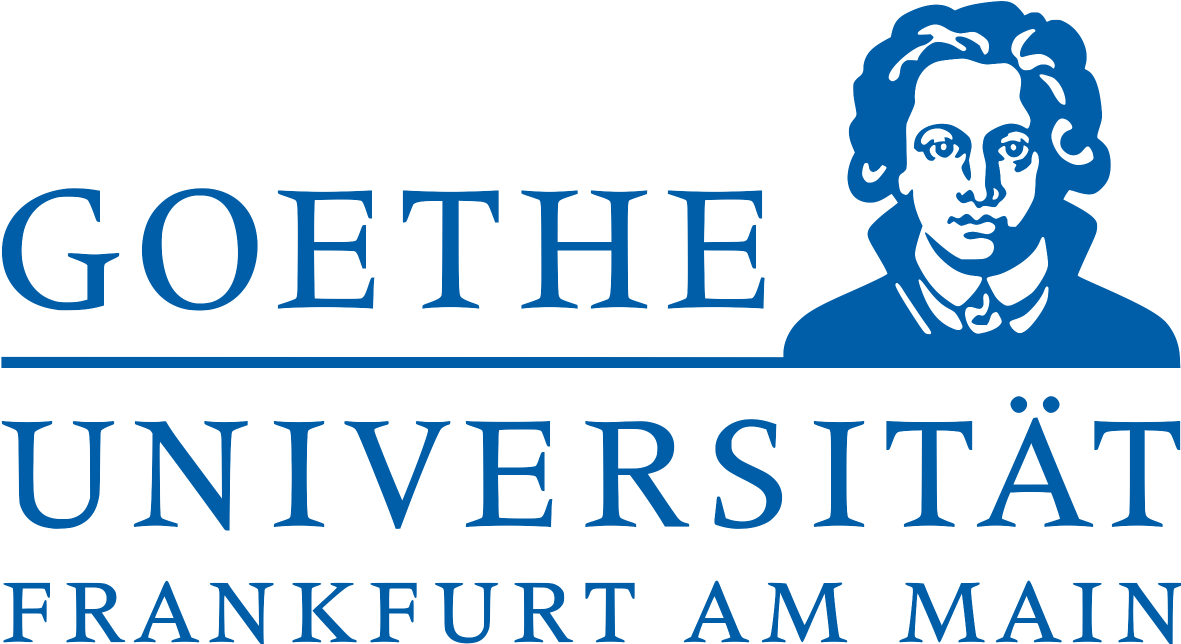}

$\:$ \\
{\large \textsf{
Johann Wolfgang Goethe-Universität \\
Frankfurt am Main \\
Institut für Theoretische Physik \\
}
$\:$ \\
$\:$ \\

Supervisor and first examiner: \\
Professor Laura Sagunski \\
Second examiner: \\
Professor Owe Philipsen \\
}
\end{center}

\hspace{0.5cm}
\includegraphics[width=0.4\textwidth]{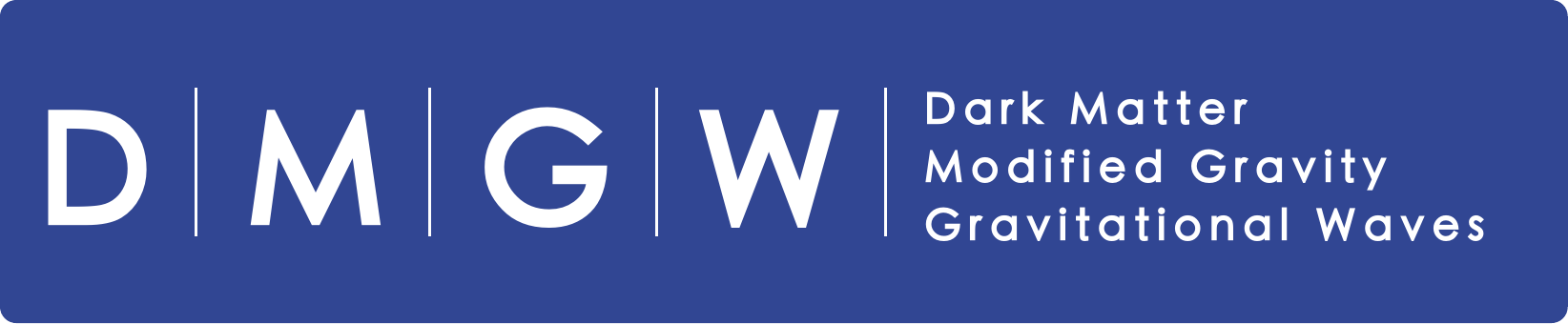}
\hspace{5.5cm}
\includegraphics[width=0.2\textwidth]{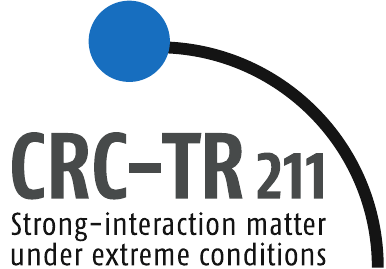}

\newpage

%%%%%%%%%%%%%%%%%%%%%%%%%%%%%%%%%%%%%%%%%%%%%%%%%%%%%%%%%%%%%%%%%%%%%%%%%%%%%%%%%%%%%%%%

% Display the table of contents
\tableofcontents
\newpage
\input{Intro.tex}

\newpage
\input{Background.tex}

\newpage
\input{fermion-boson-stars.tex}

\newpage
\input{numerics.tex}

\newpage
\input{results.tex}

\newpage
\input{conclusions.tex}

\newpage
\input{Appendix.tex}
\newpage
\markright{References}
\bibliographystyle{cedric-bibliostyle}
\bibliography{Biblio.bib}
\end{document}

%% file: Intro.tex
\markright{Foreword}
\section*{Foreword}

\begin{center}
"You just don't get it, do you, Jean-Luc? The trial never ends."
\end{center}
\begin{flushright}
- \textit{Q} (Star Trek: The Next Generation, S7, E25-26 "All Good Things")
\end{flushright}
In \textit{Star Trek: The Next Generation}, there is an omnipotent being called \textit{Q}, who visits the main characters on several occasions under the pretext of assessing and testing the justification of the human race's right to exist. In the first episode of the series, \textit{Q} puts humanity on trial and accuses the main cast of being barbarous representatives of a "dangerous, savage, child-race". During the course of the show, \textit{Q} begins to see the true potential of humanity and slowly becomes an advocate (of sorts) for humanity in this metaphorical trial. In the last episode, \textit{Q} tests humanity again and pushes the protagonists until he obtains "a glimpse" of what humanity might be capable of. The quote above is from this exact moment. I think that this trial is not only a judgment but also a goal to strive for. In the science-fiction world of \textit{Star Trek}, humanity is no longer preoccupied with quarrels for power, greed or material wants. Instead, the search for truth and knowledge, and the will to constantly self-improve to become the best possible version of ourselves are the main motivating force of human endeavour. I hope that in becoming a scientist I can contribute my part in pushing the boundaries of our knowledge. I hope that in the future, generations of people will look back on us and deem us to be the beginning of a more enlightened, wise and more human, race.

\newpage
\markright{Acknowledgments}
\section*{Acknowledgments}

This last year has been an unforgettable journey and a time of large personal and professional growth. I want to thank all the people who helped and accompanied me along this way. I want to thank my parents and family, my friends and everyone in the DMGW group for their support and the plethora of thoughtful discussions we had over the past year. My deepest gratitude goes to my supervisor Professor Laura Sagunski who welcomed me into her group last year and helped me in more ways I could recite within the confines of this page. During our first meeting, before I joined the group, we met in her office and I saw that she had a framed picture of all group members. This was the moment when I knew that I had made the right decision to join her and I am extremely grateful to have worked with this wonderful person! I also thank Professor Owe Philipsen for being my second examiner and in particular for helping me with his letter of recommendation when I applied for my PhD position. I could not have succeeded without the close cooperation of Niklas Becker and Robin Fynn Diedrichs, whom I worked closely with, in our joint paper which eventually lead to be an integral part of my thesis. Thank you! I further want to thank Professor Jürgen Schaffner-Bielich for acting as my de-facto second supervisor and for his help in my PhD applications through his letter of recommendation. Last but definitely not least, I thank everyone who has taken the time to proof-read my thesis. Thank you so much Martina Jung, Danial Jam, Edwin Genoud-Prachex and Henri Schmekies. After finishing my thesis, I will join the Albert Einstein Institute in Potsdam for my PhD and I look forward to the many new challenges to come. I hope that in the future all our paths may cross again. \\

I acknowledge support by the Deutsche Forschungsgemeinschaft (DFG, German Research Foundation) through the CRC-TR 211 `Strong-interaction matter under extreme conditions'– project number 315477589-TRR 211 and by the Hermann-Wilkomm-Stiftung 2023.

\newpage
\markright{Abstract}
\section{Abstract}

It is believed that dark matter (DM) could accumulate inside neutron stars and significantly change their masses, radii and tidal properties. We study what effect bosonic dark matter, modelled as a massive and self-interacting scalar or vector field, has on neutron stars. We derive equations to compute the tidal deformability of the full Einstein-Hilbert-Klein-Gordon system self-consistently, and probe the influence of the scalar field mass and self-interaction strength on the total mass and tidal properties of the combined system, called fermion boson stars (FBS). We are the first to combine Proca stars with neutron stars to mixed systems of fermions and a vector field in Einstein-Proca theory, which we name fermion Proca stars (FPS). We construct equilibrium solutions of FPS, compute their masses, radii and analyse them regarding their stability and higher modes. We find that FPS tend to be more massive and geometrically larger than FBS for equal boson masses and self-interaction strengths. Both FBS and FPS admit DM core and DM cloud solutions and we find that they can produce degenerate results. Core solutions compactify the neutron star component and lower their tidal deformability, cloud solutions have the inverse effect. Electromagnetic observations of certain cloud-like configurations would appear to violate the Buchdahl limit. The self-interaction strength is found to significantly affect both mass and tidal deformability. We discuss observational constraints and the connection to anomalous detections. We also show how models with an effective equation of state compare to the self-consistent solution of FBS and find the self-interaction strength where both solutions converge sufficiently. \\

\textbf{Keywords:} dark matter, ultralight bosons, self-interacting dark matter, neutron stars, equation of state, gravitational waves, tidal deformability, boson stars, Proca stars, scalar field, vector field, fermion boson stars, fermion Proca stars

\newpage
\markright{Zusammenfassung}
\section*{Zusammenfassung}

Es wird angenommen, dass sich dunkle Materie (DM) im Inneren von Neutronensternen ansammeln und deren Masse, Radius und Gezeiteneigenschaften erheblich verändern könnte. Wir untersuchen, welche Auswirkungen bosonische dunkle Materie, die als massereiches und selbstwechselwirkendes Skalar- oder Vektorfeld modelliert wird, auf Neutronensterne hat. Wir leiten Gleichungen her, um die Gezeitenverformbarkeit des vollständigen Einstein-Hilbert-Klein-Gordon-Systems selbstkonsistent zu berechnen, und untersuchen den Einfluss der Skalarfeldmasse und der Stärke der Selbstwechselwirkung auf die Gesamtmasse und die Gezeiteneigenschaften des kombinierten Systems, das wir Fermionen-Bosonen-Sterne (FBS) nennen. Wir sind die ersten, die Proca-Sterne mit Neutronensternen zu gemischten Systemen aus Fermionen und einem Vektorfeld in der Einstein-Proca-Theorie kombinieren, die wir Fermionen-Proca-Sterne (FPS) taufen. Wir konstruieren Gleichgewichtslösungen von FPS, berechnen ihre Massen, Radien und analysieren sie bezüglich ihrer Stabilität und höherer Moden. Wir finden heraus, dass FPS bei gleichen Bosonenmassen und Selbstwechselwirkungsstärken tendenziell massereicher und geometrisch ausgedehnter als FBS sind. Sowohl FBS als auch FPS bilden DM-Kern- und DM-Wolkenlösungen, und wir stellen fest, dass sie entartete Ergebnisse produzieren können. Kernlösungen verdichten die Neutronensternkomponente und verringern ihre Gezeitenverformbarkeit, Wolkenlösungen haben den umgekehrten Effekt. Elektromagnetische Beobachtungen bestimmter wolkenartiger Konfigurationen scheinen die Buchdahl-Grenze zu verletzen. Es wurde herausgefunden, dass die Stärke der Selbstwechselwirkung sowohl die Masse als auch die Gezeitenverformbarkeit erheblich beeinflusst. Wir diskutieren Einschränkungen durch Beobachtungen und die Verbindung zu anomalen Beobachtungen. Außerdem zeigen wir, wie Modelle mit einer effektiven Zustandsgleichung mit der selbstkonsistenten Lösung der FBS zu vergleichen sind, und finden die Stärke der Selbstwechselwirkung, bei der beide Lösungen hinreichend konvergieren. \\

\textbf{Schlüsselwörter:} dunkle Materie, ultraleichte Bosonen, selbstwechselwirkende dunkle Materie, Neutronensterne, Zustandsgleichung, Gravitationswellen, Gezeitenverformbarkeit, Bosonen-Sterne, Proca-Sterne, Skalarfeld, Vektorfeld, Fermionen-Bosonen-Sterne, Fermionen-Proca-Sterne

\newpage
\markright{List of Publications}
\section{List of Publications}

This thesis is based on the publication \cite{Diedrichs:2023trk} listed below. A brief summary of it is provided and the author's contributions are highlighted. The author wants to note that this thesis was created in conjunction with the listed publication in the framework of the \textit{B09-project} of the \textit{CRC-TR 211} collaboration\footnote{URL: \href{https://crc-tr211.org/}{crc-tr211.org}.} .

\begin{itemize}
\item[\cite{Diedrichs:2023trk}] Robin Fynn Diedrichs, Niklas Becker, Cédric Jockel, Jan-Erik Christian, Laura Sagunski, and Jürgen Schaffner-Bielich \textit{Tidal Deformability of Fermion-Boson Stars: Neutron Stars Admixed with Ultra-Light Dark Matter}. ArXiv: \href{https://arxiv.org/abs/2303.04089}{[2303.04089]}, accepted for publication in Phys. Rev. D \vspace{2mm} \\
The tidal deformability of a neutron star admixed with dark matter is investigated. The dark matter is modelled as a massive, self-interacting, complex scalar field. The equations to compute the tidal deformability of the full Einstein-Hilbert-Klein-Gordon system self-consistently are derived and the influence of the scalar field mass and self-interaction strength on the total mass and tidal properties of the combined system are probed. Observational constraints are discussed and the connection to anomalous detections are highlighted. The fully self-consistent model is also compared to models which use an effective bosonic equation of state. The analysis is performed using the code developed by the authors, which is also linked in the work. The code is an ODE\footnote{ODE: ordinary differential equation}-solver with shooting-method solving capabilities. \\
This publication forms the basis for roughly half of this thesis. This affects Chapters \ref{sec:fermion-boson-stars:fermion-boson-stars}, \ref{sec:numerics:numerical-methods} and \ref{subsec:results:fermion-boson-stars}. The author of this thesis contributed to all parts of the publication, including analysis and draft. In particular the author contributed a large share of the code development used for the analysis. The model using the effective equation of state discussed in Section V as well as the numerical analysis was implemented and performed by the author of this thesis. All authors contributed to the published manuscript.
\item[\cite{Diedrichs-Becker-Jockel}] Robin Fynn Diedrichs, Niklas Becker, Cédric Jockel, Jan-Erik Christian, Laura Sagunski, and Jürgen Schaffner-Bielich \textit{FBS-Solver}. URL: \href{https://github.com/DMGW-Goethe/FBS-Solver}{Github/DMGW-Goethe/FBS-Solver} \vspace{2mm} \\
The code used in \cite{Diedrichs:2023trk} was in large parts developed by the author of this thesis. It is a solver for ordinary differential equations with shooting-method solving capabilities and is capable of integrating arbitrary fermion boson star configurations and computing global quantities such as their mass, radius and tidal deformability. The code was further expanded by the author of this thesis to solve for fermion boson stars which feature a vector field (fermion Proca stars). The code is explained in detail in Chapter \ref{sec:numerics:numerical-methods}. Fermion Proca stars (FPS) are the main subjects of Chapter \ref{subsec:fermion-boson-stars:fermion-proca-stars} and \ref{subsec:results:fermion-proca-stars}. Fermion Proca stars are not part of \cite{Diedrichs:2023trk}.
\end{itemize}

If not mentioned otherwise, the figures presented in this thesis were created by the author. Figures that were produced by collaborators are indicated explicitly. Some of the figures created by collaborators were recreated for this thesis to obtain a more consistent colour scheme.

\newpage
\markright{Introduction}
\section{Introduction}

Neutron stars (NS) are highly compact remnants of heavy stars. Due to their high densities, they are excellent laboratories to probe physics under extreme conditions such as nuclear matter at high densities or gravitational physics in the strong gravity regime. The nuclear matter equation of state (EOS) describes the relation between pressure and energy density of the matter found inside neutron stars and is an active field of study. The EOS is needed to close the Tolman-Oppenheimer-Volkoff (TOV) equations \cite{Tolman:1939jz,Oppenheimer:1939ne}, which describe the density distribution of a spherically symmetric static neutron star as well as the spacetime curvature self-consistently. Neutron stars are therefore useful systems to constrain the properties of nuclear matter at high densities. A significant constraint on the EOS is the ability to produce NS with masses of over two solar masses $2\,M_\odot$. There are numerous known NS in this mass range, the most massive NS known to date is PSR J0952\ensuremath{-}0607 with a mass of $M=2.35^{+0.17}_{-0.17}\,M_\odot$ \cite{Romani:2022jhd}. The lighter companion of the binary system observed in the GW190814 gravitational wave event \cite{LIGOScientific:2020zkf} was also proposed to be the heaviest NS with a mass of around $2.6\,M_\odot$, but there is some evidence \cite{Most:2020bba} that it might be the lightest known black hole instead. High maximum NS masses require stiff EOS, where the nuclear matter is difficult to compress and the energy density rises sharply with increasing pressure. Other constraints include the measurements of the pulsars PSR J0030+0451 \cite{Riley:2019yda} and J0740+6620 \cite{Riley:2021pdl} by the NICER telescope, which also favour a stiff EOS. In contrast, the gravitational wave event GW170817 \cite{Abbott:2018exr,Abbott:2018wiz} favours softer EOS which produce smaller NS that are more compact and more difficult to tidally disrupt. \\

It is additionally possible that dark matter (DM) could accumulate inside or around neutron stars in sufficient amounts to modify their properties such as mass, radius and tidal deformability. These properties have been measured using telescopes such as NICER and the gravitational wave detectors LIGO and Virgo. This allows us to probe the properties of dark matter. The current status of research suggests that dark matter is a particle, which is only interacting gravitationally and weakly with standard model (SM) particles. It is also abundant, constituting roughly $26.8\,\%$ of the total energy density of the universe \cite{Planck:2015fie}, making
DM an integral part of the standard model of cosmology ($\Lambda$CDM). There exist numerous candidates for dark matter particles. A possible contender is that DM consists of an additional bosonic field (scalar field or vector field), as was studied in \cite{Khlopov:1985jw,Ferreira:2020fam,Biswas:2022tcw,An:2023mvf}. Dark matter could arrange itself around neutron stars as a cloud or inside neutron stars as a core. Neutron stars with DM cores could form 
\begin{itemize}
\itemsep0em
\item[$1)$] from an initial DM 'seed' through accretion of baryonic matter \cite{Diedrichs:2023trk,Meliani:2016rfe,Ellis:2017jgp,Kamenetskaia:2022lbf},
\item[$2)$] through mergers of neutron stars and boson stars \cite{Diedrichs:2023trk},
\item[$3)$] through accretion of DM onto a NS and subsequent accumulation in the centre \cite{Diedrichs:2023trk,Brito:2015yga,Kouvaris:2010vv,Goldman:1989nd,Kouvaris:2007ay},
\item[$4)$] through the decay of standard model particles inside the neutron star into DM \cite{Baym:2018ljz,Motta:2018rxp,Motta:2018bil,Husain:2022bxl,Berryman:2022zic}.
\end{itemize}

The presence of dark matter inside the NS will then affect the observable quantities, thus making them indirect laboratories for DM properties such as the DM mass and self-interaction strength. Mixed systems of NS and bosonic fields are also motivated through modified theories of gravity, where scalar and vector fields appear. The bosonic fields could then form through superradiance \cite{Brito:2015oca,Siemonsen:2022ivj} or spontaneous scalarization \cite{Liebling:2012fv,Silva:2017uqg} (or spontaneous vectorization \cite{Silva:2021jya,Ramazanoglu:2017xbl}). Some of these systems are mathematically equivalent to NS with bosonic fields, making them also relevant for modified gravity. \\

In this work, we model dark matter as a minimally coupled complex bosonic field (scalar field or vector field), which only interacts with the standard model through gravity. First proposed by \cite{Henriques:1989ar,Henriques:1989ez}, the mixed systems of a bosonic field and fermionic NS matter are called fermion boson stars (FBS). Until now, FBS have been modelled using a classical complex scalar field for the bosonic component and an ideal fluid for the fermionic NS component. Mathematically, this system is realized as the Einstein-Hilbert-Klein-Gordon (EHKG) system of equations minimally coupled to a matter term. FBS have since been connected to current constraints on the mass and radii of NS by \cite{DiGiovanni:2021ejn} and to their dynamical properties by \cite{DiGiovanni:2020frc,DiGiovanni:2021vlu}. FBS are also related to boson stars \cite{Kaup:1968zz,Ruffini:1969qy} as they can be thought of as a boson star that coexists with a neutron star at the same location in space. Boson stars which are constructed from a complex vector field are called Proca stars \cite{Brito:2015pxa}. We are the first to combine Proca stars with neutron stars to mixed systems of fermions and a vector field, which we name fermion Proca stars (FPS). Fermion Proca stars are described mathematically using the Einstein-Proca (EP) equations coupled minimally to a matter term. We will construct equilibrium solutions of FPS and compute their properties such as mass and radius. \\

We additionally study the tidal properties of fermion boson stars (with a scalar field). The tidal deformability of FBS was first investigated by \cite{Nelson:2018xtr}, where the authors considered scalar bosonic DM in the mass range of $MeV$ to $GeV$. They constructed an effective EOS for the bosonic particles and used a two-fluid formalism where they modelled the FBS using modified TOV equations. This model was further investigated by \cite{Rutherford:2022xeb,Giangrandi:2022wht} and the tidal properties were computed. Other authors \cite{Karkevandi:2021ygv, Leung:2022wcf} studied scalar DM that has a quartic self-interaction potential. They used an effective EOS derived by \cite{Colpi:1986ye}, which is only valid for strong self-interactions. We will use the two-fluid model with the effective bosonic EOS to compare the results to the results obtained by solving the full EHKG equations self-consistently. \\

Our models for FBS and FPS are applicable for scalar and vector bosons with arbitrary potentials, including for strong and weak self-interactions. We construct equilibrium solutions of FBS and FPS and show the effects that the additional scalar or vector field has on the NS properties. We derive the equations necessary to compute the tidal deformability of FBS and solve them numerically. We subsequently present our results regarding the tidal deformability. We also derive the equations describing FPS and solve them numerically. We further derive an analytical bound on the vector field amplitude and find scaling relations between the metric and vector field components. We find that both FBS and FPS admit DM core and DM cloud solutions. Core solutions compactify the NS component and lower their tidal deformability of the combined system. The inverse was observed for cloud solutions. When only observing the fermionic component, some solutions would appear to violate the Buchdahl limit. Even small amounts of DM can significantly affect the properties of the combined system. The tidal deformability of FBS changes significantly for small DM masses and could be observed using current gravitational wave detectors. FPS tend to be more massive and geometrically larger than FBS for equal boson masses and self-interaction strengths. For a given measurement, this would favour larger vector DM masses (compared to scalar DM), as they produce smaller and less massive objects. We find a significant amount of degenerate solutions between different choices of FBS, FPS, the DM properties and the EOS. For different boson masses and DM-fractions, FPS and FBS can both be degenerate with each other and also be degenerate with pure NS with a different EOS. Using scaling relations for boson stars and Proca stars, we show that FBS and FPS are virtually indistinguishable if the boson masses differ by a factor of $1.671$ and have no self-interactions. We also confirmed the existence of FPS in higher modes which are stable under linear radial perturbations. \\

This thesis is structured as follows: In Chapter \ref{sec:background:theoretical-background} we introduce all relevant concepts needed to understand fermion boson stars and offer a wide overview of neutron stars, gravitational waves, the tidal deformability, dark matter and boson stars. In Chapter \ref{sec:fermion-boson-stars:fermion-boson-stars} we present the mathematical framework behind FBS and derive the equations of motion as well as the first-order perturbations needed to compute their tidal deformability. We also present the equations of motion of fermion Proca stars and derive some analytical bounds. In Chapter \ref{sec:numerics:numerical-methods} we focus on the algorithms used to solve the equations of motion and give an overview of the numerical code used for this work. In Chapter \ref{sec:results:results} we present our results regarding fermion boson stars with scalar fields and vector fields. In Chapter \ref{subsec:results:mass-radius-relations-and-tidal-deformability} we present mass-radius relations and the tidal deformability of FBS and compare it to observational constraints in Chapter \ref{subsec:results:comparison-to-observational-constraints}. In Chapter \ref{subsec:results:comparison-to-effecive-eos} we compare the EHKG solutions to the two-fluid model. In Chapter \ref{subsec:results:fermion-proca-stars} we present all results regarding fermion Proca stars including radial profiles (Chapter \ref{subsec:results:radial-profiles}), stability considerations and mass-radius relations (Chapter \ref{subsec:results:stability-and-MR-relations}) as well as a comparison of FPS with different EOS (Chapter \ref{subsec:results:study-of-higher-modes-and-different-eos}). Finally in Chapter \ref{sec:conclusions:conclusions-and-outlook} we summarize our findings and give an outlook on possible future research projects. \\
Throughout this work, we make use of the Einstein summation convention for tensors. Greek indices $\mu,\nu,\alpha,\beta$ run over all spacetime coordinates ($t,x,y,z$) and Latin indices $i,j,k,l$ run purely over spatial coordinates ($x,y,z$). If not specified otherwise, we use units in which the gravitational constant, the speed of light and the solar mass are $G = c = M_\odot = 1$. See also Appendix \ref{sec:appendix:units} for more information on the unit conventions. Prior knowledge of general relativity is assumed. While not strictly necessary, knowledge about quantum field theory and the standard model of particle physics is highly recommended to better understand the particle-physical implications behind our models.

%% file: Background.tex
\section{Theoretical Background} \label{sec:background:theoretical-background}

In this work, we study exotic compact objects known as fermion boson stars. However, before we start the analysis, we must first develop a solid theoretical groundwork and understanding of constituent concepts such as neutron stars (NS), gravitational waves (GW), dark matter (DM) and boson stars (BS). Chapter \ref{subsec:background:neutron-stars} provides a lightweight introduction to the most important properties of NS and is based upon the reviews \cite{Lattimer:2021emm,Burgio:2021vgk} and references therein. Following that, Chapter \ref{subsec:background:gravitational-waves-and-tidal-deformability} introduces the notion of gravitational waves, including their current uses and applications. The tidal deformability is also covered in detail. Next, in Chapter \ref{subsec:background:dark-matter}, the status of the search for dark matter is reviewed and a specific type of DM with properties interesting for this work, namely ultralight dark matter (ULDM), is especially highlighted (see \cite{Bertone:2016nfn,Baryakhtar:2022hbu,Hui:2021tkt} for reviews). This chapter is closed in Chapter \ref{subsec:background:boson-stars} by an introduction to boson stars including the theoretical description, properties and possible formation mechanisms.

%%%%%%%%%%%%%%%%%%%%%%%%%%%%%%%%%%%%%%%%%%%%%%%%%%%%%%%%%%%%%%%%%%%%%%%%%%%%%%%%%%%%%%%%%%%%%%%%%%%%%%%%%%%%%%%%%%%%

\subsection{Neutron Stars} \label{subsec:background:neutron-stars}

\begin{figure}[h]
	\centering
	\includegraphics[width=0.542\textwidth]{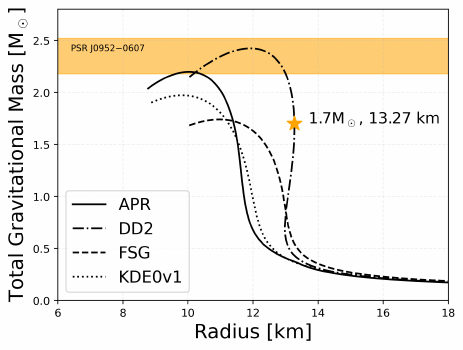}
	\includegraphics[width=0.45\textwidth]{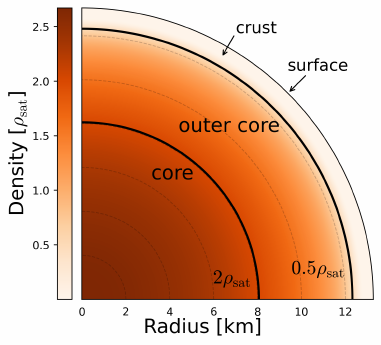}
	\caption{\textbf{Left panel:} The mass-radius relation for four different EOS (APR\cite{Schneider:2019vdm}, DD2\cite{Hempel:2009mc}, FSG\cite{Hempel:2009mc} and KDE0v1\cite{Schneider:2017tfi}). The DD2 EOS is a stiff EOS (the fluid is difficult to compress) and produces NS with larger radii than the softer APR EOS. The NS marked with a star ($\star$) has a mass of $M=1.7\,M_\odot$ and a radius of $R=13.27\,km$. The mass measurement of PSR J0952\ensuremath{-}0607 \cite{Romani:2022jhd} is shown by the orange band.
	\textbf{Right panel:} Inner structure of the NS marked with a star ($\star$) in the left panel. Densities are marked relative to the nuclear saturation density $\rho_{\mathrm{sat}} = 2.7 \e{14}\,g/cm^3$. Radii where the restmass density reaches $0.5\rho_{\mathrm{sat}}$ and $2\rho_{\mathrm{sat}}$ have been marked by black lines. The different inner phases are labelled accordingly.}
	\label{fig:background:neutron-stars:MR-diagram-and-inner-NS-structure-example}
\end{figure}

Neutron stars (NS) are highly compact and dense stellar remnants of heavy stars. They are often rapidly rotating and emit a jet, which in some cases aligns with our field of view such that the jet is observed as a regular pulse -- hence why the term pulsar (short for ''pulsating star'') is also used for these types of NS. These properties make NS excellent laboratories for physics under extreme conditions such as nuclear matter at high densities or gravitational physics in the strong gravity regime. They are formed in the aftermath of core-collapse supernova explosions of stars with masses of more than $8 M_\odot$ \cite{Janka:2012wk}. During stellar evolution, main-sequence stars burn hydrogen into increasingly heavier elements until the nuclear fusion reaction stops and the radiation pressure provided by the reaction is no longer sufficient to stabilize the star against its own gravity. Gravitational collapse ensues and the stellar core consisting of heavy elements is strongly compressed until the nuclear repulsion force repels the infalling matter, resulting in a violent supernova explosion. The highly compressed stellar core then radiates neutrinos at a high rate, through which it gradually cools down to temperatures where finite-temperature effects become largely unimportant \cite{Lattimer:2021emm}. What remains is a highly compact and neutron-rich object, with masses of roughly $1-2\,M_\odot$ and radii of around $10-15\,km$ \cite{Lattimer:2021emm}. \\
The structure of NS can be divided into the atmosphere, the crust, the outer core and the inner core \cite{Lattimer:2021emm,Schaffner-Bielich:2020psc} (also see the right panel of \autoref{fig:background:neutron-stars:MR-diagram-and-inner-NS-structure-example}). The atmosphere is a thin envelope of a few $cm$ in size with densities of $\rho \approx 1\,g/cm^3$, which consists of a mix of light and heavy nuclei. Below the surface in the upper crust, neutron-rich nuclei form a lattice which is permeated by a free-flowing degenerate electron gas \cite{Burgio:2021vgk}. With increasing depth, the density also increases. At around $4 \e{11}\,g/cm^3$, the lattice structure starts to gradually break up and nuclei become increasingly deformed. At roughly a quarter nuclear saturation density ($\rho_{\mathrm{sat}} = 2.7 \e{14}\,g/cm^3$), the neutron-rich nuclei start to arrange in macroscopic tube- and sheet-like structures, commonly referred to as nuclear pasta \cite{Lattimer:2021emm,Burgio:2021vgk}. The outer core starts at densities above $\rho \approx 0.5 \rho_{\mathrm{sat}}$. The nuclear matter forms a homogenous phase in beta-equilibrium consisting of mainly neutrons and a small fraction of protons, electrons and muons \cite{Lattimer:2021emm}. At even higher densities, the inner core starts. To date, this region of neutron stars is largely unknown, as it requires knowledge of the behaviour of nuclear matter at densities far above nuclear saturation density. \\
The composition and physical properties of matter inside NS cores is an active field of research. Models range from baryons with strange quarks (e.g. hyperons) \cite{Burgio:2021vgk} to deconfined quarks or phase transitions with colour-superconductive phases \cite{Alford:2007xm,Weissenborn:2011qu}. As such, researching the properties of NS is also linked to understanding strongly interacting matter at densities, where perturbative calculations using quantum chromodynamics (QCD) are not possible and where lattice-QCD simulations break down \cite{Philipsen:2021vgp}. \\
All these (proposed) microphysical properties of nuclear matter can be described macroscopically using the nuclear matter equation of state (EOS). The equation of state is a function  $P=P(e,T)$ which relates the pressure $P$ to the energy density $e$ and temperature $T$ of a fluid. In some cases, other quantities such as the electron fraction $Y_e$ may additionally be used to characterize the EOS. The search for the EOS at densities $\rho < \rho_{\mathrm{sat}}$ can be carried out on Earth using heavy-ion experiments \cite{Lattimer:2021emm}. Theoretically, efforts are ongoing to describe nuclear matter using chiral effective field theory ($\chi$EFT). $\chi$EFT is a framework to describe the interactions between nuclei while also taking into account their inner quark structure \cite{Tews:2020hgp}. This has enabled us to study the EOS up to $2\rho_{\mathrm{sat}}$ with $25\%$ theoretical uncertainty \cite{Lattimer:2021emm}. \\
The amount of different models for the EOS is vast and possibilities for different microphysical considerations are accordingly rich. One project aimed at aggregating EOS models is the online database CompOSE \cite{Typel:2013rza}. CompOSE allows to choose different EOS based in the microphysical properties and provides a code package to compute macroscopic quantities -- like pressure, temperature and energy density -- for each EOS. CompOSE was used in this work to obtain all the equations of state employed in the analysis. \\
Observationally, neutron stars are well established. One historic example is the pulsar inside the crab nebula \cite{Reifenstein:PhysRevLett.22.311,Cocke:1969tu}. Other NS detections include the Hulse-Taylor binary pulsar \cite{Hulse:1974eb} and the observation of the binary-NS merger GW170817 \cite{LIGOScientific:2017vwq,LIGOScientific:2017ync} by the LIGO collaboration. Some quantitative measurements of NS masses and radii have been taken of PSR J0030+0451 with $M=1.34^{+0.15}_{-0.16}\,M_\odot$ and $R=12.71^{+1.14}_{-1.19}\,km$ \cite{Riley:2019yda} and of J0740+6620 with $M=2.072^{+0.067}_{-0.066}\,M_\odot$ and $R=12.39^{+1.30}_{-0.98}\,km$ \cite{Riley:2021pdl} using the Neutron Star Interior Composition Explorer (NICER) telescope. The most massive NS known to date is PSR J0952\ensuremath{-}0607 with $M=2.35^{+0.17}_{-0.17}\,M_\odot$ \cite{Romani:2022jhd}. The lightest NS where mass and radius are known is HESS J1731\ensuremath{-}347 with $M=0.77^{+0.20}_{-0.17}\,M_\odot$ and $R=10.4^{+1.86}_{-0.78}\,km$ \cite{Doroshenko2022:aeq}. \\
\newline
Theoretically, neutron stars can be modelled as self-gravitating relativistic fluids. We can thus use general relativity (GR) and hydrodynamics to describe them (see e.g. \cite{Schaffner-Bielich:2020psc,Schutz:1985jx,Glendenning:1997wn,Misner:1973prb}). The associated action is the Einstein-Hilbert (EH) action that is minimally coupled to a term $\mathcal{L}_{m}$ describing the fluid. Minimally coupling means that there exist no interactions between gravity and the relativistic fluid outside of the back-reaction onto spacetime due to the fluid's energy-momentum content. The action assumes the form
\begin{align}
	S = \int_\mathcal{M} \sqrt{-g} \left( \frac{1}{2\kappa} R - \mathcal{L}_{m} \right)\, dx^4 \: , \label{eq:background:neutron-stars:einstein-hilbert-action}
\end{align}

where $R$ is the Ricci curvature scalar, $g$ is the determinant of the spacetime metric $\tensor{g}{_\mu_\nu}$, $\kappa = 4\pi G/c^4$ is a constant and the integral is performed over the whole spacetime manifold $\mathcal{M}$. The corresponding equations of motion can be obtained by computing the variation with respect to the inverse spacetime metric $\delta \tensor{g}{^\mu^\nu}$. One then obtains the Einstein equations (in units with $c=G=1$):
\begin{align}
	\tensor{G}{_\mu_\nu} = 8\pi \, \tensor{T}{_\mu_\nu} \: . \label{eq:background:neutron-stars:einstein-equation}
\end{align}

Here $\tensor{G}{_\mu_\nu} = \tensor{R}{_\mu_\nu} - \frac{1}{2} R\, \tensor{g}{_\mu_\nu}$ is the Einstein tensor, $\tensor{R}{_\mu_\nu}$ is the Ricci tensor and  $\tensor{T}{_\mu_\nu}$ is the energy-momentum tensor of the relativistic fluid. Formally, the energy-momentum tensor is defined through the relation
\begin{align}
	\tensor{T}{_\mu_\nu} := \frac{2}{\sqrt{-g}} \frac{\delta \mathcal{L}_{m}}{\delta \tensor{g}{^\mu^\nu}} \: . \label{eq:background:neutron-stars:definition-energy-momentum-tensor}
\end{align}
The first solution to the Einstein equations \eqref{eq:background:neutron-stars:einstein-equation} with a self-gravitating spherically symmetric static matter distribution was found by Tolman, Oppenheimer and Volkoff \cite{Tolman:1939jz,Oppenheimer:1939ne}. They assumed a spherically symmetric ansatz for the spacetime metric\footnote{Note that in the original works of Tolman \cite{Tolman:1939jz} and of Oppenheimer and Volkoff \cite{Oppenheimer:1939ne} a different convention was used. However, both conventions are quantitatively equivalent.}
\begin{align}
	ds^2 = \tensor{g}{_\mu_\nu} \tensor{dx}{^\mu} \tensor{dx}{^\nu} = - \alpha^2(r) dt^2 + a^2(r) dr^2 + r^2 d\theta^2 + r^2\sin^2(\theta) d\phi^2 \: , \label{eq:background:neutron-stars:metric-ansatz}
\end{align}

with the metric functions $g_{tt}=-\alpha^2(r)$, $g_{rr}=a^2(r)$, and for the relativistic fluid they assumed a perfect fluid with an energy-momentum tensor of the form
\begin{align}
	\tensor{T}{_\mu_\nu} = (e(r) + P(r)) \tensor{u}{_\mu} \tensor{u}{_\nu} + P(r) \tensor{g}{_\mu_\nu} \: . \label{eq:background:neutron-stars:energy-momentum-tensor-ideal-fluid}
\end{align}

$P$ and $e$ are the pressure and the energy density of the fluid respectively. $\tensor{u}{_\mu}$  is the four-velocity of the fluid. For a static fluid, the four-velocity takes the form
\begin{align}
\tensor{u}{^\mu} = \left(- \frac{1}{\alpha}, 0, 0, 0 \right) \:\: , \:\: \tensor{u}{_\mu} = (\alpha, 0, 0, 0) \: . \label{eq:background:neutron-stars:four-velocity-static-fluid}
\end{align}

In addition, the conservation of the energy-momentum tensor $ \tensor{T}{_\mu_\nu}$ (conservation of energy and momentum) and of the fluid flow $ \tensor{J}{^\mu} := \rho  \tensor{u}{^\mu}$ (conservation of restmass) hold. The restmass density $\rho$ is related to the energy density $e$ through $e=\rho(1+\epsilon)$, where $\epsilon$ is the internal energy. The relevant conservation equations are written as the covariant derivatives
\begin{align}
	\tensor{\nabla}{_\mu} \tensor{T}{^\mu^\nu} = 0 \:\: , \:\: \tensor{\nabla}{_\mu} \tensor{J}{^\mu} = 0  \: . \label{eq:background:neutron-stars:energy-momentum-and-restmass-conservation}
\end{align}

Combining the above expressions, one obtains the Tolman-Oppenheimer-Volkoff (TOV) equations:
\begin{subequations}
\begin{align}
	a' = \frac{da}{dr} &=  \frac{a}{2} \, \left[ \frac{(1-a^2)}{r} + 8\pi r a^2 e \; \right] \: , \label{eq:background:neutron-stars:TOV-equations-grr} \\
	\alpha' = \frac{d \alpha}{dr} &= \frac{\alpha}{2} \left[ \frac{(a^2 -1)}{r} + 8\pi r a^2 P \right] \: , \label{eq:background:neutron-stars:TOV-equations-gtt} \\
	P' = \frac{dP}{dr} &= - (e+P) \frac{\alpha'}{\alpha\,} \: . \label{eq:background:neutron-stars:TOV-equations-P}
\end{align}
\end{subequations}
This system of equations is closed using the EOS $P(e)$. The full TOV equations \eqref{eq:background:neutron-stars:TOV-equations-grr}-\eqref{eq:background:neutron-stars:TOV-equations-P} together with the EOS therefore provide a way of computing the properties of NS such as pressure $P(r)$, energy density $e(r)$, mass $M_\mathrm{tot}$ and radius $R$. The radius $R_\mathrm{NS}$ is defined as the point, where the pressure becomes zero: $P(R_\mathrm{NS}) = 0$. The total gravitational mass is defined by the expression
\begin{align}
	M_\mathrm{tot} := \frac{R_\mathrm{NS}}{2} \left( 1 - \frac{1}{(a(R_\mathrm{NS}))^2 } \right) \: . \label{eq:background:neutron-stars:definition-NS-total-gravitational-mass}
\end{align}

It makes use of the fact that outside of the non-rotating NS, the spacetime metric is equal to that of the Schwarzschild solution (Birkhoff-Jebsen theorem \cite{VojeJohansen:2005nd}). Another related quantity to consider is the total restmass $M_\mathrm{rm}$. It is obtained by integrating the conservation equation for the fluid flow $\tensor{J}{_\mu}$ \eqref{eq:background:neutron-stars:energy-momentum-and-restmass-conservation} over the spatial volume of the NS:
\begin{align}
	M_\mathrm{rm} := \int \sqrt{-g}\; \tensor{g}{^t^\mu} \tensor{J}{_\mu} dx^3 \: . \label{eq:background:neutron-stars:definition-NS-restmass}
\end{align}
The difference between the restmass and the total gravitational mass is the binding energy $E_B := M_\mathrm{rm} - M_\mathrm{tot}$. It can be used as a measure of stability of a given self-gravitating object and quantifies the potential energy released during collapse \cite{Lattimer:2021emm}. \\
A common way of representing NS is a mass-radius (MR) diagram. By graphing the masses and radii of neutron stars for different central densities $\rho_c$, one obtains a unique line for any given EOS. Since the EOS is intrinsically linked to the TOV equations \eqref{eq:background:neutron-stars:TOV-equations-grr}-\eqref{eq:background:neutron-stars:TOV-equations-P}, measurements of the masses and radii of different NS provide a way to constrain the EOS and thus the microphysical properties of neutron matter. Mass-radius relations of NS have therefore been intensively studied in the last years and are part of the standard analysis techniques when researching neutron stars. \autoref{fig:background:neutron-stars:MR-diagram-and-inner-NS-structure-example} shows an example of MR curves of NS with four different EOS.

Neutron stars, however, are not isolated systems. Many of them appear in close binary systems, where orbital decay due to the radiation of gravitational waves is relevant. The inspiral and merger of systems of neutron stars have shown to produce a rich phenomenology that can be observed using gravitational waves. Lately, NS have also come into the spotlight as a place to search for dark matter. For example, dark matter might accumulate in and around neutron stars, modifying their global properties such as mass and radius. Both of these topics will be discussed in the following chapters.

%%%%%%%%%%%%%%%%%%%%%%%%%%%%%%%%%%%%%%%%%%%%%%%%%%%%%%%%%%%%%%%%%%%%%%%%%%%%%%%%%%%%%%%%%%%%%%%%%%%%%%%%%%%%%%%%%%%%

\subsection{Gravitational Waves and Tidal Deformability} \label{subsec:background:gravitational-waves-and-tidal-deformability}

Another field of research closely related to neutron stars and compact objects is the study of gravitational waves (GW). Gravitational waves are ripples in spacetime, which propagate at the speed of light (for detailed introductions see e.g. \cite{Misner:1973prb,Maggiore:2007ulw}). Since their first detection in 2015 \cite{LIGOScientific:2016aoc}, GW have become an important tool for probing gravity in the strong field regime \cite{Krishnendu:2021fga,Pretorius:2023hwx}, testing modifications of gravity \cite{Shankaranarayanan:2022wbx}, cosmology \cite{Caprini:2018mtu,Christensen:2018iqi} or searching for dark matter \cite{Diedrichs:2023trk,Cardoso:2019rvt,Liddle:1993fq,Kamionkowski:1999qc}. \\
Mathematically, GW arise from the linearized Einstein equations as small perturbations $\tensor{h}{_\mu_\nu}$ (with $|\tensor{h}{_\mu_\nu}| \ll 1$) on top of an approximately flat background $\approx \tensor{\eta}{_\mu_\nu}$. The linearized Einstein equations (to linear order in the perturbation $\tensor{h}{_\mu_\nu}$) read \cite{Misner:1973prb,Maggiore:2007ulw}
\begin{align}
	\square \tensor{h}{_\mu_\nu} = - 4 \kappa \, \tensor{T}{_\mu_\nu} \:\: (\textrm{inside a source}) \:\:\:\: , \:\:\:\: \square \tensor{h}{_\mu_\nu} = 0 \:\: (\textrm{in vacuum}) \: . \label{eq:background:gravitational-waves-and-tidal-deformability:linearized-einstein-equations}
\end{align}

\begin{figure}[h]
	\centering
	\includegraphics[width=0.7\textwidth]{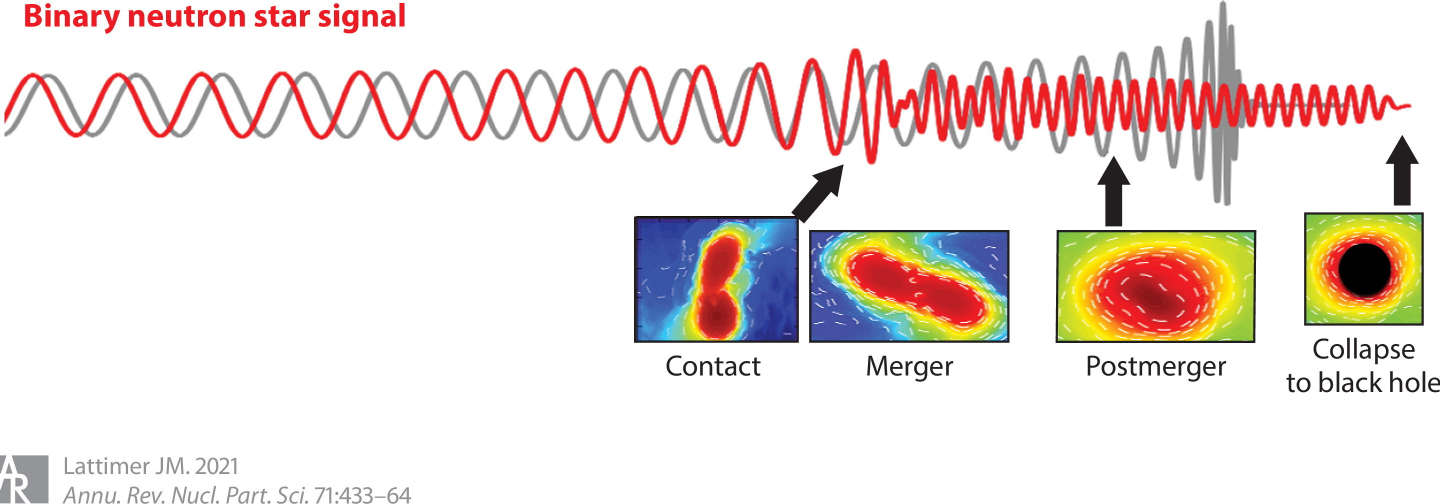}
	\caption{Illustration of the GW signal of a merger of a NS-NS binary (red) and a BH-BH binary (grey). Both systems have equal chirp-masses. The different merger phases are labelled accordingly. The figure was taken from \cite{Lattimer:2021emm} in accordance with their ''Creative Commons Attribution 4.0 International'' License. Figure originally provided by T. Dietrich.}
	\label{fig:background:gravitational-waves-and-tidal-deformability:BNS-BBH-merger-signal}
\end{figure}

In vacuum, equation \eqref{eq:background:gravitational-waves-and-tidal-deformability:linearized-einstein-equations} has the form of a wave equation. It admits -- in a suitable gauge -- planar wave solutions, which propagate with the speed of light and are polarized along two linearly independent directions (degrees of freedom). These polarization modes are commonly called ''plus'' and ''cross'' polarization of the gravitational wave. A full GW signal thus consists of an amplitude, a phase, and the GW polarization. \\
The sources of gravitational waves are plentiful \cite{Maggiore:2007ulw}: from two objects in a binary system, rotating neutron stars, to cosmological sources such as phase transitions in the early universe \cite{Caprini:2018mtu,Mazumdar:2018dfl}. But in general, any accelerated object in a gravitational field with a quadrupole moment will emit GW. In this work, we focus on astrophysical sources and specifically on binary compact objects such as black holes (BH), neutron stars (NS) and exotic compact objects close in size to BH and NS. A typical merger event will have three main phases (see \autoref{fig:background:gravitational-waves-and-tidal-deformability:BNS-BBH-merger-signal}): The inspiral phase, the merger phase and the post-merger phase. The merger and post-merger phases are to date only accessible using numerical simulations \footnote{Although note that analytic calculations have been done of post-merger black holes using black-hole perturbation theory \cite{LeTiec:2014oez}.}. 

The inspiral phase is modelled in the context of the linearized theory of gravity \eqref{eq:background:gravitational-waves-and-tidal-deformability:linearized-einstein-equations}. To that end, it is assumed that the gravitational field produced by the sources is sufficiently weak and that the velocities are small compared to the speed of light $v/c \ll 1$. Then, the background metric can be assumed as approximately flat and Newtonian forces can be used to compute the motion of the inspiralling objects. These considerations lead to the quadrupole formula \cite{Maggiore:2007ulw}, which describes the gravitational radiation emitted by a source with a quadrupole moment (e.g. a binary system of compact objects):
\begin{align}
P = \frac{dE}{dt} = \frac{G}{5c^5} \left< \tensor{\dddot{Q}}{_i_j} \tensor{\dddot{Q}}{^i^j} \right> \: . \label{eq:background:gravitational-waves-and-tidal-deformability:quadrupole-formula}
\end{align}

\begin{figure}[h]
	\centering
	\includegraphics[width=0.395\textwidth]{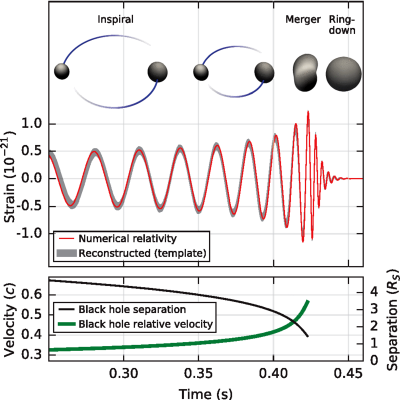}
	\hspace{0.5cm}
	\includegraphics[width=0.51\textwidth]{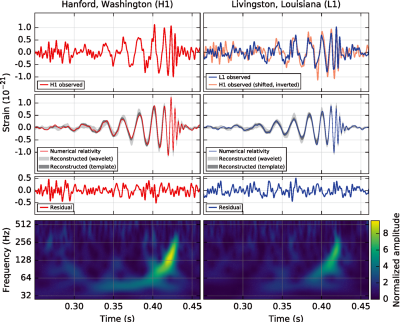}
	\caption{\textbf{Left panel:} GW signal amplitude over time for the inspiral of binary BH, including velocity (in units of $c$), separation (in units of the Schwarzschild radius) and images of numerical models of the BH event-horizons during coalescence. The red line was obtained using a numerical relativity simulation and the grey band corresponds to GW signal predictions using PN expansions.
	\textbf{Right panel:} Measurements of the GW event GW150914 \cite{LIGOScientific:2016aoc} as observed by the two LIGO interferometers in 2015. The upper panels show the measured amplitude. The next panels show waveform estimates using numerical simulations (red) and using PN expansions (grey). The residuals between the measurement and numerical simulations are shown below. The bottom panels show the measured GW frequencies over time.
	Both figures were taken from \cite{LIGOScientific:2016aoc} in accordance with their ''Creative Commons Attribution 3.0'' License.}
	\label{fig:background:gravitational-waves-and-tidal-deformability:LIGO-BBH-signal-template}
\end{figure}

Here, $\tensor{Q}{_i_j}$ is the quadrupole tensor of the source and the three dots denote the third time-derivative. Corresponding expressions for the radiated angular and linear momentum also exist \cite{Maggiore:2007ulw}. Equation \eqref{eq:background:gravitational-waves-and-tidal-deformability:quadrupole-formula} works well for sources in wide orbits with moderate orbital frequencies and has been confirmed to high accuracy using e.g. the orbital decay of the Hulse-Taylor pulsar \cite{Hulse:1974eb,Weisberg:2004hi}. To extract the GW signal of close binary objects moving at significant fractions of the speed of light, it has become necessary to go beyond the quadrupole formula \eqref{eq:background:gravitational-waves-and-tidal-deformability:quadrupole-formula}. This can be done using the post-Newtonian (PN) formalism \cite{Maggiore:2007ulw}. The PN formalism aims to include general-relativistic (GR) corrections so that the movement of objects can be described as point-masses ($m_1$, $m_2$, with $M:=m_1+m_2$) using Newtonian dynamics. In the case of circular orbits, corrections to the change of the orbital frequency $\omega$ (note that orbital frequency and velocity are related through $v^3 = GM\omega$) have been computed, yielding \cite{Buonanno:2002fy}:
\begin{align}
\frac{\dot{\omega}}{\omega^2} = \frac{96}{5} \left( \frac{v}{c} \right)^5 \frac{m_1 m_2}{(m_1 + m_2)^2} \sum_{k=0} \, \omega_{(k/2)PN}  \left( \frac{v}{c} \right)^{k} \: . \label{eq:background:gravitational-waves-and-tidal-deformability:post-newtonian-expansion}
\end{align}
The coefficients $\omega_{(k/2)PN}$ include the GR corrections of $(k/2)$PN order. The $0$PN order corresponds to Newtonian gravity. At increasingly higher PN order, additional terms enter the PN expansion. For example, the mass-ratio of the two binary objects enters at $1$PN order; the spin and angular momentum enter at $1.5$PN order. For binary neutron stars, finite-size effects such as the deformation of the NS due to tidal forces enter the PN expansion at $5$PN order \cite{JimenezForteza:2018rwr}. Using increasingly higher PN orders, the inspiral can be accurately modelled up to times close before the merger phase. The GW energy flux \eqref{eq:background:gravitational-waves-and-tidal-deformability:quadrupole-formula} for non-spinning objects has been computed up to $4.5$PN order \cite{Blanchet:2023soy,Blanchet:2023bwj,Blanchet:2023sbv}. The GW phase is known to even higher orders such as $7.5$PN, where effects related to spin-tidal coupling of rotating NS enter \cite{JimenezForteza:2018rwr}\footnote{The above list of currently known PN orders and effects should not be regarded as exhaustive. For more information, the interested reader is referred to the cited sources and references therein.}. In practice, the PN formalism is used with the desired degree of accuracy (PN orders) and a large amount of GW waveforms are computed. These calculated waveforms are then used as templates to compare to real measurements of GW (see \cite{Buonanno:2009zt} for a comparison of different PN expansion methods). The parameters of the inspiralling objects such as mass, spin, direction of the rotational axis and tidal deformability are encoded in parameters in the PN expansion. They can then be inferred by comparing the calculated waveforms to the measured data, as seen in \autoref{fig:background:gravitational-waves-and-tidal-deformability:LIGO-BBH-signal-template} (also see \cite{LIGOScientific:2016aoc} for more information).

The tidal deformability is a measure of how easily an object can be deformed by an external tidal field. It also appears as a parameter in the post-Newtonian expansion \eqref{eq:background:gravitational-waves-and-tidal-deformability:post-newtonian-expansion} at $5$PN order. This makes it interesting in the context of neutron stars and the search for the EOS since different EOS lead to different internal matter distributions, masses and radii of the NS and thus also alter the tidal properties. \\
To obtain the tidal deformability, one has to consider the effects of an external tidal field $\tensor{\mathcal{E}}{_i_j}$ on a spherically symmetric neutron star (we hereafter follow the procedures used by \cite{Diedrichs:2023trk,Hinderer:2007mb,Thorne:1997kt}). This tidal field will then induce a quadrupolar moment $\tensor{Q}{_i_j}$ in the neutron star. For a distant and static tidal field, the induced quadrupolar moment will be proportional to the field such that $\tensor{Q}{_i_j} = - \lambda_{\mathrm{tidal}} \tensor{\mathcal{E}}{_i_j}$ \footnote{The tidal deformability $\lambda_{\mathrm{tidal}}$ is related to the tidal Love number by $k_2 = \frac{3}{2}G \lambda_{\mathrm{tidal}} R^{-5}$.}. This approach is therefore also relevant for inspiralling binary NS since, in the large-distance limit, the orbital time scales are larger than the time needed for the NS to adapt its internal structure to the external tidal field (adiabatic limit). The induced quadrupolar moment modifies the $\tensor{g}{_t_t}$-component of the metric. At leading order in the asymptotic rest frame of the NS and at large radii, it can be written as
\begin{align}
	\tensor{g}{_t_t} = - \left( 1 - \frac{2 M}{r} \right) - \tensor{\mathcal{E}}{_i_j} \tensor{x}{^i} \tensor{x}{^j} \left( 1 + \frac{3 \lambda_{\mathrm{tidal}}}{r^5} \right) \: , \label{eq:background:gravitational-waves-and-tidal-deformability:perturbed-metric-external-tidal-field}
\end{align}

where $M$ is the total gravitational mass of the neutron star and the $\tensor{x}{^i}$ are the position vectors in a Cartesian coordinate system with $\tensor{x}{^i} \tensor{x}{_i} = r^2$. The exact magnitude of this perturbation can be computed using linear perturbations from the Einstein equations. Thereby we consider a small perturbation $\tensor{h}{_\mu_\nu}$ on top of the unperturbed metric $\tensor{\overline{g}}{_\mu_\nu}$ such that
\begin{align}
	\tensor{g}{_\mu_\nu} = \tensor{\overline{g}}{_\mu_\nu} + \tensor{h}{_\mu_\nu} \: . \label{eq:background:gravitational-waves-and-tidal-deformability:perturbed-metric-general-formula}
\end{align}

For the static, even-parity and quadrupolar ($l=2$) metric perturbations in the Regge-Wheeler gauge, $\tensor{h}{_\mu_\nu}$ can be written in terms of the spherical harmonic $Y_{20}(\theta,\phi)$ and radial functions $H_0, H_2, K$ which describe the radial dependence of the perturbed metric components:
\begin{align}
	\tensor{h}{_\mu_\nu} = Y_{20}(\theta,\phi) \times \mathrm{diag}\left( - \alpha^2(r) H_0(r),\: a^2(r) H_2(r),\: r^2 K(r),\: r^2 \sin^2(\theta) K(r) \right) \: . \label{eq:background:gravitational-waves-and-tidal-deformability:perturbed-metric-small-tidal-perturbation}
\end{align}
The tidally perturbed metric \eqref{eq:background:gravitational-waves-and-tidal-deformability:perturbed-metric-general-formula} is then inserted into the perturbed Einstein equations $\delta \tensor{G}{_\mu_\nu} = 8\pi \delta\tensor{T}{_\mu_\nu}$, which is expanded to linear order in $\tensor{h}{_\mu_\nu}$. The perturbed energy-momentum tensor is given by $\delta\tensor{T}{_\mu_\nu} = \mathrm{diag}\left( -\delta P/c_s^2,\: \delta P,\: \delta P,\: \delta P \right)$, where we used the relationship $\delta e = \delta P\, \partial e/\partial P = \delta P /c_s^2$ between the energy density $e$, pressure $P$ and the local speed of sound in the medium $c_s$ \cite{Diedrichs:2023trk}. Through a series of algebraic manipulations of the perturbed Einstein equations, one then arrives at $H_2 (r) = - H_0(r)$ and finally one obtains an ordinary differential equation for $H_0(r)$:
\begin{align}
    H_0'' = \left[ \frac{a'}{a} - \frac{\alpha'} {\alpha} - \frac{2}{r}\right]  H_0' +  \left[ - 2 \frac{\alpha''}{\alpha} + 2 \frac{\alpha' a'}{ \alpha  a} + 4 \frac{\alpha'^2}{\alpha^2} - \frac{a'}{ra} \left( 3+\frac{1}{c_s^2} \right) - \frac{\alpha'}{r\alpha} \left( 7+ \frac{1}{c_s^2} \right) + 6 \frac{a^2}{r^2} \right] H_0 \: . \label{eq:background:gravitational-waves-and-tidal-deformability:perturbed-H-parameter-1-H-ode}
\end{align}

The primes denote a derivative with respect to the radial coordinate. The second derivative of the metric component $\alpha''$ is given by
\begin{align}
    \alpha'' =\left[ 4 \pi r a^2 P + \frac{a^2 - 1}{2r} \right] \alpha' + \left[ 4 \pi r \left( 2 P a a' + a^2 P' \right) + 4 \pi a^2 P + \frac{aa'}{r} + \frac{1 - a^2}{2r^2} \right] \alpha \: . \label{eq:background:gravitational-waves-and-tidal-deformability:perturbed-H-parameter-2-d2adr2-ode}
\end{align}

Equations \eqref{eq:background:gravitational-waves-and-tidal-deformability:perturbed-H-parameter-1-H-ode} and \eqref{eq:background:gravitational-waves-and-tidal-deformability:perturbed-H-parameter-2-d2adr2-ode} are the same as the ones used in \cite{Diedrichs:2023trk}. However, they were changed to a different convention for the spacetime metric so that the notation is consistent within this work. The above equations can be simplified for radii outside of the neutron star, which leads to
\begin{align}
	H_0'' + \left( \frac{2}{r} + a^2 \frac{2 M}{r^2} \right) H_0' - \left( \frac{6 a^2}{r^2} + a^4 \frac{4 M^2}{r^4} \right) H_0 = 0 \: . \label{eq:background:gravitational-waves-and-tidal-deformability:perturbed-H-parameter-1-H-ode-ourside-of-source}
\end{align}

This equation \eqref{eq:background:gravitational-waves-and-tidal-deformability:perturbed-H-parameter-1-H-ode-ourside-of-source} has a solution in terms of the associated Legendre polynomials (see \cite{Diedrichs:2023trk,Hinderer:2007mb,Thorne:1997kt}), which can be expanded in $r/M$ and matched to the metric component \eqref{eq:background:gravitational-waves-and-tidal-deformability:perturbed-metric-external-tidal-field} to obtain an expression for the tidal deformability:
\begin{align}
	\lambda_\mathrm{tidal} =& \frac{16}{15} M^5 (1 - 2 C)^2 [2 + 2 C (y - 1) - y] \times \{3 (1 - C)^2 [2 - y + 2 C (y - 1)] \log (1 - 2 C) \nonumber \\
    &+ 2 C [6 - 3y + 3 C (5 y - 8)] + 4 C^3 [13 - 11 y + C (3y - 2) + 2 C^2 (1 + y)] \}^{-1} \: . \label{eq:background:gravitational-waves-and-tidal-deformability:tidal-deformablity-equation}
\end{align}

Here, $C:= M/R$ is the NS compactness. The parameter $y = R\, H'_0(R)/H_0(R)$ can be obtained by integrating equation \eqref{eq:background:gravitational-waves-and-tidal-deformability:perturbed-H-parameter-1-H-ode}, together with the TOV equations \eqref{eq:background:neutron-stars:TOV-equations-grr}-\eqref{eq:background:neutron-stars:TOV-equations-P}, from the centre of the star at $r=0$ to the surface at $r=R$. It is also useful to define the dimensionless tidal deformability $\Lambda_\mathrm{tidal} := \lambda_\mathrm{tidal} / M^5$. \\
Knowledge about the tidal deformability can now be used to gain information about the NS properties, especially the radius. Since different EOS produce NS with different radii (see \autoref{fig:background:neutron-stars:MR-diagram-and-inner-NS-structure-example}), the tidal deformability is also a useful tool to constrain the nuclear matter EOS. For example, a given EOS is said to be stiff, if the NS matter is not easily compressible. This EOS subsequently produces NS with larger radii and the NS will have a larger tidal deformability since it is more easily affected by external gravitational forces. The inverse is true for soft EOS. This can be tested using kilonova events, where a neutron star gets tidally disrupted by a black hole. From the amount and the brightness of the ejected material, it is possible to infer the stiffness of the EOS and in turn obtain estimates for the NS radius (see e.g. \cite{Gupta:2023evt}). \\
The tidal deformability can also be used in the search for dark matter. If it accumulates inside of a NS in sufficient amounts, dark matter can change the tidal properties of the NS (e.g. by compactifying it). One can compute the tidal deformability using different dark matter models and then use observations to indirectly shed light on dark matter. Due to the still unknown nature of dark matter, the amount and variety of models is vast, which is why the next chapter will provide an overview of different models and gradually introduce the dark matter models most relevant to this work.

\subsection{Dark Matter} \label{subsec:background:dark-matter}

In this chapter, we review the history and current status of dark matter (DM) research relevant to this work. Even though dark matter has a long history as an observational science, its properties remain largely unknown. The current status of research suggests that dark matter is a particle, that is only interacting gravitationally and weakly with standard model (SM) particles. Most importantly, it is invisible through electromagnetic radiation, hence the name. Large-scale structure formation in the universe suggests that DM is mostly cold, i.e. slowly moving \cite{Bertone:2016nfn,White:1983fcs,Blumenthal:1984bp,Davis:1985rj}. It is also abundant, constituting roughly $26.8\%$ of the total energy density of the universe \cite{Planck:2015fie}, making DM an integral part in the standard model of cosmology ($\Lambda$CDM). The following historic overview is loosely based on \cite{Bertone:2016nfn}, the subsequent discussion of ultralight dark matter (ULDM) and wave dark matter (WDM) is based upon \cite{Baryakhtar:2022hbu} and \cite{Hui:2021tkt}. \\

\begin{figure}[h]
	\centering
	\includegraphics[width=0.4535\textwidth]{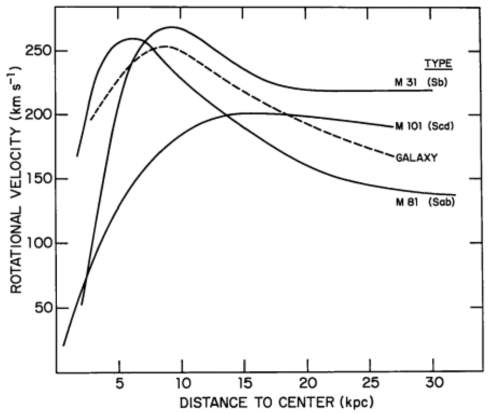} %width=0.4535
	%\hspace{0.5cm}
	\includegraphics[width=0.534\textwidth]{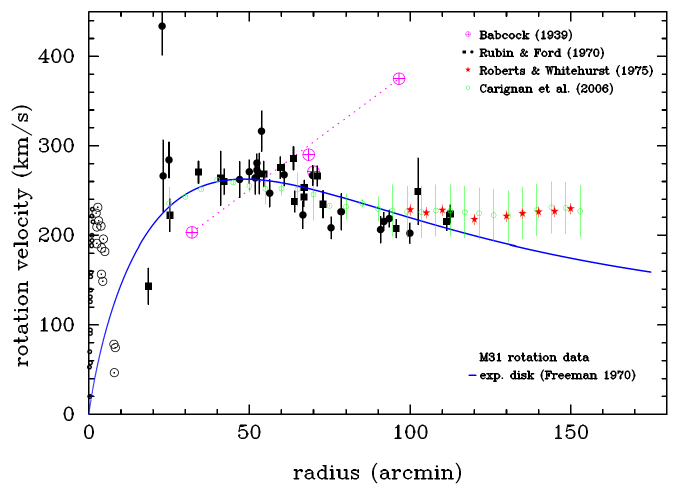} %width=0.534
	\caption{\textbf{Left panel:} Rotation curves of the galaxies M31, M101, M81 and the Milky Way galaxy (denoted by ''GALAXY''). The data for M31, M101 and M81 was originally published by \cite{Roberts:1973A&A....26..483R}, the data for the Milky Way galaxy was included by the authors of \cite{Bertone:2016nfn}.
	\textbf{Right panel:} The rotation curve of the galaxy M31 as measured using emission lines and the $21\,cm$ line of hydrogen. The measurements come from \cite{Babcock1939TheRO} (pink circles), \cite{Rubin:1970ApJ...159..379R} (black circles and squares), \cite{Roberts1975TheRC} (red stars) and \cite{Carignan:2006bm} (green circles). The blue solid line corresponds to the rotation curve of the ''exponential disk'' model from \cite{Freeman:1970ApJ...160..811F}. Figure was originally produced by Albert Bosma.
	Both figures were taken from \cite{Bertone:2016nfn}.}
	\label{fig:background:dark-matter:galactic-rotation-curves}
\end{figure}

The history of dark matter goes back more than a few centuries \cite{Bertone:2016nfn}. However, the modern history of DM kicked off in the beginning of the last century with early models of the galactic neighbourhood and considerations about the local matter density in and around the solar system by Poincaré, Öpik, Kapteyn, Oort and others \cite{Bertone:2016nfn,Poincare:1906PA.....14..475P,Oort:1932BAN.....6..249O}. It was concluded that the unseen ''dark'' matter was likely small compared to visible matter. At that time, dark matter was still considered to consist of faint objects such as planets, low-mass stars, cold interstellar gas and asteroids, which could not yet be observed using telescopes. Although the past notion of dark matter does not match the modern understanding, concepts like the local DM density are important to this day (see \cite{Read:2014qva} for a modern review), for example in DM direct-detection experiments \cite{Bertone:2016nfn}. \\
In 1933, Zwicky studied the velocity dispersion of galaxies in the Coma cluster and found that the observed velocities corresponded to a mass that is multiple times larger than the visible matter in the cluster \cite{Zwicky:1933gu}. Over time, explanations of DM such as clouds of gas and faint stars were gradually ruled out \cite{Bertone:2016nfn,Meekins:1971-wi}. Historically, the study of galactic rotation curves -- i.e. the angular velocity profile of stars and gas in a galaxy with respect to the distance to the galactic centre -- made the largest contribution to establish the idea that galaxies contain large amounts of dark matter \cite{Bertone:2016nfn}. Measurements of galactic rotation curves were first performed in 1917 using the spectral lines of the Andromeda galaxy \cite{Bertone:2016nfn}. In the following two decades, these and similar measurements were used to infer several galactic masses \cite{Bertone:2016nfn,Oort:1932BAN.....6..249O,Hubble:1926yw,Lundmark:1930}. The quality of rotation curves drastically increased with the discovery of the $21\,cm$ line of hydrogen in 1951 \cite{Bertone:2016nfn}. Measurements using the $21\,cm$ line culminated in the first accurate measurements of rotation curves \cite{van_de_Hulst:1957BAN....14....1V}, with a clearly flat rotation profile at large galactic radii \cite{Rogstad:1972ApJ...176..315R,Roberts:1973A&A....26..483R}. With each new measurement, it became increasingly clear that most galaxies feature flat rotation curves which extend further out than the optical size of the galaxies. Examples of measured rotation curves can be seen in \autoref{fig:background:dark-matter:galactic-rotation-curves}. Purely from the visible matter, the rotation velocities are expected to decrease with increasing distance from the galactic centre. The observed flat rotation curves subsequently imply the existence of additional gravitating mass in the outer regions of most galaxies \cite{Roberts:1973A&A....26..483R}. This discrepancy between expected and observed velocity came to be known as the ''missing mass problem'' (see \cite{Faber1979} for a contemporary review).

As the amount of evidence for the existence of dark matter increased, physicists gradually began to investigate the nature of dark matter. The most natural explanation attempt at that time was that DM might consist of faint macroscopic compact objects of planetary or stellar sizes \cite{Bertone:2016nfn} (interstellar and intergalactic gas was already ruled out previously \cite{Bertone:2016nfn}). These so-called MACHOS (massive astrophysical compact halo objects) were conjectured to be abundant in the outer regions of galaxies, making up the missing mass needed to explain the measured rotation curves (see \autoref{fig:background:dark-matter:galactic-rotation-curves}). There are a number of candidate objects for MACHOS, but the most popular ones are massive planets, brown dwarves and primordial black holes (PBH) -- BH of varying sizes produced in the early universe. The main way of detecting MACHOS is via gravitational microlensing \cite{Bertone:2016nfn}. Gravitational microlensing makes use of the fact that light bends around massive objects. If a massive object passes between the observer and a distant star, the light is magnified by the gravitational lens. The amount of magnification depends on the mass of the lensing object. Using the Large Magellanic Cloud as a background, a number of surveys were performed to measure the frequency of microlensing events, thus obtaining insight into the abundance of MACHOS in the Milky Way DM halo. It was found that MACHOS in the sub-solar mass scale could at most account for $8\%$ of the needed dark matter mass \cite{Lasserre:2000xw,EROS-2:2006ryy}. The current understanding is that MACHOS are excluded on most mass scales \cite{Lasserre:2000xw,EROS-2:2006ryy} (see \cite{Carr:2020gox,Carr:2020xqk,Green:2020jor} for constraints on PBH specifically) as dark matter candidates, since they can only account for a small fraction of the needed mass. In fact, through the measurement of the cosmic microwave background, it was later found that baryonic matter -- that is all visible matter consisting of standard-model particles \footnote{However, note that primordial black holes are often not counted as baryonic matter \cite{Bertone:2016nfn}.} -- can only account for less than $20\%$ of the matter-energy density in the universe (i.e. all dark matter and visible matter combined) \cite{Planck:2015fie}. \\
Another popular approach was -- and in some parts still is -- to avoid the postulation of additional invisible matter by modifying how gravity works on large scales instead \cite{Bertone:2016nfn}. The best known framework is that of modified Newtonian dynamics (MOND), which changes the mechanisms behind the gravitational force at large distances (see \cite{Famaey:2011kh} for a review). MOND initially became popular because it was able to accurately explain the rotation curves of a large amount of galaxies without the need of additional dark matter. Generalizing the MOND framework to be consistent with other predictions such as the equivalence principle and gravitational lensing proved to be more difficult and was only later achieved using the new framework of Tensor-Vector-Scalar (TeVeS) gravity \cite{Bekenstein:2004ne}. TeVeS gravity however contains a significant amount of new fields and parameters, losing the simplicity that MOND offered initially. In 2006, modified theories of gravity came under pressure due to the measurement of gravitational lensing of the Bullet Cluster \cite{Clowe:2006eq}. The authors found that the mass distribution due to gravitational lensing did not match the observed matter distribution of visible matter, bringing empirical evidence of DM as an independent massive matter distribution, rather than being a consequence of modified gravity. Furthermore, some galaxies without -- or with negligible amounts of -- dark matter were found \cite{vanDokkum:2018vup,vanDokkum:2022zdd} (i.e. the star's motion could be explained solely by the visible matter). In total, these findings disfavour the hypothesis that dark matter can be explained by modifying gravity alone. \\
With baryonic matter being largely ruled out as an explanation of dark matter, and modified gravity failing to explain gravitational lensing in the bullet cluster as well as the apparent lack of dark matter in some galaxies, the notion of dark matter as a collection of subatomic particles has received an increasing amount of traction. As most alternatives have been ruled out, this view has become the leading paradigm regarding dark matter \cite{Bertone:2016nfn}. Standard model particles, such as neutrinos, were ruled out quite early as they were not capable to produce the necessary density abundances needed to explain all DM observations \cite{Bertone:2016nfn} and failed to reproduce the observed large-scale structures in the universe in simulations \cite{White:1983fcs}. One of the reasons neutrinos were ruled out is their high velocity compared to the speed of light. The simulations suggest that dark matter must be slow-moving (also referred to as ''cold dark matter''), to correctly reproduce the structure formation in the universe \cite{Bertone:2016nfn,White:1983fcs}. \\

\begin{figure}[h]
	\centering
	\includegraphics[width=0.9\textwidth]{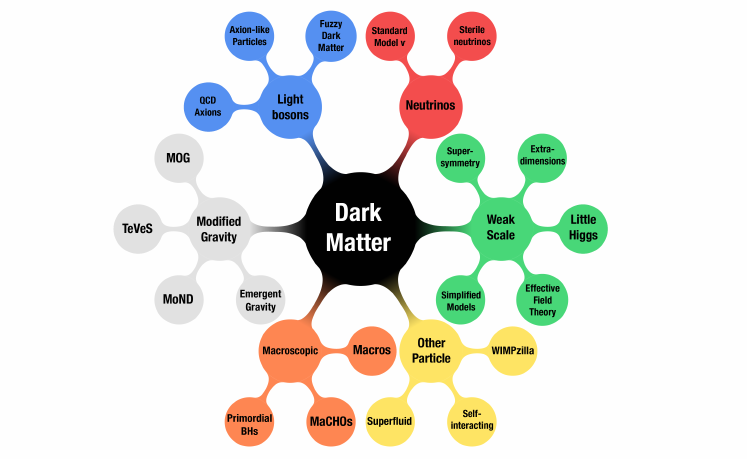}
	\caption{Visualization of possible solutions to the dark matter problem. The main strains of the search for DM are shown: standard model DM candidates, extensions of the standard model and BSM physics, modified theories of gravity, macroscopic dark objects and other exotic particles. The figure was originally produced by \cite{Bertone:2018krk} and was modified by Ana Lopes \cite{Lopes:2020cern}. For details regarding every specific model mentioned, see \cite{Bertone:2018krk} and references therein.}
	\label{fig:background:dark-matter:dark-matter-problem-visualization}
\end{figure}

\begin{figure}[h]
	\centering
	\includegraphics[width=0.85\textwidth]{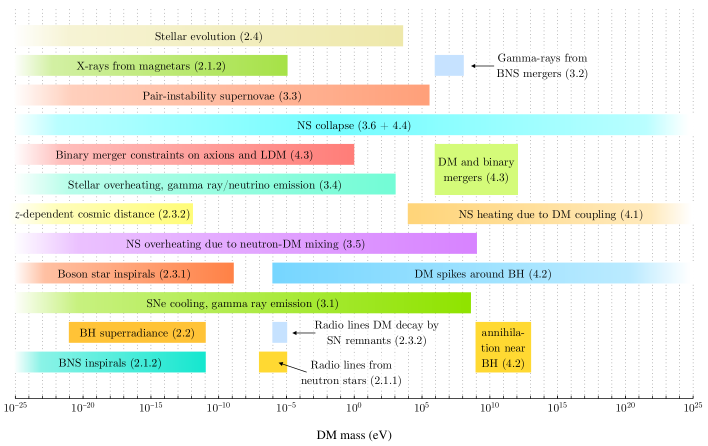}
	\caption{A summary of the dark matter mass ranges probed by different astrophysical systems. The numbers in parentheses refer to the sections in \cite{Baryakhtar:2022hbu}. The figure was taken from \cite{Baryakhtar:2022hbu}.}
	\label{fig:background:dark-matter:dark-matter-mass-ranges-summary}
\end{figure}

The search for dark matter has thus increasingly taken the form of the search for physics beyond the standard model (BSM) of particle physics (see \autoref{fig:background:dark-matter:dark-matter-problem-visualization}). Not much is known about the properties of the supposed DM particle, apart from that it must have a neutral electric charge, is ''cold'', interacts gravitationally and interacts, at most, weakly with the standard model \cite{Bertone:2016nfn}. These postulated properties have led to a collective class of particles known as weakly interacting massive particles (WIMP). Various WIMP candidates have been proposed over the years, a large number of them motivated through extensions of the standard model. Supersymmetry (SUSY), a specific extension of the standard model, was studied in detail with respect to DM candidates \cite{Bertone:2016nfn,Bertone:2004pz}. SUSY introduces a new symmetry between fermions and bosons and thus predicts a number of new particles, called ''superpartners''. Especially the gravitino, neutralino and the sneutrino received a large amount of attention as DM candidates \cite{Bertone:2016nfn,Pagels:1981ke,Ellis:1983ew}. With the failure of SUSY particles to emerge in particle-collider experiments at the predicted energy scales \cite{ParticleDataGroup:2018ovx}, supersymmetry gradually fell out of consideration for BSM physics. \\
Current studies of WIMP range from the study on galactic scales through rotation curves, to stellar scales. The interaction properties of DM with itself -- also dubbed self-interacting dark matter (SIDM) -- are also studied extensively, in particular in the context of explaining the inner regions of galactic rotation curves (see e.g. \cite{Tulin:2017ara,Bullock:2017xww}). The current consensus is that a certain amount of self-interaction of the DM particle with itself is needed to correctly explain some galactic rotation curves and galaxy clusters. However, the self-interaction must not be too strong as to have a significant effect on the large-scale structure formation of the universe, since it is well described using collisionless (i.e. non-self-interacting) dark matter \cite{Tulin:2017ara,Bullock:2017xww}. The total range of possible DM parameters is vast. In \autoref{fig:background:dark-matter:dark-matter-mass-ranges-summary} an overview of possible mass ranges for the DM particle, and which range can be probed by which astrophysical system, is provided.

Another BSM dark matter particle candidate actively researched is the QCD axion \cite{Diedrichs:2023trk,Bertone:2016nfn,Baryakhtar:2022hbu,Bertone:2018krk}, which has been proposed to solve the strong CP problem of quantum chromodynamics. The strong CP problem refers to the experimental finding that QCD is invariant under simultaneous charge conjugation (C) and parity (P) transformation. However, purely from the mathematical formulation of QCD, violations of the CP-symmetry are in principle possible. The relevant CP-symmetry violating term in the QCD Lagrangian is
\begin{align}
\mathcal{L}_{QCD} \: \supset \: \theta \frac{g^2}{32 \pi^2} \tensor{F}{^\mu^\nu^a} \tensor{\tilde{F}}{_\mu_\nu_a} \: , \label{eq:background:dark-matter:strong-CP-lagrangian}
\end{align}

where $g$ is the strong coupling constant, $\tensor{F}{^\mu^\nu^a}$ and $\tensor{\tilde{F}}{_\mu_\nu_a}$ are the gluon field-strength tensor and its dual respectively. $\theta$ is a scalar number, related to the phase of the QCD vacuum, which has to be determined experimentally. The term \eqref{eq:background:dark-matter:strong-CP-lagrangian} gives rise to an electric dipole moment of the neutron. Through experimental measurements, it was found that the parameter $\theta$ must be smaller than $10^{-10}$ to be consistent with observations. While there is nothing wrong conceptually with the parameter $\theta$ being small, some scientists have been searching for a way to explain this observational fact theoretically. It was discovered by \cite{Peccei:1977ur,Peccei:1977hh} that by including an additional global $U(1)$-symmetry, the parameter $\theta$ can dynamically go to zero. The scalar parameter $\theta$ is then promoted to a pseudo scalar field $\theta \rightarrow a(x^\mu)$ -- the axion field -- that arises as a pseudo-Goldstone boson when the new $U(1)$-symmetry is spontaneously broken. The mass of the axion is of the order $m_a \approx \Lambda_{\mathrm{QCD}}^2 / f_a$, where $\Lambda_{\mathrm{QCD}}$ is the perturbative QCD energy scale and $f_a$ is the energy scale at which the $U(1)$-symmetry is broken. To be consistent with current observational bounds like decay rates of mesons and the cooling of stars \cite{Bertone:2016nfn}, the axion must be very light with a mass of $m_a < 10^{-3}\,eV$. Direct detection experiments \cite{ADMX:2009iij} have also imposed bounds on the axion mass and coupling to the standard model (see \cite{Klasen:2015uma,Irastorza:2018dyq} for reviews). It was additionally shown that axions of masses $m_a \approx 10^{-5}\,eV$ could have formed in the early universe in sufficient amounts to account for the entirety of the dark matter abundance \cite{Bertone:2016nfn}. \\
Due to their weak coupling to the standard model, axions -- and similar axion-like particles (ALP) (for a review see \cite{Galanti:2022ijh}) -- have received a large amount of attention as dark matter candidates. Ultralight particles with masses $m \ll 10^{-5}\,eV$ in particular lead to a rich phenomenology due to their macroscopic wave-behaviour: they can form Bose-Einstein condensates on astrophysical scales, similarly sized to stars or even galaxies \cite{Ferreira:2020fam,Hui:2021tkt}. Since the Compton wavelength $\lambda_{\mathrm{c}}$ is inversely proportional to the particle mass, a small mass leads to large wavelengths. For ultralight particles, it can therefore happen that $\lambda_{\mathrm{c}}$ gets so large, that the mean distance between particles is similar to $\lambda_{\mathrm{c}}$ even on large length scales. Such ultralight dark matter (ULDM) particles can then be described using a wave function of macroscopic size. When the ULDM mass is of the order $10^{-9} - 10^{-11} \,eV$ \cite{Diedrichs:2023trk}, their Compton wavelength is of the order of kilometres. If such dark matter were to accumulate in sufficient amounts to form a Bose-Einstein condensate of similar size, it could have significant effects on neutron stars. The DM condensate could even become self-gravitating and form gravitationally stable structures on its own, if the total mass of the ULDM present is large enough. Thus, understanding the macroscopic behaviour of self-gravitating bosonic fields is relevant to both neutron stars and the search for dark matter. Keeping bosonic ULDM as the main motivator of this work in mind, we dedicate the following chapter to these self-gravitating systems and their properties.

\subsection{Boson Stars} \label{subsec:background:boson-stars} % / Proca stars / Fermion boson stars

Boson stars (BS) can be thought of as macroscopically sized Bose-Einstein condensates. They have been intensively studied in the past (see \cite{Liebling:2012fv}) as, e.g., a possible explanation for dark matter, as a source of gravitational waves, as a consequence of modified theories of gravity or as an alternative to black holes. Formally, boson stars arise as a solution to the Einstein-Hilbert-Klein-Gordon (EHKG) equations such that they are described by a bosonic scalar (or vector) field that is stabilized by its own gravity. Boson stars with a vector field are referred to as Proca stars (PS) \cite{Liebling:2012fv,Brito:2015pxa}, although in this chapter, we will mostly use the term boson star to refer to both cases. If interpreted from a particle-physics viewpoint, BS are either a collection of stable fundamental bosonic particles or a configuration of decaying particles, which have a reverse-decay channel efficient enough to form a stable equilibrium when inside a gravitational potential \cite{Liebling:2012fv}. The following chapter is loosely based on the review of \cite{Liebling:2012fv} and references therein. Other reviews about boson stars cover their astrophysical relevance \cite{Jetzer:1991jr}, formation scenarios \cite{Liddle:1992fmk} or relevance for dark matter \cite{Jetzer:1991jr} such as, e.g., the axion \cite{Braaten:2019knj} (see Chapter \ref{subsec:background:dark-matter}). \\

Boson stars have been formulated on scales ranging from ultra compact objects to galactic sizes and are likewise motivated in a wide range of astrophysical and particle-physical scenarios. In this work, we focus on astrophysical motivations that are relevant in the wider context of dark matter in neutron stars (although a brief overview of adjacent topics shall be given here as well). We thus give an overview of some different types of BS and, in the end, focus on the cases most relevant for this work. \\
There are a number of proposed ways how BS could form. On large scales, the DM halos of galaxies could themselves be considered boson stars, given that the particle constituting the halo is a boson \cite{Liebling:2012fv}. On stellar scales, bosonic particles could clump together due to gravity, or accrete onto stars and then form boson stars inside them \cite{Diedrichs:2023trk,Brito:2015yga}. If the accreting boson is a DM candidate, the amount of accreting material would depend on the local DM density as inferred from the galactic rotation curve of the host galaxy. Another widely considered formation channel is that of gravitational collapse of local over-dense regions of bosons in the early universe \cite{Liddle:1992fmk}. In this case, boson stars could serve as a kind of MACHOS. Other formation scenarios are motivated through modified gravity \cite{Liebling:2012fv}. For example, boson stars could form around black holes due to superradiance \cite{Brito:2015oca}. Superradiance is a process included in some theories of modified and quantum gravity through which rotational energy of a rotating (Kerr) BH can be transferred into a scalar (or vector) field, then forming a boson star as a cloud around the BH. A closely related type of objects are gravitational atoms \cite{Nielsen:2019izz,Arvanitaki:2014wva}, an accumulation of a scalar field with discrete energy levels around a BH, which take their name from mathematical similarities to the orbital model describing electrons around atomic nuclei. Yet another process to produce a large enough abundance of scalar field is spontaneous scalarization, which is present in some theories of modified gravity such as TeVeS theory \cite{Bekenstein:2004ne} (see Chapter \ref{subsec:background:dark-matter}) or in general scalar-tensor theories \cite{Silva:2017uqg,Burikham:2016cwz,Yazadjiev:2019oul,Doneva:2019krb,Brihaye:2016lin}. These theories feature a scalar field (or multiple scalar fields), which is used as a parameter alongside a spacetime metric to describe the dynamics of gravity. By virtue of the Einstein equations, this field can however either be interpreted as a modification to curvature or as an additional term to the energy-momentum tensor. Spontaneous scalarization in these theories is a process in which the scalar component of the gravity theory can be dynamically produced and then transition to a non-trivial localized configuration. The scalar field itself could then form a BS since it is mathematically equivalent to the fields used to describe boson stars \cite{Liebling:2012fv}. \\

\begin{figure}[h]
	\centering
	\includegraphics[width=0.8\textwidth]{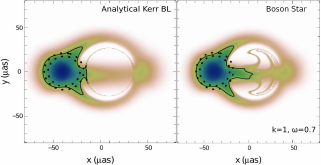} %width=0.4535
	\caption{Comparison between the shadow images as would be seen by the EHT, computed for a highly rotating Kerr black hole (left) and a rotating boson star (right) each surrounded by an accretion torus \cite{Vincent:2015xta}. The x- and y-axis show the angular sizes of these objects as seen from Earth if they were placed at the same position as Sgr A* in the centre of the Milky Way galaxy. The similarity in these images suggests that ruling out boson stars in the centre of galaxies might prove difficult using shadow images alone. The figures were taken from \cite{Vincent:2015xta} and slightly adapted for use in this work.}
	\label{fig:background:boson-stars:BS-and-Kerr-shadow}
\end{figure}

Depending on the formation scenario and the specific BS model used, a number of astrophysical consequences regarding boson stars arise \cite{Liebling:2012fv}. They can be relevant on astrophysical scales such as in main-sequence stars and in compact objects, if for example the bosonic component accretes over time and settles inside the gravitational well of the host object. This capture of bosonic DM in NS has been studied in the past \cite{Guver:2012ba}. It can also be constrained by studying the capture and annihilation rates of hypothesized bosonic particles \cite{Kouvaris:2010vv}. Boson stars can also be used as a way to describe dark matter clouds or accumulations of bosonic DM -- either as single coherent galaxy-sized configurations to, e.g., explain galactic rotation curves at large \cite{Lee:2008ab,Sharma:2008sc,Urena-Lopez:2010zva,Rindler-Daller:2011afd} and small \cite{Marsh:2015wka,DellaMonica:2022kow} distances, or as smaller objects so that the DM halo consists of many smaller boson stars \cite{Liebling:2012fv,Mielke:2000mh}. \\
BS are also studied in the context of black hole mimickers \cite{Liebling:2012fv,Guzman:2009zz}. Since in general, boson stars do not interact with light (unless a coupling to electromagnetism is explicitly included) and do not have a hard surface, they could serve as an alternative explanation for highly localized compact objects such as black holes \cite{Liebling:2012fv}. Some BS configurations also feature photon spheres around them, which is otherwise seen as a characteristic feature of black hole spacetimes \cite{Horvat:2013plm}. According to a number of simulations \cite{Meliani:2016rfe,Guzman:2009zz,Meliani:2015zta}, they would however show distinguishable torus accretion dynamics compared to black holes. Since BS do not have an event horizon, matter could accumulate in the centre of boson stars, leaving clear imprints of radiating matter inside the BS \cite{Olivares:2018abq,Shnir:2022lba,Vincent:2015xta}, for example in form of thermal or X-ray radiation \cite{Cao:2016zbh}. Due to their non-emissivity of light, the absence of an event horizon, and their ability to cover wide ranges of mass and compactness, BS have also been studied as an option for supermassive black holes in the galactic centres \cite{Torres:2000dw}. In fact, realistic models of rotating boson stars have shown that they can produce shadow images \cite{Vincent:2015xta,Rosa:2022tfv,Herdeiro:2021lwl} that are close to the images obtained of M87* \cite{EventHorizonTelescope:2019dse} and Sgr A* \cite{EventHorizonTelescope:2022wkp} using the Event Horizon Telescope (EHT). In \autoref{fig:background:boson-stars:BS-and-Kerr-shadow}, two shadow images from a Kerr black hole and from a rotating boson star as expected from the EHT are shown. Due to the similarity between those images, it might be  difficult to definitely rule out boson stars as alternatives to black holes.

\begin{figure}[h]
	\centering
	\includegraphics[width=0.467\textwidth]{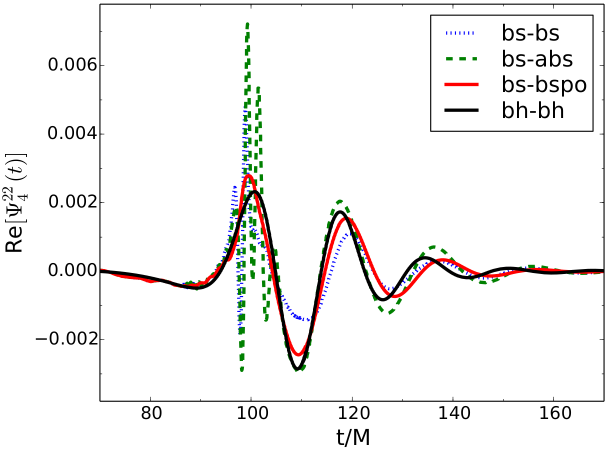} %width=0.4535
	%\hspace{0.5cm}
	\includegraphics[width=0.5257\textwidth]{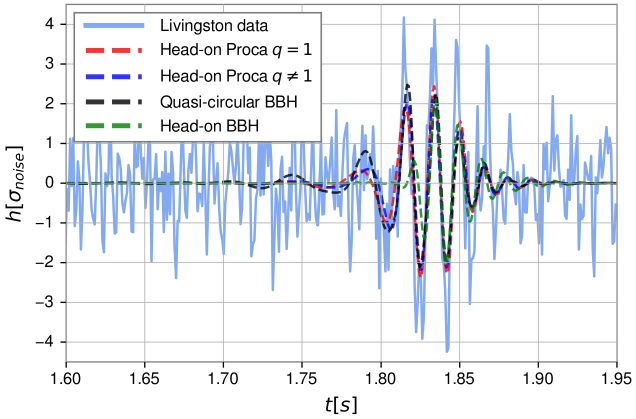} %width=0.534
	\caption{\textbf{Left panel:} Gravitational wave signal, as represented by the $l=m=2$ mode of the Newman-Penrose scalar $\Psi_4$, as a function of time, emitted during the head-on collision of (different types of) two boson stars. All configurations eventually collapse into a BH. The GW signal from a BH-BH collision is shown as a reference. The boson star signals are seen to differ significantly from the black hole signal, since the scalar fields interact before the collision. For more information about the models used, refer to \cite{Cardoso:2016oxy}. Figure was taken from \cite{Cardoso:2016oxy}.
	\textbf{Right panel:} GW strain (i.e the GW amplitude) for the event GW190521 as seen from the LIGO Livingston GW detector, together with the best fitting waveforms for a head-on merger of two black holes (green), two Proca stars (i.e boson star with a vector field) of equal ($q=1$) and unequal ($q \neq 1$) masses (red and blue respectively) and for a quasi-circular BH merger (black). Here, the GW signals from boson stars produce a similar waveform to the BH case. Figure was taken from \cite{CalderonBustillo:2020fyi}.}
	\label{fig:background:boson-stars:BS-and-BH-waveform}
\end{figure}

Nevertheless, a boson star will always differ from a black hole because it has a non-zero tidal deformability \cite{Diedrichs:2023trk,Liebling:2012fv,Johnson-Mcdaniel:2018cdu}. Due to their tidal deformation, they would produce a distinct signal in binary mergers \cite{Cardoso:2017cfl,Bezares:2018qwa,Palenzuela:2007dm}. Thus, boson stars could also serve as a source for gravitational waves. An example for possible GW signals of boson stars is given in \autoref{fig:background:boson-stars:BS-and-BH-waveform}. If admixed with neutron stars, they could also significantly change the tidal properties of the neutron star \cite{Diedrichs:2023trk}. Then, through the comparison of measured GW signals to computed GW waveforms of (inspiralling) boson stars, one can constrain the parameters of the boson star such as the boson mass and self-interaction \cite{Diedrichs:2023trk,Karkevandi:2021ygv,Pacilio:2020jza} or coupling parameters of specific theories of modified gravity \cite{Liebling:2012fv}, in which boson stars arise. Boson stars could also serve as a part in a gravitational wave background \cite{Croon:2018ftb}, if they are numerous and massive enough.
%%%

Boson stars were first conceived by \cite{Kaup:1968zz} and \cite{Ruffini:1969qy}. Mathematically, boson stars are described using a complex scalar field $\varphi(x^\mu) \in \mathbb{C}$ coupled minimally to gravity (see \cite{Liebling:2012fv} for a more detailed review). The resulting system of equations are the Einstein-Hilbert-Klein-Gordon (EHKG) equations. Since the Klein-Gordon equation is a wave equation, it will lead to a dispersion of the scalar field. Boson stars thus can also be understood as a system where the natural dispersion of the field is balanced by its self-gravity. The EHKG equations are sourced from the Lagrangian
\begin{align}
	S = \int_\mathcal{M} \sqrt{-g} \left( \frac{1}{2\kappa} R - \tensor{\nabla}{_\alpha} \bar{\varphi} \tensor{\nabla}{^\alpha} \varphi - V(\bar{\varphi} \varphi) \right)\, dx^4 \: , \label{eq:background:boson-stars:einstein-klein-gordon-action}
\end{align}

where $R$ is the Ricci curvature scalar, $g$ is the determinant of the spacetime metric $\tensor{g}{_\mu_\nu}$, $\kappa = 4\pi G/c^4$ is a constant and the integral is performed over the whole spacetime manifold $\mathcal{M}$. $\varphi$ and $\bar{\varphi}$ are the scalar field and its complex conjugate respectively and $V(\bar{\varphi} \varphi)$ is the potential depending only on the magnitude of the scalar field. By construction, the action \eqref{eq:background:boson-stars:einstein-klein-gordon-action} is invariant under a global $U(1)$-transformation (i.e. multiplication by a complex phase $\varphi \rightarrow \varphi e^{i\delta}$) of the field. The corresponding equations of motion for the spacetime can be obtained by computing the variation with respect to the inverse spacetime metric $\delta \tensor{g}{^\mu^\nu}$:
\begin{align}
\begin{split}
	\tensor{G}{_\mu_\nu} =& \kappa \, T_{\mu \nu}^{(\varphi)} \: , \\
	T_{\mu \nu}^{(\varphi)} =& \tensor{\partial}{_\mu} \bar{\varphi} \tensor{\partial}{_\nu} \varphi + \tensor{\partial}{_\mu} \varphi \tensor{\partial}{_\nu} \bar{\varphi} - \tensor{g}{_\mu_\nu} \left( \tensor{\partial}{_\alpha} \bar{\varphi} \tensor{\partial}{^\alpha} \varphi + V(\bar{\varphi} \varphi) \right) \: . \label{eq:background:boson-stars:einstein-equation}
\end{split}
\end{align}

Here $\tensor{G}{_\mu_\nu}$ is the Einstein tensor and $T_{\mu \nu}^{(\varphi)}$ is the effective energy-momentum tensor of the scalar field. Equation \eqref{eq:background:boson-stars:einstein-equation} therefore describes the coupling of the energy-momentum content of the scalar field to gravity. The equation of motion for the scalar field is obtained by varying equation \eqref{eq:background:boson-stars:einstein-klein-gordon-action} with respect to the scalar field $\varphi$ (and analogously for its complex conjugate $\bar{\varphi}$):
\begin{align}
\tensor{\nabla}{_\mu} \tensor{\nabla}{^\mu} \varphi = V'(\bar{\varphi} \varphi) \varphi \:\: , \:\: \text{with} \:\: V'(\bar{\varphi} \varphi) := \frac{d V}{d |\varphi|^2} \: . \label{eq:background:boson-stars:boson-star-klein-gordon-eq}
\end{align}

According to Noether's theorem, each continuous symmetry admits a conserved quantity. The invariance of the Lagrangian \eqref{eq:background:boson-stars:einstein-klein-gordon-action} under $U(1)$-symmetry thus implies the existence of a Noether current (see \cite{Diedrichs:2023trk,Liebling:2012fv}):
\begin{align}
\tensor{j}{^\mu} = i \, \tensor{g}{^\mu^\nu} \left( \bar{\varphi} \tensor{\nabla}{_\nu} \varphi - \varphi \tensor{\nabla}{_\nu} \bar{\varphi} \right) \: . \label{eq:background:boson-stars:noether-current}
\end{align}

The conserved quantity (i.e. the Noether charge) associated with the Noether current is obtained by integrating it over space:
\begin{align}
N_\mathrm{b} := \int \sqrt{-g} g^{t\mu} j_\mu dx^3 \: . \label{eq:background:boson-stars:conserved-boson-number-definition}
\end{align}

$N_\mathrm{b}$ is called the boson number and is related to the total number of bosons present in the system. Equivalently, it can also be interpreted as the total restmass energy of the boson star. \\
There is a large variety of different types of boson stars that have been studied (see \cite{Liebling:2012fv} for a more complete overview). They can be formulated using multiple fields, have rotation, or even have an electric charge. Boson stars have also been studied in Newtonian gravity, in modified theories of gravity, or without gravity at all (in this case they are sometimes called Q-balls). We here focus on boson stars in the context of general relativity. In general, however, boson star solutions are characterized by the potential $V(\bar{\varphi} \varphi)$ that they have. The simplest case is that of the potential consisting of a mass term $V(\bar{\varphi} \varphi) = m^2 \bar{\varphi} \varphi$ ($m$ being the boson mass). Such boson stars are often called mini boson stars due to their maximum mass of the order $M_\mathrm{max} \approx 0.633 M^2_p / m$ (where $M_p$ is the Planck mass) being much smaller than the maximum mass $M_\mathrm{Ch} \approx M^3_p / m^2$ (Chandrasekhar mass) found for stars made out of fermions \cite{Liebling:2012fv}. Boson stars made out of axions -- axion stars -- have also been proposed. In that case, the potential describing axions takes the form of the instanton potential \cite{Liebling:2012fv,Braaten:2019knj}
\begin{align}
V (\varphi) = \left( m_a f_a \right)^2 \, \left| 1 - \cos\left( \varphi / f_a \right) \right| \: , \label{eq:background:boson-stars:axion-instanton-potential}
\end{align}

where $m_a$ is the mass of the axion, $f_a$ is the axion symmetry breaking scale and $\varphi$ is the scalar field describing the axion. For more information about axion stars, we refer to the review \cite{Braaten:2019knj}. One of the ways to increase the maximum total mass of mini boson stars is to generalize the potential by including self-interactions. To retain the $U(1)$-symmetry, all additional terms must be functions of the magnitude $\bar{\varphi} \varphi = |\varphi|^2$. The simplest option is then to include a quartic self-interaction term with the self-interaction parameter $\lambda$:
\begin{align}
V(\bar{\varphi} \varphi) = m^2 \bar{\varphi} \varphi + \frac{\lambda}{2} (\bar{\varphi} \varphi)^2 \: . \label{eq:background:boson-stars:quartic-self-interaction-potential}
\end{align}

In this case the maximum mass increases to  $M_\mathrm{max} \approx 0.22 \sqrt{\Lambda_\mathrm{int}}\, M^2_p / m$ \cite{Liebling:2012fv,Colpi:1986ye} for large $\Lambda_\mathrm{int} := \lambda / 8\pi m^2$. Other potentials with even higher orders in $\bar{\varphi} \varphi$ were considered in the past, such as the sixth-order soliton potential (see \cite{Liebling:2012fv,Lee:1986ts,Friedberg:1986tq}). The fourth-order potential \eqref{eq:background:boson-stars:quartic-self-interaction-potential} however remains perhaps the most interesting scalar field potential, especially in a particle-physics context. This is due to it also being able to describe a fundamental scalar particle in a renormalizable scalar field theory. In this work, in accordance to \cite{Diedrichs:2023trk}, we restrict our study to the potential \eqref{eq:background:boson-stars:quartic-self-interaction-potential} with a mass term and a quartic self-interaction term. \\

To solve the boson star equations, a harmonic phase ansatz is used for the scalar field:
\begin{align}
\varphi(x^\mu) = \varphi(x^i) e^{-i\omega t} \: , \label{eq:background:boson-stars:scalar-field-harmonic-ansatz}
\end{align}
where the time-dependence was made explicit through a complex phase with frequency $\omega$ and the amplitude depends solely on the spatial coordinates. Although this makes the field explicitly time-dependent, since the energy-momentum tensor \eqref{eq:background:boson-stars:einstein-equation} is only dependent on the absolute value of the scalar field and its gradients, the final solution for the spacetime metric will not depend on time. The reasoning behind this ansatz is Derrick’s theorem \cite{Liebling:2012fv,Derrick:1964ww}, which states that no regular, static, localized scalar field configuration can exist in three (spatial) dimensional flat spacetime. Derrick’s theorem has since been generalized to disallow static scalar field configurations without Noether charges \cite{Diez-Tejedor:2013sza} and to hold when including general relativity \cite{Carloni:2019cyo}. This constraint is thus avoided by adopting the harmonic ansatz \eqref{eq:background:boson-stars:scalar-field-harmonic-ansatz} for the complex scalar field and by working with gravity. \\
The equations of motion for the boson star with generic potential $V(\bar{\varphi} \varphi)$ are \eqref{eq:background:boson-stars:einstein-equation} and \eqref{eq:background:boson-stars:boson-star-klein-gordon-eq}. They can be solved when assuming a spherically symmetric metric (i.e. making the same ansatz as in \eqref{eq:background:neutron-stars:metric-ansatz} for the metric) 
\begin{align}
	ds^2 = \tensor{g}{_\mu_\nu} \tensor{dx}{^\mu} \tensor{dx}{^\nu} = - \alpha^2(r) dt^2 + a^2(r) dr^2 + r^2 d\theta^2 + r^2\sin^2(\theta) d\phi^2 \: , \label{eq:background:boson-stars:metric-ansatz}
\end{align}

and a spherically symmetric scalar field (i.e. substituting $\varphi(x^i) \rightarrow \varphi(r)$ in \eqref{eq:background:boson-stars:scalar-field-harmonic-ansatz}) and be cast into the following system of differential equations (see e.g. \cite{Liebling:2012fv})
\begin{subequations}
\begin{align}
a' = \frac{da}{dr} &=  \frac{a}{2} \, \left[ \frac{(1-a^2)}{r} + 8\pi r a^2  \left(\frac{\omega^2 \varphi^2}{\alpha^2} + \frac{{\varphi'}^2 }{a^2} + V(\bar{\varphi} \varphi) \right) \right] \: , \label{eq:background:boson-stars:TOV-equations-grr} \\
\alpha' = \frac{d \alpha}{dr} &= \frac{\alpha}{2} \left[ \frac{(a^2 -1)}{r} + 8\pi r a^2 \left(\frac{\omega^2 \varphi^2}{\alpha^2} + \frac{{\varphi'}^2 }{a^2} - V(\bar{\varphi} \varphi) \right) \right] \: , \label{eq:background:boson-stars:TOV-equations-gtt} \\
\varphi'' = \frac{d^2 \varphi}{dr^2} &= \phantom{\frac{\alpha}{2}} \left[ a^2 V'(\bar{\varphi} \varphi) - \frac{\omega^2 a^2}{\alpha^2} \right] \varphi + \left[\frac{a'}{a\,} - \frac{\alpha'}{\alpha\,} - \frac{2}{r} \right] \varphi' \: . \label{eq:background:boson-stars:TOV-equations-phi}
\end{align}
\end{subequations}

Here $V'$ is defined as in \eqref{eq:background:boson-stars:boson-star-klein-gordon-eq}. The boundary conditions for the system \eqref{eq:background:boson-stars:TOV-equations-grr}-\eqref{eq:background:boson-stars:TOV-equations-phi} are obtained by imposing asymptotic flatness at $r \rightarrow \infty$ and regularity at the origin (i.e. no divergence at $r\rightarrow0$):
\begin{align}
    \begin{split}
        \lim_{r \rightarrow \infty} a(r) &= 1 \:\: , \:\: a(0) = 1 \: , \\
        \lim_{r \rightarrow \infty} \alpha(r) &= 1 \:\: , \:\: \alpha(0) = \alpha_0 \: , \\
        \lim_{r \rightarrow \infty} \varphi(r) &= 0 \:\: , \:\: \varphi(0) = \varphi_c \: , \\
        \lim_{r \rightarrow \infty} \varphi'(r) &= 0 \:\: , \:\: \varphi'(0) = 0 \: . \\
    \end{split} \label{eq:background:boson-stars:TOV-initial-conditions}
\end{align}

\begin{figure}[h]
	\centering
	\includegraphics[width=0.49\textwidth]{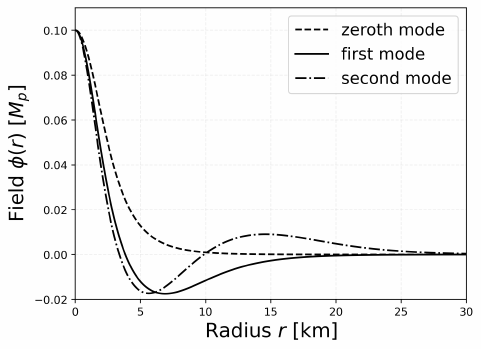}
	\includegraphics[width=0.49\textwidth]{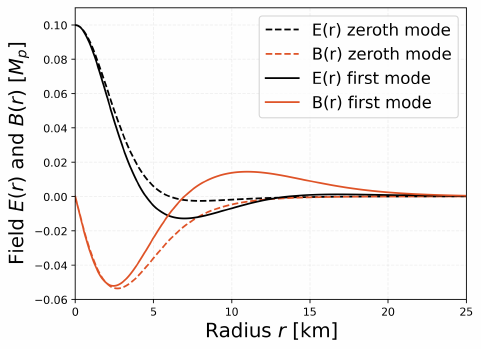}
	\caption{\textbf{Left panel:} Radial scalar field profiles of the zeroth, first and second mode of a boson star described by equations \eqref{eq:background:boson-stars:TOV-equations-grr}-\eqref{eq:background:boson-stars:TOV-equations-phi}, where the potential $V(\bar{\varphi} \varphi) = m^2 \bar{\varphi} \varphi$ corresponds to that of a mini-boson star. The scalar field mass is $m= 1.34\e{-10}\,eV$. The field is seen to asymptotically reach zero from positive (negative) values for even (odd) modes.
	\textbf{Right panel:} The first two modes of a Proca star with potential $V(A_\mu \bar{A}^\mu ) = m^2 A_\mu \bar{A}^\mu$ where $m= 1.34\e{-10}\,eV$. The radial profiles show the vector field components $A_t = E(r)$ and the $A_r = i B(r)$. In contrast to boson stars with a scalar field, the lowest mode of a Proca star always has one zero-crossing in the $E(r)$ component. Both $E(r)$ and $B(r)$ asymptotically approach zero from negative (positive) values for even (odd) modes.}
	\label{fig:background:boson-stars:boson-stars-radial-profiles}
\end{figure}

Asymptotic flatness requires tuning the initial condition for $\alpha_0$ to some finite non-zero value. However, since the system of equations \eqref{eq:background:boson-stars:TOV-equations-grr}-\eqref{eq:background:boson-stars:TOV-equations-phi} is invariant when re-scaling the frequency to $\tilde{\omega} = \omega \alpha_0$, it is possible to absorb the initial value for $\alpha_0$ so that $\alpha(0)$ may be set equal to one. After integrating the equations, one can obtain the physical value of $\omega$ by performing the inverse transformation back to $\omega$ using the asymptotic value for $\alpha(r\rightarrow\infty)$ (see also \cite{Diedrichs:2023trk}). \\
The only parameter which is not directly constrained by the equations of motion \eqref{eq:background:boson-stars:TOV-equations-grr}-\eqref{eq:background:boson-stars:TOV-equations-phi} is the scalar field frequency $\omega$. For a given value of $\varphi_c$, one has to adjust the value of $\omega$ so that the boundary conditions \eqref{eq:background:boson-stars:TOV-initial-conditions} are fulfilled. This requires tuning $\omega$ to specific quantized values, which are called modes. These modes are characterized by the amount of zero-crossings (i.e. radial positions with $\varphi(r)=0$) the field $\varphi(r)$ has before eventually converging to zero at infinity. This can be done using a shooting-algorithm, where the system of equations \eqref{eq:background:boson-stars:TOV-equations-grr}-\eqref{eq:background:boson-stars:TOV-equations-phi} is integrated until the matching value for $\omega$ is found (also see \cite{Diedrichs:2023trk,Liebling:2012fv} for more information). In the left panel of \autoref{fig:background:boson-stars:boson-stars-radial-profiles}, an example for radial scalar field profiles of different modes of a boson star is shown.

As is the case for neutron stars (see eq. \eqref{eq:background:neutron-stars:definition-NS-total-gravitational-mass}), it is also possible to define the total gravitational mass of a boson star. However, boson stars do not have a clearly defined surface because the scalar field only asymptotically goes to zero (see \eqref{eq:background:boson-stars:TOV-initial-conditions}). The total gravitational mass can thus only be defined in the limit of large radii:
\begin{align}
	M_\mathrm{tot} := \lim_{r \rightarrow \infty} \frac{r}{2} \left( 1 - \frac{1}{(a(r))^2 } \right) \: . \label{eq:background:boson-stars:definition-BS-total-gravitational-mass}
\end{align}

Although a surface does not exist, the radius of a boson star $R_\mathrm{b}$ is often defined as the radius, in which $99\%$ of the restmass (i.e. boson number \eqref{eq:background:boson-stars:conserved-boson-number-definition}) is included (see e.g. \cite{DiGiovanni:2021ejn}). \\

Just as boson stars with a scalar field can be thought of as condensates of spin-0\footnote{The reason why spin-0 and spin-1 particles are described using Lorentz scalar and vector fields respectively is due to the spin representations of the rotational part of the Lie-algebra of the Lorentz group $SO(3,1)$. The Lorentz group is the fundamental symmetry group of general and special relativity. The transformation properties of a field under the Lie-algebra $\mathfrak{so}(3,1)$ can be related to its quantum-mechanical spin. Scalar fields transform under the $(0,0)$ representation and vector fields under the $(\frac{1}{2},\frac{1}{2})$ representation.} bosonic particles, condensates of spin-1 particles have been conceived as well \cite{Liebling:2012fv}. These objects are called Proca stars (PS) and are modelled by a complex vector field $\tensor{A}{_\mu}(x^\mu) \in \mathbb{C}$ and were first proposed by \cite{Brito:2015pxa}. The corresponding action is the Proca action which is coupled minimally to gravity
\begin{align}
S = \int \sqrt{-g} \left( \frac{1}{2\kappa} R - \frac{1}{2} \tensor{F}{_\mu_\nu} \tensor{\bar{F}}{^\mu^\nu} - V(\tensor{A}{_\mu} \tensor{\bar{A}}{^\mu}) \right) dx^4 \: , \label{eq:background:boson-stars:einstein-proca-action}
\end{align}
where $\tensor{F}{_\mu_\nu} = \tensor{\nabla}{_\mu} \tensor{A}{_\nu} - \tensor{\nabla}{_\nu} \tensor{A}{_\mu}$ is the field strength tensor and $V(\tensor{A}{_\mu} \tensor{\bar{A}}{^\mu})$ is the vector field potential that depends solely on the magnitude $\tensor{A}{_\mu} \tensor{\bar{A}}{^\mu}$. Most of the concepts established for boson stars with a scalar field also apply to Proca stars, such as the existence of a conserved Noether current due to the global $U(1)$-symmetry and the presence of quantized frequency modes due to the harmonic time-dependence. An example of radial profiles of a Proca star with the potential $V(A_\mu \bar{A}^\mu ) = m^2 A_\mu \bar{A}^\mu$ is shown in \autoref{fig:background:boson-stars:boson-stars-radial-profiles}. It can be seen that, in contrast to boson stars, Proca stars have no mode where the field components do not cross zero. \\
Although being relatively new as a concept, Proca stars have been studied by a number of groups analytically \cite{Cardoso:2021ehg,Aoki:2022woy,Brihaye:2017inn} and numerically \cite{Sanchis-Gual:2017bhw}, such as in merger simulations \cite{CalderonBustillo:2020fyi,Sanchis-Gual:2018oui}. Different types of Proca stars with charge \cite{SalazarLandea:2016bys} and rotation \cite{Brito:2015pxa} were also considered. Other works \cite{Rosa:2022tfv,Herdeiro:2021lwl,Rosa:2022toh} studied shadow images of Proca stars in different scenarios (similarly to \autoref{fig:background:boson-stars:BS-and-Kerr-shadow}). Proca stars with a quartic self-interaction potential \cite{Minamitsuji:2018kof} were also studied. It was found that the maximum Proca star mass scales like $M_\mathrm{max} \approx \sqrt{\Lambda_\mathrm{int}} \ln(\Lambda_\mathrm{int})\, M^2_p / m$, for large self-interaction strengths $\Lambda_\mathrm{int} := \lambda / 8\pi m^2$. This is different to boson stars. \\

%The following chapter will further elaborate on Proca stars and their solutions and analytical properties.
As we have seen, boson stars and Proca stars are interesting objects to study in a variety of physical scenarios. Especially in the context of dark matter, these objects can provide quantitative measures to probe the dark matter mass and self-interaction parameters. In astrophysical scenarios it is likely that dark matter might not accumulate and self-gravitate solely by itself, but that it forms alongside baryonic matter. If for example a neutron star accumulates a significant amount of bosonic dark matter particles, the presence of DM might change the NS properties. A combined system of a boson star and a neutron star would subsequently form. We dedicate the remainder of this work to study these combined systems of fermions (e.g. neutrons, quarks) and bosons (e.g. bosonic DM), called fermion boson stars. We will especially investigate how the presence of a bosonic field affects the NS properties and how this can be used to infer the boson properties.

%% file: fermion-boson-stars.tex
\section{Fermion Boson Stars} \label{sec:fermion-boson-stars:fermion-boson-stars}

Building upon the knowledge introduced in Chapter \ref{sec:background:theoretical-background}, this chapter is dedicated to fermion boson stars (FBS). Fermion boson stars are mixed systems of bosons and fermions, that can be seen as a boson star that coexists with a neutron star at the same location in space. This chapter as a whole aims to build the analytical framework of fermion boson stars. First in Chapter \ref{subsec:fermion-boson-stars:motivation-and-formation} we will consider the physical motivation behind FBS and introduce mechanisms through which they might form. The part about fermion boson stars with a scalar field (Chapter \ref{subsec:fermion-boson-stars:scalar-fermion-boson-stars}) is built in large parts upon \cite{Diedrichs:2023trk}. We discuss the formulation and derivation of the equations of motion of the Einstein-Hilbert-Klein-Gordon system, including how to compute the tidal deformability of FBS. The following chapter \ref{subsec:fermion-boson-stars:fermion-proca-stars} about fermion Proca stars (FPS) -- i.e. fermion boson stars with a vector field -- is completely original to this work. In addition to the theoretical framework, some analytical results constraining FPS will be discussed as well.

\subsection{Motivation and Formation} \label{subsec:fermion-boson-stars:motivation-and-formation}

The idea that an astrophysical object consists of a mixture of fermionic and bosonic matter goes back to \cite{Henriques:1989ar,Henriques:1989ez}. Since then, a multitude of different models of these fermion boson stars (FBS) have been investigated (see e.g. \cite{Kouvaris:2010vv,Liebling:2012fv,Bertone:2007ae,Zurek:2013wia,Kouvaris:2013awa} for reviews). In the simplest case, the fermionic and bosonic components interact only gravitationally (i.e. they are minimally coupled), which is why FBS are regarded as interesting objects in the context of dark matter (DM) research (see e.g. \cite{Diedrichs:2023trk,DiGiovanni:2021ejn,Nelson:2018xtr}). Fermion boson stars have been studied in connection to neutron stars (NS), where the NS provides the fermionic component and a bosonic field provides the bosonic component of the FBS \cite{Diedrichs:2023trk,DiGiovanni:2021ejn}. In that sense, FBS are also related to boson stars (see Chapter \ref{subsec:background:boson-stars}), since the bosonic component can be modelled via scalar and vector fields. FBS have been studied with regard to their stability \cite{Henriques:1989ez} and their dynamical properties were explored as well \cite{DiGiovanni:2020frc,DiGiovanni:2021vlu,Valdez-Alvarado:2012rct,Valdez-Alvarado:2020vqa,Ruter:2023uzc,Bezares:2019jcb}. Numerical simulations aiming to understand the gravitational wave (GW) signals have also been done \cite{Bezares:2019jcb}. In all these cases, the NS component was modelled using a perfect fluid for the nuclear matter and a classical complex scalar field was used for the bosonic component. To our knowledge, no study of FBS with a vector field for the bosonic component has been done to date. \\

A number of processes have been proposed with regard to their formation (see e.g. references in \cite{Diedrichs:2023trk}), but the problem in essence is how one can accumulate a large amount of scalar (vector) field in and around a neutron star. One large motivator is bosonic dark matter. It is possible that neutron stars could accumulate DM in a sufficient abundance to modify their observables, such as the mass, radius, and tidal deformability. This would make FBS an adequate description of NS with admixed DM and could be used to constrain DM properties such as its mass and its self-interaction strength (see e.g. \cite{Diedrichs:2023trk,Goldman:1989nd,Rutherford:2022xeb,Bertone:2007ae,Ellis:2018bkr}) on various scales (see \autoref{fig:background:dark-matter:dark-matter-mass-ranges-summary}). Dark matter could arrange itself around neutron stars as a cloud or inside neutron stars as a core. Neutron stars with DM cores could form 
\begin{itemize}
\itemsep0em
\item[$1)$] from an initial DM 'seed' through accretion of baryonic matter \cite{Diedrichs:2023trk,Meliani:2016rfe,Ellis:2017jgp,Kamenetskaia:2022lbf},
\item[$2)$] through mergers of neutron stars and boson stars \cite{Diedrichs:2023trk},
\item[$3)$] through accretion of DM onto a NS and subsequent accumulation in the centre \cite{Diedrichs:2023trk,Brito:2015yga,Kouvaris:2010vv,Goldman:1989nd,Kouvaris:2007ay},
\item[$4)$] through the decay of standard model (SM) particles inside the neutron star into DM \cite{Baym:2018ljz,Motta:2018rxp,Motta:2018bil,Husain:2022bxl,Berryman:2022zic}.
\end{itemize}

Neutron stars with clouds could form in a similar way, given that either the dark matter is the dominant contribution to the FBS or that the DM properties only allow low-compactness configurations (e.g. when the particle mass is small \cite{Diedrichs:2023trk}). The fermionic and bosonic components could conceivably be separated from one another e.g. during a supernova NS-kick \cite{Arzoumanian:2001dv,Holland-Ashford:2017tqz,Bezares:2022obu,Kulkarni:2020fhv}, where the stellar remnant gets ejected and rapidly moves away from the remaining stellar envelope. The dark matter particles most interesting for fermion boson stars are generally (self-interacting) ultralight DM (ULDM) particles, various WIMP candidates, dark photons \cite{Biswas:2022tcw,An:2023mvf} (as a candidate for vector DM) and axions (see Chapter \ref{subsec:background:dark-matter}). \\
Another formation channel is motivated through theories of modified gravity. As was already established in Chapter \ref{subsec:background:boson-stars}, one way of producing large amounts of scalar (or vector) fields is superradiance \cite{Brito:2015oca,Siemonsen:2022ivj}. Likewise, spontaneous scalarization \cite{Liebling:2012fv,Silva:2017uqg} provides a way of producing significant scalar \cite{Liebling:2012fv,Damour:1996ke} and vector\footnote{In the case of vector fields, the process is also called spontaneous vectorization.} \cite{Silva:2021jya,Ramazanoglu:2017xbl} field amplitudes as well. Spontaneous scalarization has also been studied explicitly in neutron stars \cite{Silva:2017uqg,Kase:2020yhw,Minamitsuji:2020pak} and could be a way of forming FBS with scalar and vector fields. In addition, scalarization might take place dynamically in the late stages of the evolution of binary NS systems \cite{Barausse:2012da}, forming either a black hole or an FBS after merger (depending e.g. on the initial masses of the binary objects). \\

Be it bosonic dark matter or a bosonic field predicted by some theory of modified gravity, the possibilities and formation channels of FBS with scalar and with vector fields are vast. The study of FBS can thus reveal a large amount of knowledge about the underlying theories. In particular, FBS can be used to constrain the properties of underlying theories using measurements of macroscopic quantities of FBS -- such as their masses, radii and tidal deformabilities. It is therefore crucial to develop a theoretical framework through which these objects can be described. In the next chapters, we will develop the mathematical framework to describe fermion boson stars self-consistently.

\subsection{Scalar Fermion Boson Stars} \label{subsec:fermion-boson-stars:scalar-fermion-boson-stars}

Fermion boson stars (FBS) are combined systems of fermions and (scalar) bosons, which interact only gravitationally. We will take the bosonic field to describe dark matter (DM) from here onwards. We model FBS using a complex scalar field for the DM component and neutron matter for the fermionic component. We here follow the derivations by \cite{Diedrichs:2023trk} but adapt our conventions and notation to be consistent within this work. The action describing FBS is the combination of the Einstein-Hilbert-Klein-Gordon (EHKG) system (see eq. \eqref{eq:background:boson-stars:einstein-klein-gordon-action}) minimally coupled to nuclear matter (see eq. \eqref{eq:background:neutron-stars:einstein-hilbert-action}):
\begin{align}
	S = \int_\mathcal{M} \sqrt{-g} \left( \frac{1}{2\kappa} R - \tensor{\nabla}{_\alpha} \bar{\varphi} \tensor{\nabla}{^\alpha} \varphi - V(\bar{\varphi} \varphi) - \mathcal{L}_{m} \right)\, dx^4 \: , \label{eq:fermion-boson-stars:scalar-fermion-boson-stars:fermion-boson-star-action}
\end{align}

where $R$ is the Ricci curvature scalar, $g$ is the determinant of the spacetime metric $\tensor{g}{_\mu_\nu}$, $\kappa = 4\pi G/c^4$ is a constant and the integral is performed over the whole spacetime manifold $\mathcal{M}$. $\mathcal{L}_{m}$ is the Lagrangian describing nuclear matter and $V(\bar{\varphi} \varphi)$ is the potential depending only on the magnitude of the scalar field. By taking the variation of \eqref{eq:fermion-boson-stars:scalar-fermion-boson-stars:fermion-boson-star-action} with respect to the inverse spacetime metric $\delta \tensor{g}{^\mu^\nu}$, one obtains the Einstein equations
\begin{align}
\tensor{G}{_\mu_\nu} =& \kappa \, \left( T_{\mu \nu}^{(NS)} + T_{\mu \nu}^{(\varphi)} \right) \: , \label{eq:fermion-boson-stars:scalar-fermion-boson-stars:einstein-equations-fbs}
\end{align}

where $T_{\mu \nu}^{(NS)}$ and $T_{\mu \nu}^{(\varphi)}$ are the energy-momentum tensors describing the neutron star matter and the scalar field matter respectively. The energy-momentum tensor of the neutron star matter, which is defined in equation \eqref{eq:background:neutron-stars:definition-energy-momentum-tensor}, is assumed to be that of a perfect fluid
\begin{align}
	T_{\mu \nu}^{(NS)} = (e + P) \tensor{u}{_\mu} \tensor{u}{_\nu} + P \tensor{g}{_\mu_\nu} \: , \label{eq:fermion-boson-stars:scalar-fermion-boson-stars:energy-momentum-tensor-perfect-fluid}
\end{align}

where $P$ and $e$ are the pressure and the energy density of the fluid respectively, where the energy density $e$ is related to the restmass density $\rho$ through $e=\rho(1+\epsilon)$ with $\epsilon$ being the internal energy. $\tensor{u}{_\mu}$ is the four-velocity of the fluid. The energy-momentum tensor \eqref{eq:fermion-boson-stars:scalar-fermion-boson-stars:energy-momentum-tensor-perfect-fluid}, as well as the fluid flow $\tensor{J}{^\mu} := \rho \tensor{u}{^\mu}$, are conserved (signifying conservation of energy-momentum and of the restmass respectively) which leads to the conservation equations
\begin{align}
	\tensor{\nabla}{_\mu} T^{\mu \nu}_{(NS)} = 0 \:\: , \:\: \tensor{\nabla}{_\mu} J^{\mu}_{\phantom{()}} = 0  \: . \label{eq:fermion-boson-stars:scalar-fermion-boson-stars:energy-momentum-and-restmass-conservation}
\end{align}

The conservation of the fluid flow $\tensor{J}{^\mu}$ allows us to define the conserved total restmass of neutron matter, which we call the fermion number $N_\mathrm{f}$. We obtain the fermion number by integrating the right part of \eqref{eq:fermion-boson-stars:scalar-fermion-boson-stars:energy-momentum-and-restmass-conservation} over space:
\begin{align}
	N_\mathrm{f} := \int \sqrt{-g}\; \tensor{g}{^t^\mu} \tensor{J}{_\mu} dx^3 \: . \label{eq:fermion-boson-stars:scalar-fermion-boson-stars:definition-NS-restmass}
\end{align}

Note that this expression is identical to \eqref{eq:background:neutron-stars:definition-NS-restmass}, validating the interpretation of $N_\mathrm{f}$ as the total restmass. The energy-momentum tensor for the scalar part is given by
\begin{align}
T_{\mu \nu}^{(\varphi)} =& \tensor{\partial}{_\mu} \bar{\varphi} \tensor{\partial}{_\nu} \varphi + \tensor{\partial}{_\mu} \varphi \tensor{\partial}{_\nu} \bar{\varphi} - \tensor{g}{_\mu_\nu} \left( \tensor{\partial}{_\alpha} \bar{\varphi} \tensor{\partial}{^\alpha} \varphi + V(\bar{\varphi} \varphi) \right) \: . \label{eq:fermion-boson-stars:scalar-fermion-boson-stars:energy-momentum-tensor-scalar-field}
\end{align}

The equations of motion (Klein-Gordon equations) of the scalar field are computed from the action \eqref{eq:fermion-boson-stars:scalar-fermion-boson-stars:fermion-boson-star-action} using the Euler-Lagrange equations for a complex scalar field:
\begin{align}
0 = \frac{\delta \mathcal{L}}{\delta \bar{\varphi}} - \nabla_\mu \frac{\delta \mathcal{L}}{\delta (\nabla^\mu \bar{\varphi})} \:\:\:\:\: \text{and} \:\:\:\:\: 0 = \frac{\delta \mathcal{L}}{\delta \varphi} - \nabla_\mu \frac{\delta \mathcal{L}}{\delta (\nabla^\mu \varphi)} \: . \label{eq:fermion-boson-stars:scalar-fermion-boson-stars:euler-lagrange-equations}
\end{align}

One then obtains the equations of motion for the field and the complex conjugate:
\begin{align}
\tensor{\nabla}{_\mu} \tensor{\nabla}{^\mu} \varphi = V'(\bar{\varphi} \varphi) \varphi \:\:\:\:\: \text{and} \:\:\:\:\:  \tensor{\nabla}{_\mu} \tensor{\nabla}{^\mu} \bar{\varphi} = V'(\bar{\varphi} \varphi) \bar{\varphi} \: , \label{eq:fermion-boson-stars:scalar-fermion-boson-stars:klein-gordon-equations}
\end{align}

where the derivative of the potential $V'(\bar{\varphi} \varphi)$ is defined as
\begin{align}
V'(\bar{\varphi} \varphi) := \frac{d V}{d |\varphi|^2} \: . \label{eq:fermion-boson-stars:scalar-fermion-boson-stars:derivative-of-scalar-potential}
\end{align}

The equations \eqref{eq:fermion-boson-stars:scalar-fermion-boson-stars:klein-gordon-equations} directly imply that the energy-momentum tensor of the scalar field \eqref{eq:fermion-boson-stars:scalar-fermion-boson-stars:energy-momentum-tensor-scalar-field} is separately conserved from the perfect fluid energy-momentum tensor \eqref{eq:fermion-boson-stars:scalar-fermion-boson-stars:energy-momentum-tensor-perfect-fluid}. Since the Lagrangian \eqref{eq:fermion-boson-stars:scalar-fermion-boson-stars:fermion-boson-star-action} is invariant under a global $U(1)$-symmetry transformation of the scalar field $\varphi$ (and $\bar{\varphi}$), Noether's theorem gives rise to a conserved Noether current
\begin{align}
\tensor{j}{^\mu} = i \, \tensor{g}{^\mu^\nu} \left( \bar{\varphi} \tensor{\nabla}{_\nu} \varphi - \varphi \tensor{\nabla}{_\nu} \bar{\varphi} \right) \: . \label{eq:fermion-boson-stars:scalar-fermion-boson-stars:noether-current}
\end{align}

The conserved quantity (i.e. the Noether charge) associated to the Noether current \eqref{eq:fermion-boson-stars:scalar-fermion-boson-stars:noether-current} is obtained by integrating the conservation equation $\nabla_\mu j^\mu = 0$ over space:
\begin{align}
N_\mathrm{b} := \int \sqrt{-g} g^{t\mu} j_\mu dx^3 \: . \label{eq:fermion-boson-stars:scalar-fermion-boson-stars:conserved-boson-number-definition}
\end{align}

$N_\mathrm{b}$ is called the boson number and is related to the total number of bosons present in the system. It can equivalently also be interpreted as the total restmass energy of the scalar component of the FBS. We also define the bosonic radius $R_\mathrm{b}$ as the radius, at which $99\%$ of the bosonic restmass energy ($N_\mathrm{b}$) is included. \\

We proceed by solving the equations of motion \eqref{eq:fermion-boson-stars:scalar-fermion-boson-stars:einstein-equations-fbs} and \eqref{eq:fermion-boson-stars:scalar-fermion-boson-stars:klein-gordon-equations} for spherically symmetric configurations in equilibrium. For that, we consider the spherically symmetric ansatz for the spacetime metric
\begin{align}
ds^2 = \tensor{g}{_\mu_\nu} \tensor{dx}{^\mu} \tensor{dx}{^\nu} = - \alpha^2(r) dt^2 + a^2(r) dr^2 + r^2 d\theta^2 + r^2\sin^2(\theta) d\phi^2 \: . \label{eq:fermion-boson-stars:scalar-fermion-boson-stars:metric-ansatz}
\end{align}

We further consider the perfect fluid to be static, such that the four-velocity can be written as
\begin{align}
\tensor{u}{^\mu} = \left(- \frac{1}{\alpha}, 0, 0, 0 \right) \:\: , \:\: \tensor{u}{_\mu} = (\alpha, 0, 0, 0) \: . \label{eq:fermion-boson-stars:scalar-fermion-boson-stars:four-velocity-static-fluid}
\end{align}

For the scalar field, we use the harmonic phase ansatz detailed in Chapter \ref{subsec:background:boson-stars}:
\begin{align}
\varphi(t,r) = \varphi(r) e^{-i\omega t} \: , \label{eq:fermion-boson-stars:scalar-fermion-boson-stars:scalar-field-harmonic-ansatz}
\end{align}

where $\omega$ is the field frequency. Using the spherical symmetric metric ansatz \eqref{eq:fermion-boson-stars:scalar-fermion-boson-stars:metric-ansatz} together with the harmonic phase ansatz \eqref{eq:fermion-boson-stars:scalar-fermion-boson-stars:scalar-field-harmonic-ansatz} for the scalar field, we solve the Einstein equations and obtain the equations of motion. One quickly sees that the $\phi\phi$-component and the $\theta\theta$-component of \eqref{eq:fermion-boson-stars:scalar-fermion-boson-stars:einstein-equations-fbs} differ only by a factor $\sin^2(\theta)$ and therefore are linearly dependent. By re-arranging the $tt$-component of \eqref{eq:fermion-boson-stars:scalar-fermion-boson-stars:einstein-equations-fbs}, one obtains an expression for the radial derivative of $a(r)$. By adding the $tt$- and $rr$-components of \eqref{eq:fermion-boson-stars:scalar-fermion-boson-stars:einstein-equations-fbs} in a certain way,
\begin{align}
G_{tt} \frac{1}{\alpha^2} + G_{rr} \frac{1}{a^2} = 8\pi (T^{NS}_{tt} + T^\varphi_{tt}) \frac{1}{\alpha^2} + 8\pi (T^{NS}_{rr} + T^\varphi_{rr}) \frac{1}{a^2} \: ,
\end{align}

we find a direct relation between the first radial derivatives of $a(r)$ and $\alpha(r)$, using which we can solve for $\alpha(r)$. Using the Klein-Gordon equation \eqref{eq:fermion-boson-stars:scalar-fermion-boson-stars:klein-gordon-equations} results in an equation describing the radial dependence of the bosonic field. It does not matter which equation of \eqref{eq:fermion-boson-stars:scalar-fermion-boson-stars:klein-gordon-equations} is used since, due to the chosen ansatz for the field, the complex phase will cancel out and will leave only the purely radial part $\varphi(r)$ in both cases. Finally, the $r$-component of the conservation equation for the energy-momentum tensor (left side of \eqref{eq:fermion-boson-stars:scalar-fermion-boson-stars:energy-momentum-and-restmass-conservation}) provides a differential equation for the pressure $P$. We find the full system of equations describing a radially symmetric static fermion boson star with a complex scalar field

\begin{subequations}
\begin{align}
a' = \frac{da}{dr} &=  \frac{a}{2} \, \left[ \frac{(1-a^2)}{r} + 8\pi r a^2  \left( \: e + \frac{\omega^2 \varphi^2}{\alpha^2} + \frac{{\Psi}^2 }{a^2} + V(\bar{\varphi} \varphi) \right) \right] \: , \label{eq:fermion-boson-stars:scalar-fermion-boson-stars:TOV-equations-grr} \\
\alpha' = \frac{d \alpha}{dr} &= \frac{\alpha}{2} \left[ \frac{(a^2 -1)}{r} + 8\pi r a^2 \left( P + \frac{\omega^2 \varphi^2}{\alpha^2} + \frac{{\Psi}^2 }{a^2} - V(\bar{\varphi} \varphi) \right) \right] \: , \label{eq:fermion-boson-stars:scalar-fermion-boson-stars:TOV-equations-gtt} \\
\Psi' = \frac{d \Psi}{dr} &= \phantom{\frac{\alpha}{2}} \left[ a^2 V'(\bar{\varphi} \varphi) - \frac{\omega^2 a^2}{\alpha^2} \right] \varphi + \left[\frac{a'}{a\,} - \frac{\alpha'}{\alpha\,} - \frac{2}{r} \right] \Psi \: , \label{eq:fermion-boson-stars:scalar-fermion-boson-stars:TOV-equations-phi} \\
\varphi' = \frac{d \varphi}{dr} &= \Psi \: , \label{eq:fermion-boson-stars:scalar-fermion-boson-stars:TOV-equations-psi} \\
P' =\frac{d P}{dr} &= -\left[ e + P \right] \frac{\alpha'}{\alpha} \: , \label{eq:fermion-boson-stars:scalar-fermion-boson-stars:TOV-equations-P}
\end{align}
\end{subequations}

where $V(\varphi \bar{\varphi})$ is the scalar field potential and the variable $\Psi(r) := \varphi'(r)$ (see eq. \eqref{eq:fermion-boson-stars:scalar-fermion-boson-stars:TOV-equations-psi}) was defined to obtain a system of first-order ordinary differential equations. Primes denote derivatives with respect to the radial coordinate $r$. This system of equations is closed by providing an EOS $P(e)$ (or $P(\rho,\epsilon)$) for the nuclear matter part. Plugging in the quartic self-interaction potential \eqref{eq:background:boson-stars:quartic-self-interaction-potential} gives the same equations as in \cite{DiGiovanni:2021ejn}, except for a different normalization of the field $\varphi$, which differs by a factor of $\sqrt{2}$. For the considered system and ansatz \eqref{eq:fermion-boson-stars:scalar-fermion-boson-stars:metric-ansatz} and \eqref{eq:fermion-boson-stars:scalar-fermion-boson-stars:scalar-field-harmonic-ansatz}, the expressions for the fermion number \eqref{eq:fermion-boson-stars:scalar-fermion-boson-stars:definition-NS-restmass} and boson number \eqref{eq:fermion-boson-stars:scalar-fermion-boson-stars:conserved-boson-number-definition} simplify to
\begin{subequations}
\begin{align}
N_\mathrm{f} =& 4\pi \int_0^{R_\mathrm{f}} a \rho r^2 dr \: , \label{eq:fermion-boson-stars:scalar-fermion-boson-stars:conserved-fermion-number-simplified} \\
N_\mathrm{b} =& 8\pi \int_0^\infty \frac{a \omega \varphi^2}{\alpha} r^2 dr \: . \label{eq:fermion-boson-stars:scalar-fermion-boson-stars:conserved-boson-number-simplified}
\end{align}
\end{subequations}

Here $R_\mathrm{f}$ denotes the fermionic radius, which is determined by the radial position at which the pressure $P$ of the neutron star component reaches zero. This is also the radius relevant to electromagnetic observations, as only the neutron matter component can be observed visually (by virtue of dark matter not interacting with photons). As is the case for pure boson stars, the total gravitational mass can only be defined in the limit of large radii, imposing that the solution asymptotically converges to the Schwarzschild solution:
\begin{align}
	M_\mathrm{tot} := \lim_{r \rightarrow \infty} \frac{r}{2} \left( 1 - \frac{1}{(a(r))^2 } \right) \: . \label{eq:fermion-boson-stars:scalar-fermion-boson-stars:definition-BS-total-gravitational-mass}
\end{align}

In order to integrate the equations \eqref{eq:fermion-boson-stars:scalar-fermion-boson-stars:TOV-equations-grr}-\eqref{eq:fermion-boson-stars:scalar-fermion-boson-stars:TOV-equations-P}, it is still necessary to provide suitable initial conditions. By imposing asymptotic flatness at $r \rightarrow \infty$ and regularity at the origin (i.e. no divergence at $r\rightarrow0$), we obtain \cite{Diedrichs:2023trk,DiGiovanni:2021ejn}
\begin{align}
    \begin{split}
        \lim_{r \rightarrow \infty} a(r) &= 1 \:\: , \:\: a(0) = 1 \: , \\
        \lim_{r \rightarrow \infty} \alpha(r) &= 1 \:\: , \:\: \alpha(0) = \alpha_0 \: , \\
        \lim_{r \rightarrow \infty} \varphi(r) &= 0 \:\: , \:\: \varphi(0) = \varphi_c \: , \\
        \lim_{r \rightarrow \infty} \Psi(r) &= 0 \:\: , \:\: \Psi(0) = 0 \: , \\
        \rho(r > R_\mathrm{f}) &= 0 \:\: , \:\: \rho(0) = \rho_c \: . \\
    \end{split} \label{eq:fermion-boson-stars:scalar-fermion-boson-stars:TOV-initial-conditions}
\end{align}

Since the scalar field frequency $\omega$ is not directly constrained by the equations of motion \eqref{eq:fermion-boson-stars:scalar-fermion-boson-stars:TOV-equations-grr}-\eqref{eq:fermion-boson-stars:scalar-fermion-boson-stars:TOV-equations-P}, it is necessary to tune its value to specific quantities, corresponding to the modes of the boson star component (see the discussion on the boundary conditions of boson stars in Chapter \ref{subsec:background:boson-stars}). The same scaling argument for the scalar field as used in Chapter \ref{subsec:background:boson-stars} holds here, which is why setting the initial condition $\alpha(0) = 1$ is possible for FBS as well (see the discussion after eq. \eqref{eq:background:boson-stars:TOV-initial-conditions}). The difference between boson stars and FBS is that the initial condition of the NS component (i.e. the central restmass density $\rho_c$) must be specified for FBS in addition to the central field amplitude $\varphi_c$. \\

Every FBS solution is characterized by the initial values for the central density $\rho_c$ and the central value of the scalar field $\varphi_c$. The question of the stability of FBS thus naturally arises, if one wants to study them in astrophysical contexts. In the case of pure boson stars and neutron stars, their stability to radial perturbations has been extensively studied in the past \cite{Liebling:2012fv} and is now well known: The stable and unstable solutions are separated by the point at which the total gravitational mass reaches its maximum with regard to the central density $\rho_c$ (for NS) and the central scalar field $\varphi_c$ (for BS). For example, this makes the NS to the left of the maximum in the MR curves in \autoref{fig:background:neutron-stars:MR-diagram-and-inner-NS-structure-example} (left panel) unstable under radial perturbation. \\
Since FBS are two-parameter solutions, the stability criterion needs to be modified. It was first presented for FBS by \cite{Henriques:1990xg} (also see \cite{Liebling:2012fv} for a review), although the criterion is more general and can also be applied to systems of two gravitationally interacting fluids. The idea behind the generalized stability criterion is to find extrema in the total number of particles (fermion number $N_\mathrm{f}$ or boson number $N_\mathrm{b}$) for a fixed total gravitational mass, depending on the central values $\rho_c$ and $\varphi_c$. The transition between stable and unstable configurations is given by the point at which
\begin{align}
    \frac{d N_\mathrm{f}}{d \sigma} = \frac{d N_\mathrm{b}}{d \sigma} = 0 \: , \label{eq:fermion-boson-stars:scalar-fermion-boson-stars:FBS-stability-criterion}
\end{align}

where $d/d\sigma$ denotes the derivative in the direction of constant total gravitational mass. Up to a normalization factor, \eqref{eq:fermion-boson-stars:scalar-fermion-boson-stars:FBS-stability-criterion} can be written as
\begin{equation}
    \frac{d N_\mathrm{f}}{d\sigma} \propto - \frac{\partial M_\text{tot}}{\partial \rho_c} \frac{\partial N_\mathrm{f}}{\partial \varphi_c} + \frac{\partial M_\text{tot}}{\partial \varphi_c} \frac{\partial N_\mathrm{f}}{\partial \rho_c} \: . \label{eq:fermion-boson-stars:scalar-fermion-boson-stars:FBS-stability-criterion-rewritten}
\end{equation}

If one is only interested in the precise points where FBS become unstable, the unspecified normalization factor in \eqref{eq:fermion-boson-stars:scalar-fermion-boson-stars:FBS-stability-criterion-rewritten} becomes irrelevant, since the whole relation is set to zero. In summary, the stability criterion \eqref{eq:fermion-boson-stars:scalar-fermion-boson-stars:FBS-stability-criterion} can be used to discriminate between astrophysically stable FBS solutions and unstable solutions. When perturbed, unstable solutions will either (also see \cite{Liebling:2012fv}) collapse to a black hole, dissipate to infinity (especially in the case of the scalar field) or migrate to a stable solution through internal re-configuration. \\

We proceed by deriving the tidal deformability of FBS as was first presented by \cite{Diedrichs:2023trk}. Thereby we follow the same procedure as was used in \cite{Hinderer:2007mb} to obtain the tidal deformability for pure NS (also see Chapter \ref{subsec:background:gravitational-waves-and-tidal-deformability}), and then combine the method with the one applied to pure boson stars \cite{Mendes:2016vdr,Sennett:2017etc}. The matter and gravitational field are first expanded around a static and spherically symmetric configuration. Then, these expansions are inserted into the equations of motion describing the FBS (Einstein equations and the Klein-Gordon equations) to obtain a system of differential equations that allows to solve for the linear perturbations, from which we then extract the tidal deformability. \\

Similar to the derivation of the tidal deformability of NS (see Chapter \ref{subsec:background:gravitational-waves-and-tidal-deformability}), we start by considering an external quadrupolar tidal field $\tensor{\mathcal{E}}{_i_j}$ acting on the FBS. This tidal field induces a quadrupolar moment $\tensor{Q}{_i_j}$ as a response, such that at linear order it can be written as $\tensor{Q}{_i_j} = - \lambda_{\mathrm{tidal}} \tensor{\mathcal{E}}{_i_j}$, where $\lambda_{\mathrm{tidal}}$ is the tidal deformability. The induced quadrupolar moment modifies the $\tensor{g}{_t_t}$ metric component. At leading order in the asymptotic rest frame of the FBS and at large radii, it can be written as
\begin{align}
	\tensor{g}{_t_t} = - \left( 1 - \frac{2 M}{r} \right) - \tensor{\mathcal{E}}{_i_j} \tensor{x}{^i} \tensor{x}{^j} \left( 1 + \frac{3 \lambda_{\mathrm{tidal}}}{r^5} \right) \: , \label{eq:fermion-boson-stars:scalar-fermion-boson-stars:perturbed-metric-external-tidal-field}
\end{align}

where $M$ is the total gravitational mass of the FBS and the $\tensor{x}{^i}$ are the position vectors in a Cartesian coordinate system with $\tensor{x}{^i} \tensor{x}{_i} = r^2$. We now consider small perturbations on top of the spacetime metric and the scalar field. For the metric, we consider a small perturbation $\tensor{h}{_\mu_\nu}$ on top of the unperturbed metric $\tensor{\overline{g}}{_\mu_\nu}$ \eqref{eq:fermion-boson-stars:scalar-fermion-boson-stars:metric-ansatz} such that
\begin{align}
	\tensor{g}{_\mu_\nu} = \tensor{\overline{g}}{_\mu_\nu} + \tensor{h}{_\mu_\nu} \: . \label{eq:fermion-boson-stars:scalar-fermion-boson-stars:perturbed-metric-general-formula}
\end{align}

For the static, even-parity and quadrupolar ($l=2$) metric perturbations in the Regge-Wheeler gauge, $\tensor{h}{_\mu_\nu}$ can be written in terms of the spherical harmonic $Y_{20}(\theta,\phi)$ and radial functions $H_0, H_2, K$ which describe the radial dependence of the perturbed metric components:
\begin{align}
	\tensor{h}{_\mu_\nu} = Y_{20}(\theta,\phi) \times \mathrm{diag}\left( - \alpha^2(r) H_0(r),\: a^2(r) H_2(r),\: r^2 K(r),\: r^2 \sin^2(\theta) K(r) \right) \: . \label{eq:fermion-boson-stars:scalar-fermion-boson-stars:perturbed-metric-small-tidal-perturbation}
\end{align}

We expand the scalar field in a similar manner by considering small perturbations $\delta \varphi$ on top of the unperturbed background $\varphi$ such that
\begin{align}
	\varphi(t,r,\theta,\phi) = \varphi(t,r) + \delta \varphi(t,r,\theta,\phi) \: , \label{eq:fermion-boson-stars:scalar-fermion-boson-stars:perturbed-scalar-field-general-formula}
\end{align}

where the unperturbed scalar field background $\varphi(t,r)$ exactly corresponds to the scalar field used for the static FBS (see eq. \eqref{eq:fermion-boson-stars:scalar-fermion-boson-stars:scalar-field-harmonic-ansatz} and \eqref{eq:fermion-boson-stars:scalar-fermion-boson-stars:TOV-equations-grr}-\eqref{eq:fermion-boson-stars:scalar-fermion-boson-stars:TOV-equations-P}). As was done in \cite{Diedrichs:2023trk,Sennett:2017etc}, we make the following ansatz for the first-order perturbations of the scalar field
\begin{align}
	\delta \varphi(t, r, \theta, \varphi) = \varphi_1(r) \frac{e^{- i \omega t}}{r} Y_{20}(\theta, \phi) \: ,
\end{align}

where the same time-dependence was chosen for the perturbations in order to ensure that the energy-momentum tensor remains static. $\varphi_1(r)$ describes the radial perturbation of the scalar field and fulfils the same role as $\tensor{h}{_\mu_\nu}$ does for the metric \eqref{eq:fermion-boson-stars:scalar-fermion-boson-stars:perturbed-metric-general-formula}. The spherical harmonic $Y_{20}(\theta,\phi)$ is chosen since it describes a dipolar deformation, which is present in the case of an external tidal field. We are now able to obtain a set of differential equations that relates the perturbations to the background solutions by expanding the Einstein equations (see \eqref{eq:fermion-boson-stars:scalar-fermion-boson-stars:einstein-equations-fbs}) and the Klein-Gordon equation \eqref{eq:fermion-boson-stars:scalar-fermion-boson-stars:klein-gordon-equations} to first order in $\tensor{h}{_\mu_\nu}$ and $\varphi_1(r)$. Inserting the expressions \eqref{eq:fermion-boson-stars:scalar-fermion-boson-stars:perturbed-metric-general-formula} and \eqref{eq:fermion-boson-stars:scalar-fermion-boson-stars:perturbed-scalar-field-general-formula} into equation \eqref{eq:fermion-boson-stars:scalar-fermion-boson-stars:klein-gordon-equations} and only keeping terms linear in the perturbations results in
\begin{align}
\begin{split}
    \varphi_1'' = \left[ \frac{a'}{a} - \frac{\alpha'}{\alpha} \right]  \varphi_1' 
     &+  \left[ - \omega^2 \frac{a^2}{\alpha^2} + 32\pi {\Psi}^2 + 2 \varphi^2 a^2 V'' + a^2 V' - \frac{a'}{ra} + \frac{\alpha'}{r\alpha} + 6\frac{a^2}{r^2} \right] \varphi_1 \\
     &+  \left[ \omega^2 r \varphi \frac{a^2}{\alpha^2} - r\Psi' + \left( r \frac{a'}{a} + r\frac{\alpha'}{\alpha} -2 \right) \Psi \right] H_0  \: ,
\end{split}
\label{eq:fermion-boson-stars:scalar-fermion-boson-stars:perturbed-equations-phi1} 
\end{align}
where $V''$ is the second derivative of the potential with the derivative as defined in \eqref{eq:fermion-boson-stars:scalar-fermion-boson-stars:derivative-of-scalar-potential}. We proceed in a similar way with the Einstein equations \eqref{eq:fermion-boson-stars:scalar-fermion-boson-stars:einstein-equations-fbs}. We insert the perturbed metric \eqref{eq:fermion-boson-stars:scalar-fermion-boson-stars:perturbed-metric-general-formula} and scalar field \eqref{eq:fermion-boson-stars:scalar-fermion-boson-stars:perturbed-scalar-field-general-formula} and only keep the terms in linear order of the perturbations $\tensor{h}{_\mu_\nu}$ and $\varphi_1(r)$. We obtain the perturbed Einstein equations
\begin{align}
 \delta \tensor{G}{_\mu_\nu} = 8\pi \left( \delta T_{\mu \nu}^{(NS)} + \delta T_{\mu \nu}^{(\varphi)} \right) \: . \label{eq:fermion-boson-stars:scalar-fermion-boson-stars:perturbed-einstein-equations}
\end{align}

Here, the perturbed energy-momentum tensor of the fermionic part is given by 
\begin{align}
	\delta\tensor{T}{_\mu_\nu} = \mathrm{diag}\left( -\delta P/c_s^2,\: \delta P,\: \delta P,\: \delta P \right) \: , \label{eq:fermion-boson-stars:scalar-fermion-boson-stars:perturbed-energy-momentum-tensor}
\end{align}

where we used the relationship $\delta e = \delta P\, \partial e/\partial P = \delta P /c_s^2$ between the energy density $e$, pressure $P$ and the local speed of sound in the medium $c_s$ (see \cite{Diedrichs:2023trk}). The perturbed energy-momentum tensor of the scalar field $\delta T_{\mu \nu}^{(\varphi)}$ is computed by expanding \eqref{eq:fermion-boson-stars:scalar-fermion-boson-stars:energy-momentum-tensor-scalar-field} to linear order in $\tensor{h}{_\mu_\nu}$ and $\varphi_1(r)$. \\
Subtracting the $\theta\theta$-component from the $\phi\phi$-component of the perturbed Einstein equations \eqref{eq:fermion-boson-stars:scalar-fermion-boson-stars:perturbed-einstein-equations} reveals $H_2(r) = - H_0(r)$. Adding the $\theta\theta$-component to the $\phi\phi$-component allows us to obtain an expression for $\delta P$, which can be substituted into the $tt$- minus the $rr$-component to obtain a differential equation for $H_0$:
\begin{align}
    & H_0'' - \left[ \frac{a'}{a} - \frac{\alpha'} {\alpha} - \frac{2}{r}\right]  H_0' \label{eq:fermion-boson-stars:scalar-fermion-boson-stars:perturbed-equations-H0} \\
    & -  \left[ 8 \pi \omega^2 \varphi^2 \frac{a^2}{ \alpha^2}  \frac{1- c_s^2}{c_s^2}  + 8\pi {\Psi}^2 \frac{1+3c_s^2}{c_s^2} - 2 \frac{\alpha''}{\alpha} + 2 \frac{\alpha' a'}{ \alpha  a} + 4 \frac{\alpha'^2}{\alpha^2} - \frac{a'}{ra} \frac{1+3c_s^2}{c_s^2} - \frac{\alpha'}{r\alpha} \frac{1+7c_s^2}{c_s^2} + 6 \frac{a^2}{r^2} \right] H_0 \nonumber \\
    &= 16\pi \left[ \omega^2 \varphi \frac{a^2}{r\alpha^2} \frac{c_s^2 -1}{c_s^2} +  \varphi V' \frac{a^2}{r} \frac{1+c_s^2}{c_s^2} - \frac{\Psi'}{ r} \frac{1+3c_s^2}{c_s^2} + \Psi \frac{a'}{r a} \frac{1+3c_s^2}{c_s^2} +   \Psi \frac{\alpha'} {r \alpha} \frac{c_s^2 -1}{c_s^2} - 2  \frac{\Psi}{r^2} \frac{1+3c_s^2}{c_s^2} \right]  \varphi_1 \: . \nonumber
\end{align}

The above equation contains a term depending on the second derivative of the metric component $\alpha''$, which is explicitly given by
\begin{align}
\alpha'' =& \frac{4\pi \omega^2}{\alpha} \left[ 2 r \varphi^2 a a' + 2r \varphi a^2 \Psi + \varphi^2 a^2 \right] + \left[ 4 \pi r a^2 \left(- \frac{\omega^2 \varphi^2}{\alpha^2} + P - V + \frac{{\Psi}^2}{a^2} \right) + \frac{a^2 - 1}{2r} \right]  \alpha' \label{eq:fermion-boson-stars:scalar-fermion-boson-stars:perturbed-equations-ddalpha} \\ 
		 +& \left[ 4 \pi r ( 2 P a a' - 2 V a a'- 2 \varphi a^2 \Psi V' + a^2 P' + 2 \Psi \Psi') + 4 \pi a^2 (P - V) + 4 \pi {\Psi}^2 + \frac{aa'}{r} + \frac{1 - a^2}{2r^2} \right] \alpha \: . \nonumber
\end{align}

As mentioned in \cite{Diedrichs:2023trk,Sennett:2017etc}, for radii larger than the typical size of the combined system, the differential equation for $H_0$ \eqref{eq:fermion-boson-stars:scalar-fermion-boson-stars:perturbed-equations-H0} reduces to
\begin{align}
	H_0'' + \left( \frac{2}{r} + a^2 \frac{2 M}{r^2} \right) H_0' - \left( \frac{6 a^2}{r^2} + a^4 \frac{4 M^2}{r^4} \right) H_0 = 0 \: . \label{eq:fermion-boson-stars:scalar-fermion-boson-stars:perturbed-H-parameter-1-H-ode-ourside-of-source}
\end{align}

This equation \eqref{eq:fermion-boson-stars:scalar-fermion-boson-stars:perturbed-H-parameter-1-H-ode-ourside-of-source} has a solution in terms of the associated Legendre polynomials (see \cite{Diedrichs:2023trk,Hinderer:2007mb,Thorne:1997kt}), which can be expanded in $r/M$ and matched to the metric component \eqref{eq:fermion-boson-stars:scalar-fermion-boson-stars:perturbed-metric-external-tidal-field} to obtain an expression for the tidal deformability:
\begin{align}
	\lambda_\mathrm{tidal} =& \frac{16}{15} M^5 (1 - 2 C)^2 [2 + 2 C (y - 1) - y] \times \{3 (1 - C)^2 [2 - y + 2 C (y - 1)] \log (1 - 2 C) \nonumber \\
    &+ 2 C [6 - 3y + 3 C (5 y - 8)] + 4 C^3 [13 - 11 y + C (3y - 2) + 2 C^2 (1 + y)] \}^{-1} \: , \label{eq:fermion-boson-stars:scalar-fermion-boson-stars:tidal-deformablity-equation}
\end{align}

where $C := M_\mathrm{ext} / r_\mathrm{ext}$ is the effective compactness of the FBS, $y := r_\mathrm{ext} H_0'(r_\mathrm{ext}) / H_0(r_\mathrm{ext})$ and $r_\mathrm{ext}$ denotes the radial position at which $\lambda_\mathrm{tidal}$ is calculated. Typically $r_\mathrm{ext}$ is situated at large distances outside the source. The dimensionless tidal deformability is defined as $\Lambda_\mathrm{tidal} := \lambda_\mathrm{tidal} / M_\text{tot}^5$. The parameter $y$ can be obtained by integrating equation \eqref{eq:fermion-boson-stars:scalar-fermion-boson-stars:perturbed-equations-H0}, together with the TOV equations \eqref{eq:fermion-boson-stars:scalar-fermion-boson-stars:TOV-equations-grr}-\eqref{eq:fermion-boson-stars:scalar-fermion-boson-stars:TOV-equations-P}, from the centre of the star at $r=0$ to the radius $r=r_\mathrm{ext}$ where the integration converges sufficiently. \\

We obtain the boundary conditions for the perturbed fields $\varphi_1(r)$ and $H_0(r)$ with the same procedure as described in \cite{Mendes:2016vdr}. First, we expand the quantities around the origin ($r=0$) such that
\begin{align}
\varphi_1(r) = \sum_{i=0} \varphi_1^{(i)} r^i \:\: , \:\: H_0(r) = \sum_{i=0} H_0^{(i)} r^i \: , \label{eq:fermion-boson-stars:scalar-fermion-boson-stars:expansions_phi_H0}
\end{align}

where the coefficients $\varphi_1^{(i)}$ and $H_0^{(i)}$ do not depend on the radius. The expansions are then inserted into equations \eqref{eq:fermion-boson-stars:scalar-fermion-boson-stars:perturbed-equations-phi1} and \eqref{eq:fermion-boson-stars:scalar-fermion-boson-stars:perturbed-equations-H0}. After solving the resulting polynomial equations, the leading-order behaviour for $\varphi_1(r)$ and $H_0(r)$ is found to be
\begin{align}
\varphi_1(r) \approx \varphi_1^{(3)} r^3 + \mathcal{O}(r^5) \:\: , \:\: H_0(r) \approx H_0^{(2)} r^2 + \mathcal{O}(r^4) \: . \label{eq:fermion-boson-stars:scalar-fermion-boson-stars:expansions_phi_H0-simplified}
\end{align}

We additionally demand that the perturbation to the scalar field vanishes at large radii (i.e. we assume asymptotic flatness). This leads to the boundary conditions
\begin{align}
\begin{split}
    \lim_{r \rightarrow 0 } H_0 \approx H_0^{(2)} r^{2} \: , \qquad  & \lim_{r\to0} H_0' \approx 2 H_0^{(2)} r \: , \\
    \lim_{r \rightarrow \infty } \varphi_1(r) = 0 \: , \qquad  & \lim_{r\to0} \varphi_1' \approx 3 \varphi_1^{(3)} r^2 \: .
\end{split} \label{eq:fermion-boson-stars:scalar-fermion-boson-stars:perturbations-initial-conditions}
\end{align}

Since the equations \eqref{eq:fermion-boson-stars:scalar-fermion-boson-stars:perturbed-equations-phi1} and \eqref{eq:fermion-boson-stars:scalar-fermion-boson-stars:perturbed-equations-H0} are invariant under a simultaneous scaling of $\varphi_1(r)$ and $H_0(r)$ by a constant, we can fix $H_0^{(2)} = 1$, which only leaves the parameter $\varphi_1^{(3)}$ to be determined. Note that the scaling of $H_0(r)$ does not affect the tidal deformability \eqref{eq:fermion-boson-stars:scalar-fermion-boson-stars:tidal-deformablity-equation}, since the parameter $y = r_\mathrm{ext} H_0'(r_\mathrm{ext}) / H_0(r_\mathrm{ext})$ is invariant when scaling $H_0(r)$ by a constant value. The value of $\varphi_1^{(3)}$ is found similarly to the frequency $\omega$ by tuning $\varphi_1^{(3)}$ so that the boundary conditions \eqref{eq:fermion-boson-stars:scalar-fermion-boson-stars:perturbations-initial-conditions} are fulfilled. \\

We have now a fully developed analytical framework to study FBS, where the bosonic component is modelled by a complex scalar field. In particle physics, scalar fields describe bosons with spin-0. The next logical step is thus to consider different types of bosons inside of an FBS. Vector bosons (spin-1 particles) in particular are interesting candidates as they appear in every gauge theory of the standard model. Some dark matter candidates, such as the dark photon (see \cite{Biswas:2022tcw,An:2023mvf,Siemonsen:2022ivj}), are vector bosons as well. If vector bosons were to condensate inside neutron stars, they would form objects similar to FBS with scalar bosons. The next chapter is dedicated to these systems, which we call fermion Proca stars (FPS).

\subsection{Fermion Proca Stars} \label{subsec:fermion-boson-stars:fermion-proca-stars}

For the first time, we present the mathematical framework behind fermion Proca stars (FPS). FPS are combined systems of fermions and vector bosons, which interact only gravitationally with each other. Conceptionally, they are similar to fermion boson stars with a bosonic field, in that they share a number of properties. In particular the physical motivation is similar, as both FBS and FPS can be seen as a macroscopic Bose-Einstein condensate which coexists inside (or around) a neutron star (NS) at the same point in space. We model the FPS using a relativistic fluid for the NS component, and a complex vector field for the bosonic component. The action describing FPS is the combination of the Einstein-Proca system \eqref{eq:background:boson-stars:einstein-proca-action} minimally coupled to a matter term $\mathcal{L}_m$:
\begin{align}
S = \int_\mathcal{M} \sqrt{-g} \left( \frac{1}{2\kappa} R - \frac{1}{2} \tensor{F}{_\mu_\nu} \tensor{\bar{F}}{^\mu^\nu} - V(\tensor{A}{_\mu} \tensor{\bar{A}}{^\mu}) - \mathcal{L}_{m} \right) dx^4 \: , \label{eq:fermion-boson-stars:fermion-proca-stars:fermion-proca-star-action}
\end{align}

where $R$ is the Ricci curvature scalar, $g$ is the determinant of the spacetime metric $\tensor{g}{_\mu_\nu}$, $\kappa = 4\pi G/c^4$ is a constant and the integral is performed over the whole spacetime manifold $\mathcal{M}$. $\tensor{F}{_\mu_\nu} = \tensor{\nabla}{_\mu} \tensor{A}{_\nu} - \tensor{\nabla}{_\nu} \tensor{A}{_\mu}$ is the antisymmetric field strength tensor and $V(\tensor{A}{_\mu} \tensor{\bar{A}}{^\mu})$ is the vector field potential that depends solely on the magnitude of the vector field $\tensor{A}{_\mu} \tensor{\bar{A}}{^\mu}$. By taking the variation of \eqref{eq:fermion-boson-stars:fermion-proca-stars:fermion-proca-star-action} with respect to the inverse spacetime metric $\delta \tensor{g}{^\mu^\nu}$, one obtains the Einstein equations
\begin{align}
\tensor{G}{_\mu_\nu} =& \kappa \, \left( T_{\mu \nu}^{(NS)} + T_{\mu \nu}^{(A)} \right) \: , \label{eq:fermion-boson-stars:fermion-proca-stars:einstein-equations-fps}
\end{align}

where $T_{\mu \nu}^{(NS)}$ and $T_{\mu \nu}^{(A)}$ are the energy-momentum tensors describing the neutron star matter and the vector field matter respectively. The energy-momentum tensor of the neutron star matter is, analogously to FBS, taken to be that of a perfect fluid (see equation \eqref{eq:fermion-boson-stars:scalar-fermion-boson-stars:energy-momentum-tensor-perfect-fluid}). The conservation equations for energy-momentum and the restmass \eqref{eq:fermion-boson-stars:scalar-fermion-boson-stars:energy-momentum-and-restmass-conservation} apply as well. We also define the fermion number $N_\mathrm{f}$ in the same way as we did for FBS (see eq. \eqref{eq:fermion-boson-stars:scalar-fermion-boson-stars:definition-NS-restmass}). The energy-momentum tensor of the vector field is given by
\begin{align}
T_{\mu \nu}^{(A)} =  \tensor{F}{_\mu_\rho} \tensor{\bar{F}}{_\nu^\rho} + \tensor{\bar{F}}{_\mu_\rho} \tensor{F}{_\nu^\rho}  - \frac{1}{2} \tensor{g}{_\mu_\nu} \tensor{F}{^\rho^\sigma} \tensor{\bar{F}}{_\rho_\sigma}
+ \tensor{g}{_\mu_\nu} V(\tensor{A}{_\rho} \tensor{\bar{A}}{^\rho}) + V'(\tensor{A}{_\rho} \tensor{\bar{A}}{^\rho}) ( \tensor{A}{_\mu} \tensor{\bar{A}}{_\nu} + \tensor{A}{_\nu} \tensor{\bar{A}}{_\mu} ) \: , \label{eq:fermion-boson-stars:fermion-proca-stars:energy-momentum-tensor-vector-field}
\end{align}

where the derivative of the potential $V$ is given by
\begin{align}
 V'(A_\rho \bar{A}^\rho ) := \frac{d V(A_\rho \bar{A}^\rho ) }{d (A_\rho \bar{A}^\rho)}  \: . \label{eq:fermion-boson-stars:fermion-proca-stars:derivative-of-vector-potential}
\end{align}

The equations of motion (Proca equations) of the vector field are computed from the action \eqref{eq:fermion-boson-stars:fermion-proca-stars:fermion-proca-star-action} using the Euler-Lagrange equations for a complex vector field:
\begin{align}
0 = \frac{\delta \mathcal{L}}{\delta A^\mu} - \nabla_\mu \frac{\delta \mathcal{L}}{\delta (\nabla^\mu A^\nu)} \:\:\:\:\: \text{and} \:\:\:\:\:  0 = \frac{\delta \mathcal{L}}{\delta \bar{A}^\mu} - \nabla_\mu \frac{\delta \mathcal{L}}{\delta (\nabla^\mu \bar{A}^\nu)} \: . \label{eq:fermion-boson-stars:fermion-proca-stars:euler-lagrange-equations}
\end{align}

One obtains the equations of motion for the field and the complex conjugate:

\begin{align}
\tensor{\nabla}{^\mu} \tensor{\bar{F}}{_\mu_\nu} = V'(\tensor{A}{_\rho} \tensor{\bar{A}}{^\rho} ) \tensor{\bar{A}}{_\nu} \:\:\:\:\: \text{and} \:\:\:\:\:  \tensor{\nabla}{^\mu} \tensor{F}{_\mu_\nu} = V'(\tensor{A}{_\rho} \tensor{\bar{A}}{^\rho} ) \tensor{A}{_\nu} \: . \label{eq:fermion-boson-stars:fermion-proca-stars:proca-equations}
\end{align} 

The covariant derivative of \eqref{eq:fermion-boson-stars:fermion-proca-stars:proca-equations} being zero -- i.e. $\tensor{\nabla}{^\mu} \tensor{\nabla}{^\nu} \tensor{F}{_\mu_\nu} = 0$ -- leads to a dynamical constraint on the field derivative, resembling the Lorentz condition used in the Maxwell and Proca equations (also see \cite{Liebling:2012fv,Brito:2015pxa}):
\begin{align}
\tensor{\nabla}{^\nu} \tensor{A}{_\nu} = - \frac{\tensor{\nabla}{^\nu} \left[ V'(\tensor{A}{_\rho} \tensor{\bar{A}}{^\rho} ) \right]}{ V'(\tensor{A}{_\rho} \tensor{\bar{A}}{^\rho} ) } \tensor{A}{_\nu} = - \tensor{\nabla}{^\nu} \left[ \ln\left( V'(\tensor{A}{_\rho} \tensor{\bar{A}}{^\rho} ) \right) \right] \tensor{A}{_\nu} \: . \label{eq:fermion-boson-stars:fermion-proca-stars:proca-constraint}
\end{align}

The global $U(1)$-symmetry in the Lagrangian \eqref{eq:fermion-boson-stars:fermion-proca-stars:fermion-proca-star-action} under the transformation of the vector field $\tensor{A}{_\mu}$ (and $ \tensor{\bar{A}}{_\mu}$) gives rise to a conserved Noether current
\begin{align}
j^\mu = i \left( \bar{F}^{\mu\nu} A_\nu - F^{\mu\nu} \bar{A}_\nu \right) \: . \label{eq:fermion-boson-stars:fermion-proca-stars:noether-current}
\end{align}

The boson number $N_\mathrm{b}$ can be defined analogously to \eqref{eq:fermion-boson-stars:scalar-fermion-boson-stars:conserved-boson-number-definition} from the conserved Noether current \eqref{eq:fermion-boson-stars:fermion-proca-stars:noether-current}. \\

We proceed by solving the Einstein equations \eqref{eq:fermion-boson-stars:fermion-proca-stars:einstein-equations-fps} and the Proca equations \eqref{eq:fermion-boson-stars:fermion-proca-stars:proca-equations} for spherically symmetric and static configurations in equilibrium. Exactly as with FBS, we consider the spherically symmetric metric ansatz \eqref{eq:fermion-boson-stars:scalar-fermion-boson-stars:metric-ansatz}, which leads to the velocity of the perfect fluid to be \eqref{eq:fermion-boson-stars:scalar-fermion-boson-stars:four-velocity-static-fluid}. For the vector field, we employ the harmonic time-dependence ansatz and a purely radial vector field (see \cite{Brito:2015pxa,Minamitsuji:2018kof}), which is similarly motivated as the harmonic phase ansatz for the scalar field \eqref{eq:fermion-boson-stars:scalar-fermion-boson-stars:scalar-field-harmonic-ansatz} (also see Chapter \ref{subsec:background:boson-stars}). The vector field is given by
\begin{align}
A_\mu (t,x) = e^{-i \omega t} (E(r), iB(r), 0, 0)  \: , \label{eq:fermion-boson-stars:fermion-proca-stars:vector-field-harmonic-ansatz}
\end{align}

where $\omega$ is the vector field frequency and $E(r)$, $B(r)$ are purely radial functions (note that $E(r)$ and $B(r)$ have nothing to do with electromagnetic fields and the similarity in notation is purely coincidental). Using the metric ansatz 
\begin{align}
ds^2 = \tensor{g}{_\mu_\nu} \tensor{dx}{^\mu} \tensor{dx}{^\nu} = - \alpha^2(r) dt^2 + a^2(r) dr^2 + r^2 d\theta^2 + r^2\sin^2(\theta) d\phi^2 \: , \label{eq:fermion-boson-stars:fermion-proca-stars:metric-ansatz}
\end{align}

together with the field ansatz \eqref{eq:fermion-boson-stars:fermion-proca-stars:vector-field-harmonic-ansatz}, one can obtain first-order differential equations for the metric components $\alpha(r)$ and $a(r)$ by following the same algebra steps as explained under equation \eqref{eq:fermion-boson-stars:scalar-fermion-boson-stars:scalar-field-harmonic-ansatz} in Chapter \ref{subsec:fermion-boson-stars:scalar-fermion-boson-stars}. The evolution equations for the vector field components can be computed from the Proca equations \eqref{eq:fermion-boson-stars:fermion-proca-stars:proca-equations}. The $\nu=r$ component yields the equation of motion for $E(r)$:
\begin{align}
 E'  = \frac{dE}{dr} = - V'(A_\rho \bar{A}^\rho ) \frac{B \alpha^2}{\omega} + \omega B  \: . \label{eq:fermion-boson-stars:fermion-proca-stars:general-eq-FBS-E}
\end{align}

Next, we consider the $\nu=t$ component of \eqref{eq:fermion-boson-stars:fermion-proca-stars:proca-equations} and obtain the equation of motion for $B(r)$:

\begin{align}
 (E'' - \omega B') - \left( \frac{a'}{a\,} + \frac{\alpha'}{\alpha\,} - \frac{2}{r} \right) (E' - \omega B) = V'(A_\rho \bar{A}^\rho ) \: a^2 E \: . \label{eq:fermion-boson-stars:fermion-proca-stars:general-eq-FBS-B}
\end{align}

This equation depends on the second radial derivative $E''$. It is possible to eliminate this dependence by taking the radial derivative of \eqref{eq:fermion-boson-stars:fermion-proca-stars:general-eq-FBS-E} and substituting it into \eqref{eq:fermion-boson-stars:fermion-proca-stars:general-eq-FBS-B}. After re-arranging the resulting equations, one obtains the final evolution equation for $B'$. A more detailed derivation with additional in-between steps can be found in Appendix \ref{sec:appendix:derivation-of-fermion-proca-stars}. Finally, the $r$-component of the conservation equation for the energy-momentum tensor (left side of \eqref{eq:fermion-boson-stars:scalar-fermion-boson-stars:energy-momentum-and-restmass-conservation}) provides a differential equation for the pressure $P(r)$. The full equations of motion for the Einstein-Proca system coupled to matter are thus:
\begin{subequations} 
\begin{align}
a' = \frac{da}{dr} &=  \frac{a}{2} \left[ \frac{(1-a^2)}{r} + 8\pi r a^2 \left( \:\: e + \frac{1}{\alpha^2 a^2} (E' - \omega B)^2 + V(A_\rho \bar{A}^\rho ) + 2 V'(A_\rho \bar{A}^\rho ) \frac{E^2}{\alpha^2}  \right)  \right] \: , \label{eq:fermion-boson-stars:fermion-proca-stars:TOV-equations-grr} \\
\alpha' = \frac{d \alpha}{dr} &= \frac{\alpha}{2} \left[ \frac{(a^2 -1)}{r} + 8\pi r a^2 \left( P - \frac{1}{\alpha^2 a^2} (E' - \omega B)^2 - V(A_\rho \bar{A}^\rho ) + 2 V'(A_\rho \bar{A}^\rho ) \frac{B^2}{a^2}  \right)  \right] \: , \label{eq:fermion-boson-stars:fermion-proca-stars:TOV-equations-gtt} \\
E'  = \frac{dE}{dr} &= - V'(A_\rho \bar{A}^\rho ) \frac{B \alpha^2}{\omega} + \omega B  \: , \label{eq:fermion-boson-stars:fermion-proca-stars:TOV-equations-E} \\
B' = \frac{dB}{dr} &= \left\{ V''(A_\rho \bar{A}^\rho ) \left( \frac{2B^2 a'}{a^3} + \frac{2E E'}{\alpha^2} - \frac{2E^2 \alpha'}{\alpha^3} \right) \frac{B \alpha^2}{\omega} - V'(A_\rho \bar{A}^\rho ) \left( a^2 E + \frac{2 B \alpha \alpha'}{\omega} \right) \right. \nonumber \\
&- \left. \left( \frac{a'}{a\,} + \frac{\alpha'}{\alpha\,} - \frac{2}{r} \right) (E' - \omega B) \right\} \left( V''(A_\rho \bar{A}^\rho ) \frac{2}{\omega} \frac{B^2 \alpha^2}{a^2} + V'(A_\rho \bar{A}^\rho ) \frac{\alpha^2}{\omega} \right)^{-1} \: , \label{eq:fermion-boson-stars:fermion-proca-stars:TOV-equations-B} \\
P' =\frac{d P}{dr} &= -\left[ e + P \right] \frac{\alpha'}{\alpha} \: . \label{eq:fermion-boson-stars:fermion-proca-stars:TOV-equations-P}
\end{align}
\end{subequations}

This system of equations is closed by providing an EOS $P(e)$ (or $P(\rho,\epsilon)$) for the nuclear matter part. Note that all equations are a first-order ODE respectively. This is different to the scalar field case where the additional variable $\Psi(r)$ \eqref{eq:fermion-boson-stars:scalar-fermion-boson-stars:TOV-equations-psi} has to be introduced to make the system first-order. Another difference is that no derivative of the potential enters the equations of motion for the metric components \eqref{eq:fermion-boson-stars:scalar-fermion-boson-stars:TOV-equations-grr}-\eqref{eq:fermion-boson-stars:scalar-fermion-boson-stars:TOV-equations-gtt} in the scalar field case, but it does for the vector field case \eqref{eq:fermion-boson-stars:fermion-proca-stars:TOV-equations-grr}-\eqref{eq:fermion-boson-stars:fermion-proca-stars:TOV-equations-gtt}. For the considered system and ansatz for the metric \eqref{eq:fermion-boson-stars:fermion-proca-stars:metric-ansatz} and vector field \eqref{eq:fermion-boson-stars:fermion-proca-stars:vector-field-harmonic-ansatz}, the expressions for the fermion number \eqref{eq:fermion-boson-stars:scalar-fermion-boson-stars:definition-NS-restmass} and boson number \eqref{eq:fermion-boson-stars:scalar-fermion-boson-stars:conserved-boson-number-definition} simplify to
\begin{subequations}
\begin{align}
N_\mathrm{f} &= 4\pi \int_0^{R_\mathrm{f}} a \rho r^2 dr \: , \label{eq:fermion-boson-stars:fermion-proca-stars:conserved-fermion-number-simplified} \\
N_\mathrm{b} &= 8 \pi \int_0^\infty B \frac{(\omega B - E')}{\alpha a} r^2 dr \: . \label{eq:fermion-boson-stars:fermion-proca-stars:conserved-boson-number-simplified}
\end{align}
\end{subequations}

$R_\mathrm{f}$ denotes the fermionic radius, which is determined by the radial position at which the pressure $P$ of the neutron star component reaches zero. The total gravitational mass is defined analogously to FBS \eqref{eq:fermion-boson-stars:scalar-fermion-boson-stars:definition-BS-total-gravitational-mass} in the limit of large radii, imposing that the solution asymptotically converges to the Schwarzschild solution:
\begin{align}
	M_\mathrm{tot} := \lim_{r \rightarrow \infty} \frac{r}{2} \left( 1 - \frac{1}{(a(r))^2 } \right) \: . \label{eq:fermion-boson-stars:fermion-proca-stars:definition-BS-total-gravitational-mass}
\end{align}

In the following, we derive the boundary conditions of equations \eqref{eq:fermion-boson-stars:fermion-proca-stars:TOV-equations-grr}-\eqref{eq:fermion-boson-stars:fermion-proca-stars:TOV-equations-P} at $r = 0$ and at $r= \infty$. The values at the origin will later serve as initial conditions for the numerical integration. We first consider the equations of motion in the limit $r \rightarrow 0$ while imposing regularity at the origin (i.e. the solution must not diverge). We first analyse equation \eqref{eq:fermion-boson-stars:fermion-proca-stars:TOV-equations-grr} in this limit. The term proportional to $1/r$ will dominate at small radii and will diverge if $r \rightarrow 0$. Thus, the only way to maintain regularity is to set $a(r=0) = 1$. It directly follows that $a'(r=0) = 0$ and also, by virtue of equation \eqref{eq:fermion-boson-stars:fermion-proca-stars:TOV-equations-gtt} that $\alpha'(r=0) = 0$. The exact value of $\alpha (r=0) = \alpha_0$ is a priori undetermined and can be chosen in a way thought suitable. \\
The initial conditions for the vector field components $E(r)$ and $B(r)$ can be obtained in a similar manner. We first consider eq. \eqref{eq:fermion-boson-stars:fermion-proca-stars:TOV-equations-B}. In the limit $r\rightarrow 0$, the term proportional to $1/r$ will dominate and regularity then demands, that $E' = \omega B$. It follows immediately that $B'(r=0) = 0$. This result can be inserted into equation \eqref{eq:fermion-boson-stars:fermion-proca-stars:TOV-equations-E}, which leads to the relation
\begin{align}
 E'  =  \omega B = - V'(A_\rho \bar{A}^\rho ) \frac{B \alpha^2}{\omega} + \omega B  \:\:\:\: \Longrightarrow \:\:\:\: 0 = V'(A_\rho \bar{A}^\rho ) B \alpha^2\: .
\end{align}

Since at $r=0$, $\alpha(r=0) \neq 0$ and $V' \neq 0$ in general, this relation can only be fulfilled if we demand that $B(r=0) = 0$. Plugging this new finding back into eq. \eqref{eq:fermion-boson-stars:fermion-proca-stars:TOV-equations-E} yields $E'(r=0) = 0$. The central value of the field $E'(r=0) = E_0$ is therefore undetermined by the equations of motion and thus is a free parameter of the theory. \\
A similar analysis at large distances reveals the boundary conditions at $r \rightarrow \infty$ for all variables. At large radii, we impose the flat spacetime limit, which necessitates that $a(r \rightarrow \infty) = \alpha(r \rightarrow \infty) = 1$. All terms proportional to $r$ in the equations \eqref{eq:fermion-boson-stars:fermion-proca-stars:TOV-equations-grr} and \eqref{eq:fermion-boson-stars:fermion-proca-stars:TOV-equations-gtt} must vanish at infinity to fulfil the flat spacetime limit. Therefore, the vector field components must vanish at infinity: $E(r \rightarrow \infty) = 0$ and $B(r \rightarrow \infty) = 0$. Also, pressure $P(r)$ and energy density $e(r)$ must be zero outside of the NS component of the FPS, as well as the restmass density $\rho$. In fact, the pressure and energy density of the neutron star component will become zero at some final radius $R_\mathrm{f}$ (the fermionic radius). We summarize all boundary conditions in the following:
\begin{align}
    \begin{split}
        \lim_{r \rightarrow \infty} a(r) &= 1 \:\: , \:\: a(0) = 1 \: , \\
        \lim_{r \rightarrow \infty} \alpha(r) &= 1 \:\: , \:\: \alpha(0) = \alpha_0 \: , \\
        \lim_{r \rightarrow \infty} E(r) &= 0 \:\: , \:\: E(0) = E_0 \: , \\
        \lim_{r \rightarrow \infty} B(r) &= 0 \:\: , \:\: B(0) = 0 \: , \\
        \rho(r > R_\mathrm{f}) &= 0 \:\: , \:\: \rho(0) = \rho_c \: .
    \end{split} \label{eq:fermion-boson-stars:fermion-proca-stars:TOV-initial-conditions}
\end{align}

The initial condition for the metric component $\alpha(0) = \alpha_0$ is fixed by its behaviour at infinity. In the case of boson stars, and subsequently FBS with a scalar field, it was possible to scale the frequency to absorb the initial value of $\alpha_0$ so that it may be set to one (see the discussion after eq. \eqref{eq:background:boson-stars:TOV-initial-conditions}). We investigate whether a similar scaling relation also exists for the metric component $\alpha(r)$ of (fermion) Proca stars. The equations of motion \eqref{eq:fermion-boson-stars:fermion-proca-stars:TOV-equations-grr}-\eqref{eq:fermion-boson-stars:fermion-proca-stars:TOV-equations-P} are in fact invariant when simultaneously scaling the following variables as 
\begin{align}
	\tilde{\alpha} = \sigma \alpha \:\: , \:\: \tilde{\omega} = \sigma \omega \:\: , \:\: \tilde{E} = \sigma E \:\: , \:\: \text{where} \:\: \sigma \in \mathbb{R} \: . \label{eq:fermion-boson-stars:fermion-proca-stars:TOV-scaling-relations}
\end{align}

The potential $V(A_\rho \bar{A}^\rho)$ is always invariant with respect to the scaling since it is a function of the magnitude of the vector field $A_\rho \bar{A}^\rho$. The expression $A_\rho \bar{A}^\rho$ reads
\begin{align}
	A_\rho \bar{A}^\rho = \left( \frac{B^2}{a^2} - \frac{E^2}{\alpha^2} \right) = \left( \frac{B^2}{a^2} - \frac{\tilde{E}^2}{\tilde{\alpha}^2} \right) \: .
\end{align}

The invariance of the equations of motion \eqref{eq:fermion-boson-stars:fermion-proca-stars:TOV-equations-grr}-\eqref{eq:fermion-boson-stars:fermion-proca-stars:TOV-equations-P} under the scaling relation \eqref{eq:fermion-boson-stars:fermion-proca-stars:TOV-scaling-relations} thus allows us to chose $\sigma$ in such a way that the initial condition for $\alpha(0) = \alpha_0$ may be set to $\alpha_0 = 1$\footnote{Or one could, in principle, also re-scale $E_0$ to always be equal to one.}. We will make use of this relation in the numerical analysis. All pre-scaling physical values can be recovered from the asymptotic behaviour of $\alpha(r \rightarrow \infty)$ by performing the inverse transformation to \eqref{eq:fermion-boson-stars:fermion-proca-stars:TOV-scaling-relations}. In contrast to the scaling relation of boson stars with a scalar field, where only the frequency $\omega$ and the metric component $\alpha$ are re-scaled, the vector field component $E$ is also affected in the case of Proca stars. To our knowledge, this is the first time the scaling relation \eqref{eq:fermion-boson-stars:fermion-proca-stars:TOV-scaling-relations} has been mentioned explicitly. \cite{Minamitsuji:2018kof} briefly mentioned scaling the frequency but not the vector field component. \\

We continue the analytical analysis of equations \eqref{eq:fermion-boson-stars:fermion-proca-stars:TOV-equations-grr}-\eqref{eq:fermion-boson-stars:fermion-proca-stars:TOV-equations-P} by studying the Proca equations \eqref{eq:fermion-boson-stars:fermion-proca-stars:TOV-equations-E} and \eqref{eq:fermion-boson-stars:fermion-proca-stars:TOV-equations-B}, which govern the dynamics of the vector field. Note that the term in the denominator of the equation of motion for $B(r)$ \eqref{eq:fermion-boson-stars:fermion-proca-stars:TOV-equations-B} could in some cases lead to singularities. We analyse the behaviour of the denominator by setting it equal to zero. We obtain:
\begin{align}
 0 \stackrel{!}{=} V''(A_\rho \bar{A}^\rho ) \frac{2 B^2}{a^2} + V'(A_\rho \bar{A}^\rho ) \: . \label{eq:fermion-boson-stars:fermion-proca-stars:B-equation-singularity-criterium}
\end{align}

This leads to a remarkable behaviour when considering a quartic self-interaction potential $V$ of the form
\begin{align}
  V(A_\mu \bar{A}^\mu ) = m^2 A_\mu \bar{A}^\mu + \frac{\lambda}{2} ( A_\mu \bar{A}^\mu )^2 \: , \label{eq:fermion-boson-stars:fermion-proca-stars:quartic-self-interaction-potential}
\end{align}

where $m$ is the mass of the vector boson and $\lambda$ is the self-interaction parameter. When inserting the potential \eqref{eq:fermion-boson-stars:fermion-proca-stars:quartic-self-interaction-potential} into \eqref{eq:fermion-boson-stars:fermion-proca-stars:B-equation-singularity-criterium}, one obtains
\begin{align}
 \lambda = - m^2 \left( \frac{3B^2}{a^2} - \frac{E^2}{\alpha^2} \right)^{-1} \: . \label{eq:fermion-boson-stars:fermion-proca-stars:B-equation-singularity-criterium-rewritten}
\end{align}

We analyse the behaviour of this expression in the limit $r \rightarrow 0$, i.e. we apply the initial conditions given in equation \eqref{eq:fermion-boson-stars:fermion-proca-stars:TOV-initial-conditions}. When inserting $B \rightarrow 0$, $E \rightarrow E_0$ for the vector field components and  $a \rightarrow 1$, $\alpha \rightarrow \alpha_0$ for the metric components, one then obtains a critical value for the central field amplitude $E_0$:
\begin{align}
E_{0,\text{crit}} = \frac{m \alpha_0}{\sqrt{\lambda}} \: . \label{eq:fermion-boson-stars:fermion-proca-stars:analytical-bound-amplitude}
\end{align}

This expression constitutes an analytical upper bound for the central amplitude of the vector field, meaning that any FPS with initial conditions for the field larger than $E_{0,\text{crit}}$ will be physically forbidden, since the ODE \eqref{eq:fermion-boson-stars:fermion-proca-stars:TOV-equations-B} will become singular and diverge. This result matches the analytical bound found by \cite{Minamitsuji:2018kof}. Equation \eqref{eq:fermion-boson-stars:fermion-proca-stars:analytical-bound-amplitude} can also be rewritten using the dimensionless interaction parameter $\Lambda_{\mathrm{int}} = \lambda / 8 \pi m^2$:
\begin{align}
E_{0,\text{crit}} = \frac{\alpha_0}{\sqrt{ 8\pi \Lambda_{\mathrm{int}}} } \: . \label{eq:fermion-boson-stars:fermion-proca-stars:analytical-bound-amplitude-rewritten}
\end{align}

The relation has as a consequence that for strong self-interaction strengths $\Lambda_{\mathrm{int}}$, the allowed range for Proca stars becomes increasingly small and vanishes in the limit of very strong interactions. This fact could conceivably be used to constrain the vector field parameters $m$ and $\lambda$, but we leave a thorough investigation for future work. \\

In this chapter, we have developed the theory behind the description of fermion boson stars. In particular, we considered FBS with a complex scalar field and FPS with a complex vector field. We derived the tidal deformability for the scalar field case. For the vector field case, we analysed the equation of motion and discovered that it obeys a scaling relation. When choosing a quartic self-interaction potential, we derived an analytical upper bound for the vector field component $E$ that depends on the interaction parameter. To complete our understanding of fermion boson stars, the next step is to solve the equations numerically and to gain insight into how these equilibrium solutions behave and what their global quantities are. To that end, we will first introduce the numerical methods in the following Chapter \ref{sec:numerics:numerical-methods}, after which we present the results in Chapter \ref{sec:results:results}.

%% file: numerics.tex
\markright{Numerical Methods}
\section{Numerical Methods} \label{sec:numerics:numerical-methods}

In this chapter, we introduce the algorithms and numerical methods employed to solve the differential equations that describe fermion boson stars (FBS) \eqref{eq:fermion-boson-stars:scalar-fermion-boson-stars:TOV-equations-grr}-\eqref{eq:fermion-boson-stars:scalar-fermion-boson-stars:TOV-equations-P} and fermion Proca stars (FPS) \eqref{eq:fermion-boson-stars:fermion-proca-stars:TOV-equations-grr}-\eqref{eq:fermion-boson-stars:fermion-proca-stars:TOV-equations-P}. The aim is to provide a pedagogical explanation of the workflow and to encourage reproducibility of this work. In particular, we will go into detail on how the frequency modes $\omega$ of FBS and FPS are found and how to extract the global quantities such as the total mass $M_\mathrm{tot}$, the fermionic radius $R_\mathrm{f}$ and the tidal deformability $\Lambda_\mathrm{tidal}$. We will also highlight some algorithmic subtleties to avoid numerical pitfalls for future scientists trying to study these, or similar, systems. Most of the methods discussed here are also found in \cite{Diedrichs:2023trk} and were implemented in the code \cite{Diedrichs-Becker-Jockel}. \\

\begin{figure}[h]
	\centering
	\includegraphics[width=0.75\textwidth]{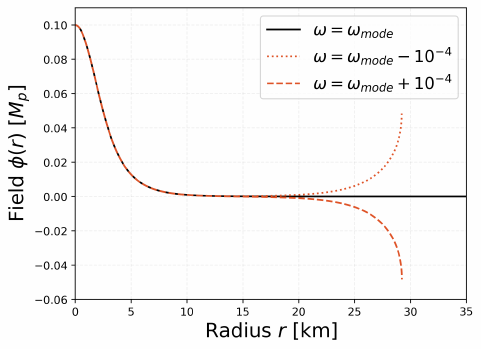}
	%\hspace{0.5cm}
	%\includegraphics[width=0.5257\textwidth]{BS-collision-waveform-2.png} %width=0.534
	\caption{Scalar field profile corresponding to a purely bosonic FBS (i.e. a boson star) with a quartic self-interaction profile \eqref{eq:background:boson-stars:quartic-self-interaction-potential} and $m= 1.34\e{-10}\,eV$, $\lambda=0$. The scalar field profile is shown for the zeroth mode (black solid line) and for two configurations for which the frequency $\omega$ is smaller (dotted orange line) and larger (dashed orange line). The solution can be seen to diverge when the frequency is not tuned to the correct value $\omega_\mathrm{mode}$ corresponding to the mode. Even small differences in $\omega$ from $\omega_\mathrm{mode}$ manifest as a divergence at relatively small radii (note that the bosonic field can easily reach sizes of several dozen kilometres or more). A divergence is thus unavoidable due to finite machine precision (in our case 64-bit floating-point numbers). Tuning $\omega$ more accurately to $\omega_\mathrm{mode}$ increases the radius where the field $\varphi(r)$ diverges. The bisection algorithm makes use of the direction of divergence to find the correct value of $\omega$.}
	\label{fig:numerics:numerical-methods:diverging-phi}
\end{figure}

We start by considering FBS with a complex scalar field. As already stated in Chapter \ref{sec:fermion-boson-stars:fermion-boson-stars}, the equations of motion of FBS \eqref{eq:fermion-boson-stars:scalar-fermion-boson-stars:TOV-equations-grr}-\eqref{eq:fermion-boson-stars:scalar-fermion-boson-stars:TOV-equations-P} have one undetermined variable, the field frequency $\omega$. In the following, we will explain the shooting-algorithm used to find $\omega$ numerically. For given $\rho_c$ and $\varphi_c$, there exist only discrete values of $\omega$, such that the boundary conditions at infinity \eqref{eq:fermion-boson-stars:scalar-fermion-boson-stars:TOV-initial-conditions} are fulfilled. These discrete values are called eigenvalues or modes. There are infinitely many of these modes and they are characterized by the number of roots (i.e. zero-crossings) the field $\varphi(r)$ has. Usually we are only interested in the lowest mode with zero roots, since only they are believed to be dynamically stable \cite{Liebling:2012fv}. The following algorithm can however be used to find any desired mode. \\
When the system of ordinary differential equations \eqref{eq:fermion-boson-stars:scalar-fermion-boson-stars:TOV-equations-grr}-\eqref{eq:fermion-boson-stars:scalar-fermion-boson-stars:TOV-equations-P} is integrated (in \cite{Diedrichs-Becker-Jockel} we use a Runge-Kutta-Fehlberg ODE solver) for some fixed value of $\omega$, the scalar field will diverge to positive or negative infinity at some finite radius. The system will only converge at infinity if any frequency mode is hit directly, which is impossible to achieve numerically with finite precision. We thus make use of this diverging property to find the wanted frequency mode. In fact, when the frequency $\omega$ is close to the wanted mode, the divergence will happen at increasingly large radii, the closer the chosen value for $\omega$ is to the mode. A higher accuracy in finding $\omega$ will therefore push the divergence to larger radii. When $\omega$ is not exactly tuned to the mode, the scalar field profile will diverge towards $+\infty$ or $-\infty$ and change its direction of divergence when $\omega$ passes a mode. In \autoref{fig:numerics:numerical-methods:diverging-phi} we have illustrated this behaviour. The direction of divergence will depend on which mode is solved for in detail: if we search for zeroth mode, the field will diverge to $+\infty$ if the frequency $\omega$ is below the mode, and it will diverge to $-\infty$ if $\omega$ is above the mode. This will be the case for every even mode, and will be reversed for every odd mode. By making use of the direction of divergence, we gain a binary criterion to find the correct mode. The value of $\omega$ can then be adapted -- increased or decreased -- based on the direction of divergence and the wanted mode. This procedure requires to integrate the system of equations multiple times with different values for $\omega$, until the correct value is found. In our code \cite{Diedrichs-Becker-Jockel}, we implement this method using a bisection algorithm, which converges exponentially fast. We start with an upper and a lower value of $\omega$, which are guaranteed to be smaller/larger than the wanted value of $\omega$ at the mode. In practice, lower and upper bounds of $\omega_\mathrm{bound} = [1,10]$ have proven to be numerically robust. Then, we perform the bisection search by taking the middle value of $\omega$ in this range and counting the number of roots in $\varphi(r)$ at each step. This also allows us to discriminate between different modes and to target specific modes by demanding a certain number of roots in the field $\varphi(r)$, corresponding to the wanted mode. The bisection is complete when the current value of $\omega$ found through bisection is close enough to the value of the mode. In our experience, the absolute accuracy needed to obtain robust solutions is on the order of $\Delta\omega = | \omega_\mathrm{mode} - \omega_\mathrm{bisection} | \approx 10^{-15}$. \\

Once a sufficiently accurate frequency $\omega$ is found, we modify the integration, such that $\varphi(r)$ is set to zero at a finite radius $r_{B}^*$. This radius $r_{B}^*$ is defined at the point where the field $\varphi(r)$ and its derivative $\varphi'(r)$ are small. This roughly corresponds to the last minimum of $\varphi(r)$ before it diverges. The condition can be summarized as the point where $\varphi (r_{B}^*)/ \varphi_c < 10^{-4}$ and $\varphi'(r_{B}^*) \ll 1$. This is necessary because the interplay of the scalar field and the neutron star matter can complicate the numerical solution. In some parts of the parameter space, especially for small initial densities $\rho_c$, the scalar field could diverge while still inside of the neutron star component, i.e. before the pressure $P(r)$ reaches zero (within numerical precision, we consider the pressure to be zero when $P < 10^{-15}$). This divergence would make finding physical values such as the fermionic radius $R_\mathrm{f}$ impossible. Therefore, we artificially set $\varphi = 0$ for $r > r_{B}^*$, which allows us to circumvent the divergence and accurately resolve the rest of the neutron star component. The condition was chosen so that the remaining contribution of the scalar field to the other quantities (i.e. the metric components) is minimized. We have checked for different thresholds and confirmed that all extracted results are the same. \\

After integrating the solution to radii outside of the matter sources, we can extract global observables such as the total gravitational mass and radius. The outside of the source is located at radii $r$ larger than the fermionic radius $R_\mathrm{f}$ and $r > r_{B}^*$, i.e. the regime where neither the NS matter nor the scalar field contribute significantly. There, we can extract the total gravitational mass $M_\mathrm{tot}$ \eqref{eq:fermion-boson-stars:scalar-fermion-boson-stars:definition-BS-total-gravitational-mass} and then compute the integrals \eqref{eq:fermion-boson-stars:scalar-fermion-boson-stars:conserved-fermion-number-simplified} and \eqref{eq:fermion-boson-stars:scalar-fermion-boson-stars:conserved-boson-number-simplified} to obtain the fermion/boson numbers $N_\mathrm{f}$, $N_\mathrm{b}$. \\
For some configurations, due to numerical precision limits, the scalar field convergence condition $\varphi (r_{B}^*)/ \varphi_c < 10^{-4}$ cannot be fulfilled. This generally happens for small initial field values $\varphi_c \lesssim 10^{-4}$, where the scalar field extends far outside the neutron star component. In these cases, we extract the total gravitational mass $M_\text{tot} = \frac{1}{2} r_\mathrm{ext} (1 - a^{-2}(r_\mathrm{ext}))$ at the point where its derivative has a global minimum. When the scalar field diverges, also the metric components do, and with it also $M_\text{tot}$. By taking the point where the derivative of the mass has a global minimum, which roughly corresponds to where the scalar field and its derivative is closest to zero, we get the best possible estimate of the mass of the system before the divergence. \\

To compute the tidal deformability $\Lambda_\mathrm{tidal}$, the system of equations \eqref{eq:fermion-boson-stars:scalar-fermion-boson-stars:perturbed-equations-phi1} and \eqref{eq:fermion-boson-stars:scalar-fermion-boson-stars:perturbed-equations-H0} needs to be solved on top of the unperturbed background described by the equations of motion of the FBS \eqref{eq:fermion-boson-stars:scalar-fermion-boson-stars:TOV-equations-grr}-\eqref{eq:fermion-boson-stars:scalar-fermion-boson-stars:TOV-equations-P}. In practice, we first find a solution to the unperturbed FBS by tuning $\omega$, and then later integrate the perturbed equations to obtain the tidal deformability. Then, we perform a bisection algorithm similar to the one to find $\omega$, but for the initial value of the perturbed field $\varphi^{(3)}_1$ (see equation \eqref{eq:fermion-boson-stars:scalar-fermion-boson-stars:expansions_phi_H0-simplified}), so that the boundary conditions \eqref{eq:fermion-boson-stars:scalar-fermion-boson-stars:perturbations-initial-conditions} are fulfilled. The perturbation $\varphi_1(r)$ converges to zero in the same way as the unperturbed field $\varphi(r)$ does. Therefore, we also set $\varphi_1(r) = 0$ for $r > r_B^*$. This allows us to avoid the divergence of the perturbed field $\varphi_1(r)$ while having no effect on the tidal deformability. This process is valid since the equations for $\varphi_1$ \eqref{eq:fermion-boson-stars:scalar-fermion-boson-stars:perturbed-equations-phi1} and $H_0$ \eqref{eq:fermion-boson-stars:scalar-fermion-boson-stars:perturbed-equations-H0} decouple when the unperturbed field $\varphi(r) \approx 0$ (see eq. \eqref{eq:fermion-boson-stars:scalar-fermion-boson-stars:perturbed-H-parameter-1-H-ode-ourside-of-source}). Since the tidal deformability is constant for any $r > r_B^*$, we can compute it directly using \eqref{eq:fermion-boson-stars:scalar-fermion-boson-stars:tidal-deformablity-equation}. \\
In the cases where the scalar field convergence condition $\varphi (r_{B}^*)/ \varphi_c < 10^{-4}$ cannot be fulfilled, we follow the same procedure as in \cite{Sennett:2017etc} and extract $y := r H_0'(r) / H_0(r)$ at a radius $r_\text{ext}$ such that $y(r_\text{ext})$ is a local maximum. Since there are two components in the fermion boson star at play, there can be multiple local maxima, of which we choose the one at the largest radius. \\

We further note that it can be numerically difficult to integrate the full system at high self-interaction strengths $\Lambda_{\mathrm{int}} \gtrsim 400$ and for small masses $m$ because 
\begin{itemize}
\itemsep0em
\item[$1)$] the frequency $\omega$ must be tuned up to higher accuracy than what is possible using 64-bit floating-point numbers,
\item[$2)$] increasingly small step-sizes are needed and
\item[$3)$] the integration has to be performed up to larger radii, to solve the equations correctly and accurately.
\end{itemize}

This may increase the run-time of the code up to more than an order of magnitude. \\

In the case of FPS with a complex vector field \eqref{eq:fermion-boson-stars:fermion-proca-stars:TOV-equations-grr}-\eqref{eq:fermion-boson-stars:fermion-proca-stars:TOV-equations-P}, we perform an algorithm which is very similar to the one for FBS (see above). We perform a bisection in the same way, with the difference that we count the roots of the vector field component $E(r)$ instead of the scalar field $\varphi(r)$. Another difference is that for FPS, the lowest mode has always one root in $E(r)$. The remaining steps are analogous to the step performed to solve the FBS with the scalar field. During our numerical analysis, we encountered the phenomenon that the bisection algorithm to find the frequency $\omega$ could fail for some specific initial conditions for $E_0$ and $\rho_c$. We found this to be the case due to the bisection algorithm jumping over multiple modes in one iteration step, thus also skipping the wanted mode, such that the mode ended up outside of the bisection bounds. The bisection then converged on an unwanted $\omega$-value, or ended up failing entirely. We solved this problem by employing a backup algorithm that activates if the bisection fails. The backup algorithm restarts the bisection for $\omega$ but with different lower and upper bounds of $\omega_\mathrm{bound}$. We tested the backup algorithm for $4800$ FPS configurations with different masses $m$ and self-interaction strengths $\Lambda_{\mathrm{int}} = \lambda/8\pi m^2$ with equally distributed initial conditions for $E_0$ and $\rho_c$. We found that $330$ ($\approx 6.8\,\%$) of all configurations needed one restart of the bisection and only $3$ ($\approx 0.06\,\%$) of all configurations needed two restarts. In none of the cases tested, the bisection had to be restarted three times or more. \\

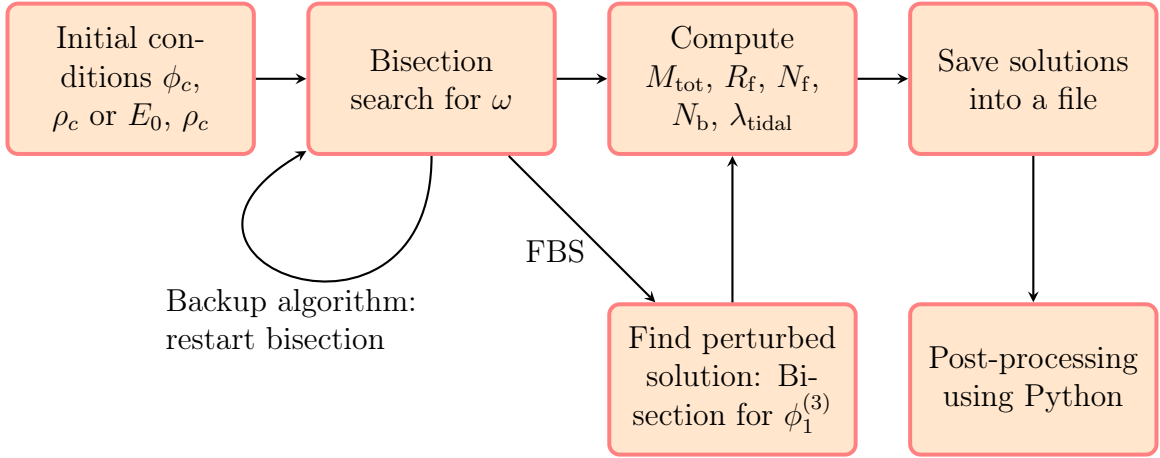
\begin{figure}[h]
\centering
\begin{tikzpicture}[node distance=4cm]

\node (start) [process,line width=1.5pt] {Initial conditions $\varphi_c$, $\rho_c$ or $E_0$, $\rho_c$};
\node (bisection) [process,line width=1.5pt, right of=start] {Bisection search for $\omega$};
\node (computeGlobalVars) [process,line width=1.5pt, right of=bisection] {Compute $M_\mathrm{tot}$, $R_\mathrm{f}$, $N_\mathrm{f}$, $N_\mathrm{b}$, $\lambda_\mathrm{tidal}$};
\node (perturbedSolution) [process,line width=1.5pt, below of=computeGlobalVars] {Find perturbed solution: Bisection for $\varphi^{(3)}_1$};
\node (SaveSols) [process,line width=1.5pt, right of=computeGlobalVars] {Save solutions into a file};
\node (PostProcessing) [process,line width=1.5pt, below of=SaveSols] {Post-processing using Python};

\draw [arrow] (start) -- (bisection);
\draw [arrow] (bisection) -- (computeGlobalVars);
\draw [arrow] (perturbedSolution) -- (computeGlobalVars);
\draw [arrow] (computeGlobalVars) -- (SaveSols);
\draw [arrow] (SaveSols) -- (PostProcessing);
\draw [arrow] (bisection) -- node[ text width=1.5cm,anchor=north] {FBS} (perturbedSolution);
\draw[arrow] (bisection) to [looseness=4.5, out= 270, in=210] node[rectangle, text width=3.5cm, below] {Backup algorithm: restart bisection} (bisection);

\end{tikzpicture}
\caption{Schematic of the total algorithm used to compute solutions of FBS and FPS. We implemented this algorithm in our C++ code \cite{Diedrichs-Becker-Jockel}. We start with initial conditions for FBS ($\varphi_c$, $\rho_c$) and FPS ($E_0$, $\rho_c$). We then use the bisection algorithm described in this chapter to find the wanted mode and the corresponding value of $\omega$. For FBS (with a scalar field), there is the option to also compute the perturbed FBS solutions. After the solutions have been found, the global quantities are computed. In the end, we save the global quantities computed for all FBS/FPS into a file. The post-processing, for example the computation of the stability curves \eqref{eq:fermion-boson-stars:scalar-fermion-boson-stars:FBS-stability-criterion-rewritten} or the creation of plots, is performed using a separate code in Python, which was developed alongside the C++ code (see \cite{Diedrichs-Becker-Jockel}).}
\label{fig:numerics:numerical-methods:code-flowchart}
\end{figure}

We close this chapter by providing a flowchart of the complete algorithm to compute FBS and FPS in \autoref{fig:numerics:numerical-methods:code-flowchart}. \\

%% file: results.tex
%\markright{Results}
\section{Results} \label{sec:results:results}

In this chapter, we present our results. We solve the equations of motion for fermion boson stars with a scalar field \eqref{eq:fermion-boson-stars:scalar-fermion-boson-stars:TOV-equations-grr}-\eqref{eq:fermion-boson-stars:scalar-fermion-boson-stars:TOV-equations-P} and with a vector field \eqref{eq:fermion-boson-stars:fermion-proca-stars:TOV-equations-grr}-\eqref{eq:fermion-boson-stars:fermion-proca-stars:TOV-equations-P} numerically using the methods described in Chapter \ref{sec:numerics:numerical-methods} with a quartic self-interaction potential. We compute a number of configurations with different masses and self-interaction strengths and analyse the solutions with respect to their masses, radii and tidal properties. The results regarding fermion boson stars with a scalar field (FBS) are congruent with the results presented in our publication \cite{Diedrichs:2023trk}. Some figures have been remade with a different colour scheme to be more consistent with the remainder of this work. The results regarding fermion Proca stars (FPS) are original to this work and are presented here for the first time.

\subsection{Fermion Boson Stars} \label{subsec:results:fermion-boson-stars}

We consider fermion boson stars with a quartic self-interaction potential with a mass term
\begin{align}
V(\bar{\varphi} \varphi) = m^2 \bar{\varphi} \varphi + \frac{\lambda}{2} (\bar{\varphi} \varphi)^2 \: , \label{eq:results:fermion-boson-stars:quartic-self-interaction-potential}
\end{align}

where $m$ is the boson mass and $\lambda$ is the self-interaction parameter. For easy comparability with previous works, we define the effective self-interaction parameter $\Lambda_\mathrm{int} = \lambda / 8\pi m^2$, which was first introduced by \cite{Colpi:1986ye}. $\Lambda_\mathrm{int}$ can be used to quantify the self-interaction strength and the effect of self-interaction on the total gravitational mass of boson stars, as illustrated by the scaling relations $M_\mathrm{max} \approx 0.633 M^2_p / m$ \cite{Liebling:2012fv} (for small $\Lambda_\mathrm{int}$) and $M_\mathrm{max} \approx 0.22 \sqrt{\Lambda_\mathrm{int}}\, M^2_p / m$ \cite{Colpi:1986ye} (for large $\Lambda_\mathrm{int}$); also see Chapter \ref{subsec:background:boson-stars}. In the regime of large $\Lambda_\mathrm{int}$, the energy-momentum tensor of the scalar field becomes approximately isotropic and can be described using an effective equation of state. A comparison of the full system and the effective system is the subject of Chapter \ref{subsec:results:comparison-to-effecive-eos}. Note that the parameter $\Lambda_\mathrm{int}$ was originally introduced in the context of pure boson stars and thus the scaling relations will not be generally valid for the mixed system. Scaling relations can however be useful to understand the limiting cases where the FBS is dominated by either the fermionic or the bosonic component.  Nonetheless, we regard $\Lambda_\mathrm{int}$ to be a useful measure to compare different choices of the mass and self-interaction strength. \\

We hereafter investigate nine different models with parameters $m = \{0.1, 1, 10\}\times 1.34 \e{-10}\,eV$ and $\Lambda_\mathrm{int} = \{0, 10, 100 \}$. This mass range is chosen so that the reduced Compton wavelength of the bosonic field is half the Schwarzschild radius of the Sun. $m=1$ then corresponds to $1.336 \e{-10}\,eV$ (see a detailed explanation in Appendix \ref{sec:appendix:units}) . The range for the self-interaction parameter was chosen so that it fulfils the observational constraints for the DM cross-section of $1\,cm^2/g$ obtained from the bullet cluster \cite{Eby:2015hsq,Sagunski:2020spe}:
\begin{align}
    \pi \Lambda_\mathrm{int}^2 m  = \frac{\lambda^2}{64 \pi m^3 } = \frac{\sigma}{m} \stackrel{!}{<} 1 \frac{\text{cm}^2}{\text{g}} \iff \Lambda_\mathrm{int} \stackrel{!}{<} 10^{50} \sqrt{\frac{1.34 \e{-10}\text{eV}}{m}} \: .
\end{align}

For the fermionic component of the FBS, we use the DD2 equation of state (with electrons) \cite{Hempel:2009mc} taken from the CompOSE database \cite{Typel:2013rza}. It was chosen due to it being widely used by a number of groups and thus being well known in the literature. The DD2 EOS is based on a relativistic mean-field model with density-dependent coupling constants, which has been fitted to the properties of nuclei and results from Brueckner-Hartree-Fock calculations for dense nuclear matter. Therefore, the DD2 EOS describes also the EOS of pure neutron matter from chiral effective field theory, see \cite{Kruger:2013kua}. For the purpose of our investigations, the particular choice of the nuclear equation of state is not of importance and has no effect on our general conclusions.

\subsubsection{Mass-Radius Relations and Tidal Deformability} \label{subsec:results:mass-radius-relations-and-tidal-deformability}

\begin{figure}[h]
\centering
    \includegraphics[width=0.499\textwidth]{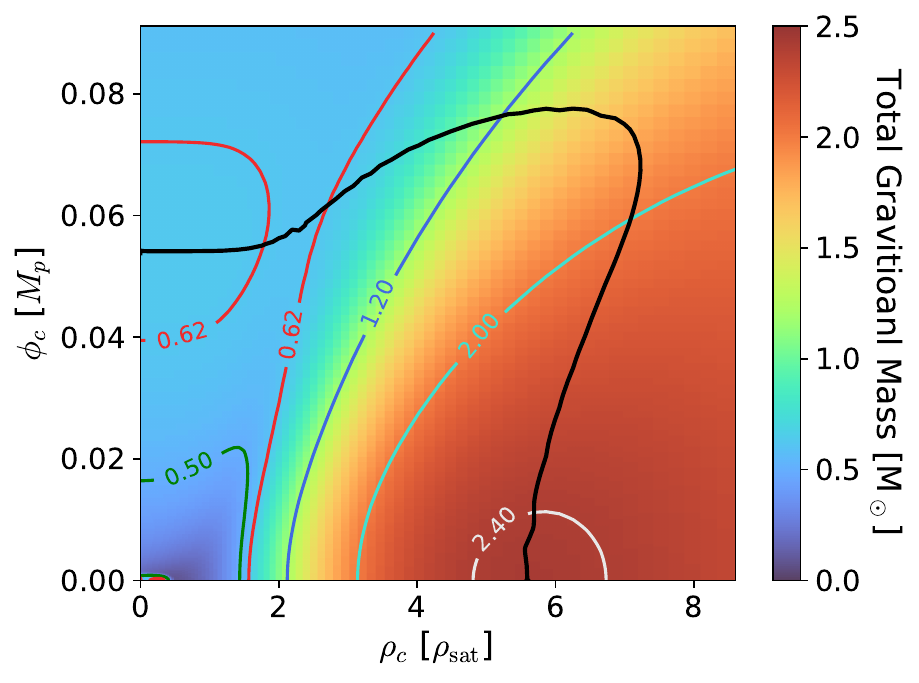}
    \includegraphics[width=0.49\textwidth]{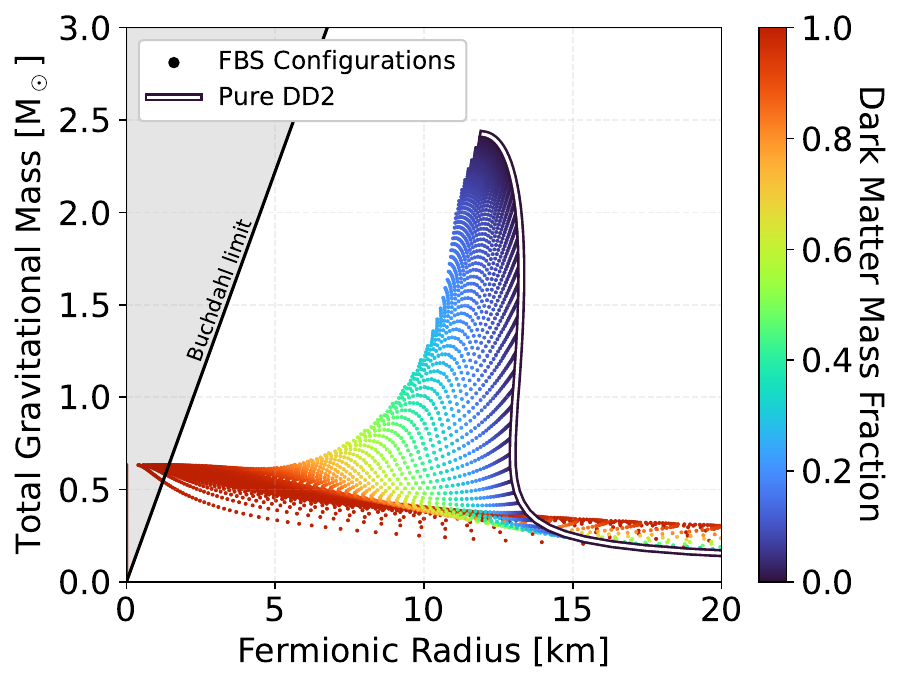}
    \caption{\textbf{Left panel:} Total gravitational mass of various FBS as a function of the restmass density $\rho_c$ and central scalar field amplitude $\varphi_c$. The black line corresponds to the stability curve computed using \eqref{eq:fermion-boson-stars:scalar-fermion-boson-stars:FBS-stability-criterion-rewritten}. All configurations within the stability curve, i.e. in the lower left corner of the parameter space, are stable with respect to linear radial perturbations. The presence of DM can increase the possible maximum central densities.
    \textbf{Right panel:} Mass-radius diagram displaying the fermionic radius vs the total gravitational mass for FBS configurations that are within the stability region displayed in the left panel. Each point corresponds to a single configuration and is colour-coded according to the restmass fraction of the dark matter component. The solid black-white line shows the mass-radius curve for pure fermionic matter. In both cases, a scalar field with a mass of $m=1.34 \e{-10}\,eV$ and no self-interactions was considered in addition to the DD2 EOS for the fermionic part. Figures adapted from \cite{Diedrichs:2023trk}.}
    \label{fig:results:fermion-boson-stars:stability-and-MR-curve-example}
\end{figure}

We compute a grid of FBS with varying values of the central restmass density $\rho_c$ and central scalar field amplitude $\varphi_c$. Using this, we compute the stability curve as explained in Chapter \ref{subsec:fermion-boson-stars:scalar-fermion-boson-stars}. The stable solutions can then be filtered and analysed further. We illustrate this method in \autoref{fig:results:fermion-boson-stars:stability-and-MR-curve-example} for an FBS with $m=1.34 \e{-10}\,eV$ and no self-interactions. The left panel of \autoref{fig:results:fermion-boson-stars:stability-and-MR-curve-example} shows a grid of FBS with different $\rho_c$ and $\varphi_c$, coloured with respect to the total gravitational mass. The black line is the stability curve as obtained from the condition \eqref{eq:fermion-boson-stars:scalar-fermion-boson-stars:FBS-stability-criterion-rewritten}. The stability curve marks the region of stable configurations, which corresponds to the lower left part of the plot, which is enclosed by the stability curve. In the bottom left corner of the diagram, the total gravitational mass can be seen to increase. This effect is related to the EOS as at those densities, the white dwarf solutions start to emerge. The presence of DM can increase the possible maximum densities in the centre of the FBS, which might be relevant for phase transitions that happen at high densities \cite{Burgio:2021vgk}. In the right panel of \autoref{fig:results:fermion-boson-stars:stability-and-MR-curve-example}, all FBS within the stability region are plotted in an MR diagram, where the fermionic radius $R_\mathrm{f}$ is the radius where the fermionic component of the FBS vanishes. Note that, instead of a mass-radius curve, the stable configurations form a mass-radius region for the FBS with different fermionic and bosonic content. The points are coloured with respect to the dark matter mass fraction, which is defined by $N_\mathrm{b}/(N_\mathrm{b} + N_\mathrm{f})$. It is therefore a measure of how much of the total restmass of the FBS is dark matter. \\

\begin{figure}
    \centering
    \includegraphics[width = 0.99\textwidth]{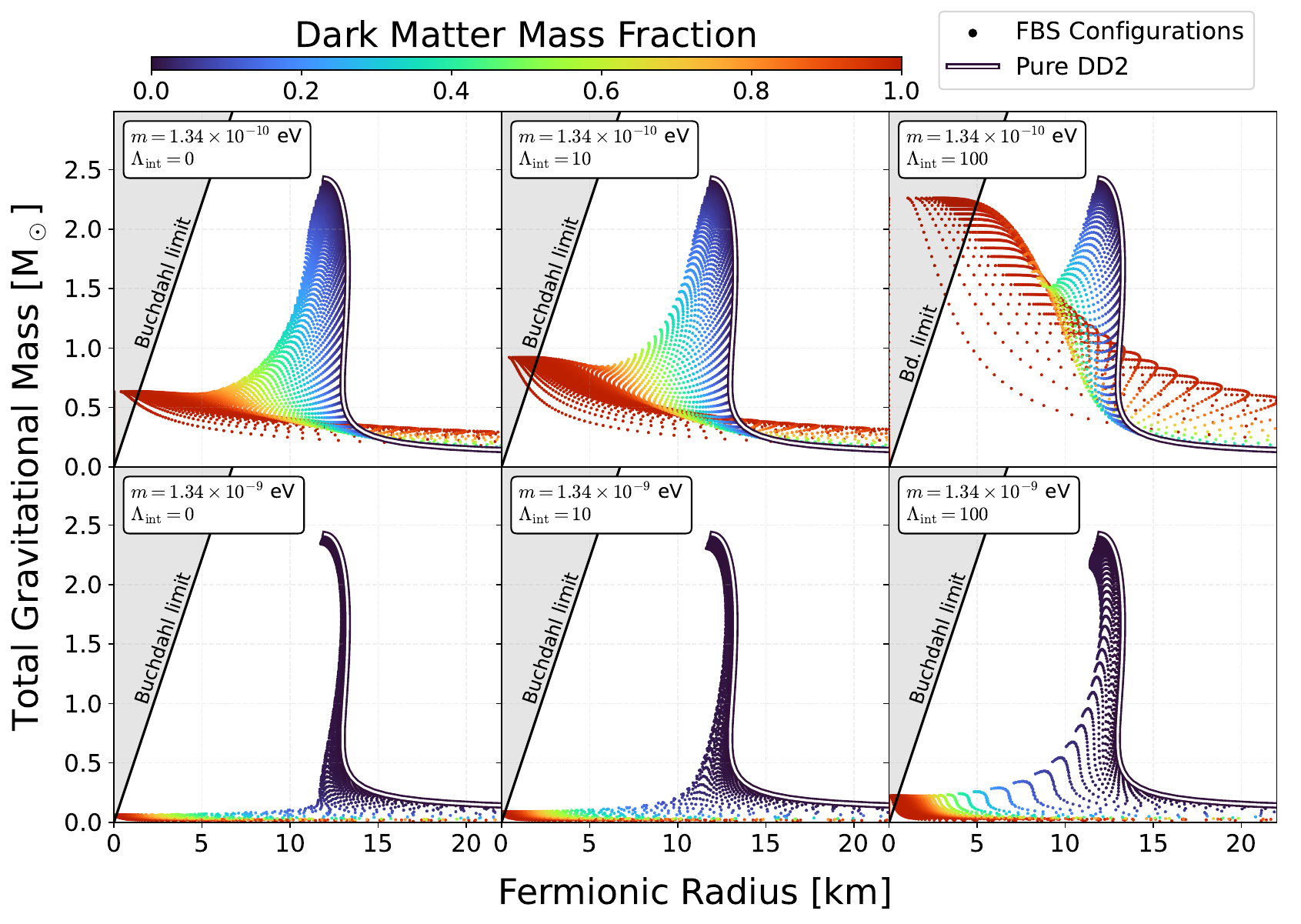}
    \includegraphics[width = 0.99\textwidth]{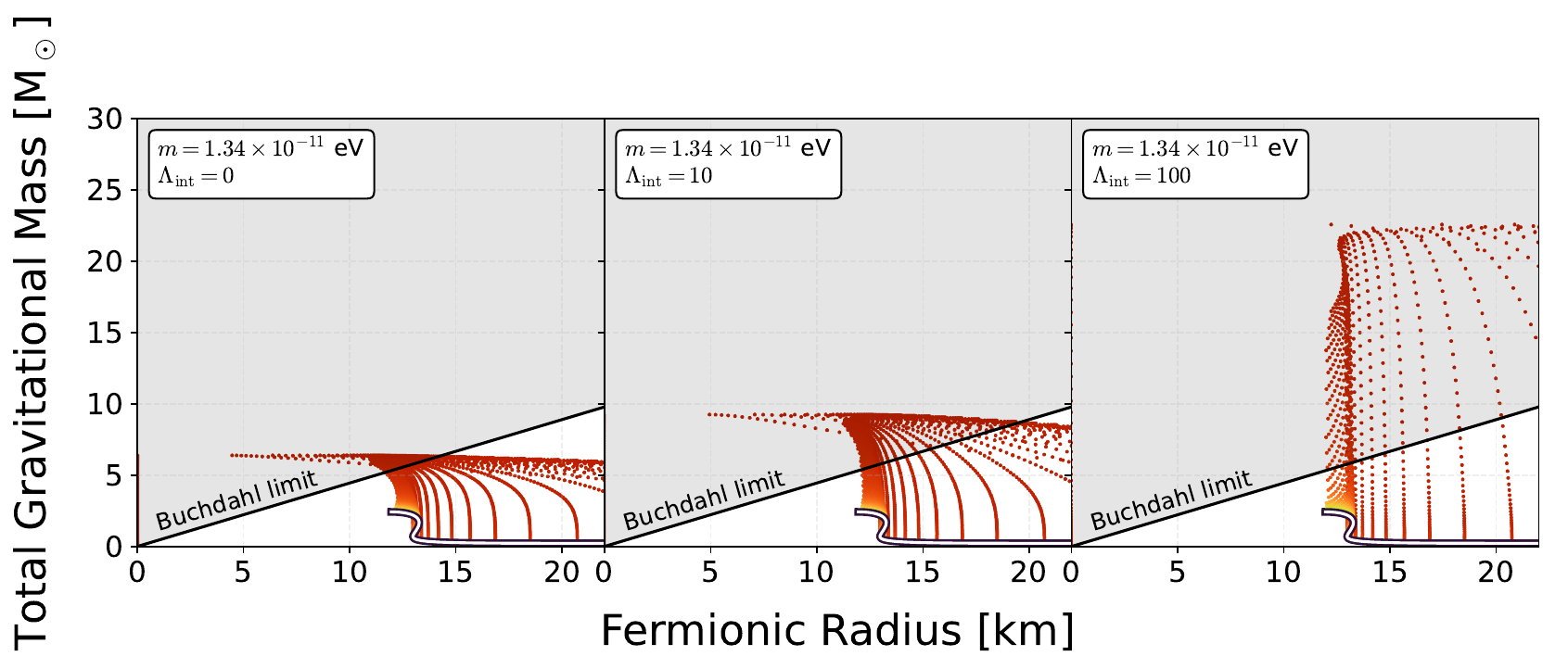}
    \caption{Relation between total gravitational mass $M_\mathrm{tot}$ and fermionic radius $R_\mathrm{f}$ for different FBS. The rows correspond to bosonic masses of $m = \{1, 10, 0.1\}\times 1.34 \e{-10}\,eV$, columns correspond to self-interactions of $\Lambda_\mathrm{int}= \{0, 10, 100\}$ respectively. We use the DD2 EOS for the fermionic part. Notice the different scale of the bottom plots. The grey region marks the Buchdahl limit, where no stable NS can exist. Observing only $R_\mathrm{f}$ of these systems would appear to violate the Buchdahl limit, even though the FBS as a whole does not. Figures adapted from \cite{Diedrichs:2023trk}.}
    \label{fig:results:fermion-boson-stars:MR-grid}
\end{figure}

\begin{figure}
    \centering
    \includegraphics[width = 0.93\textwidth]{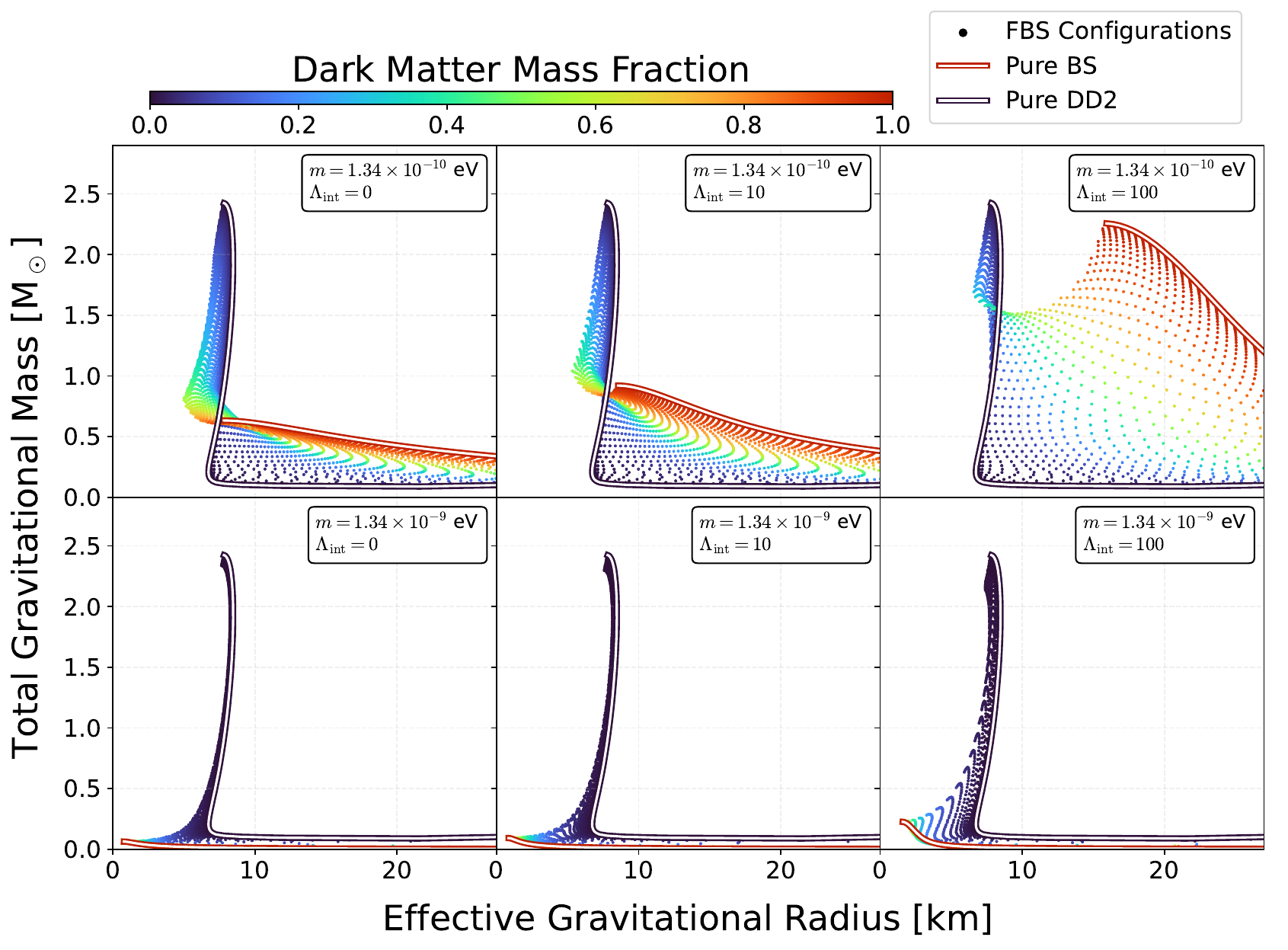}
    \includegraphics[width = 0.93\textwidth]{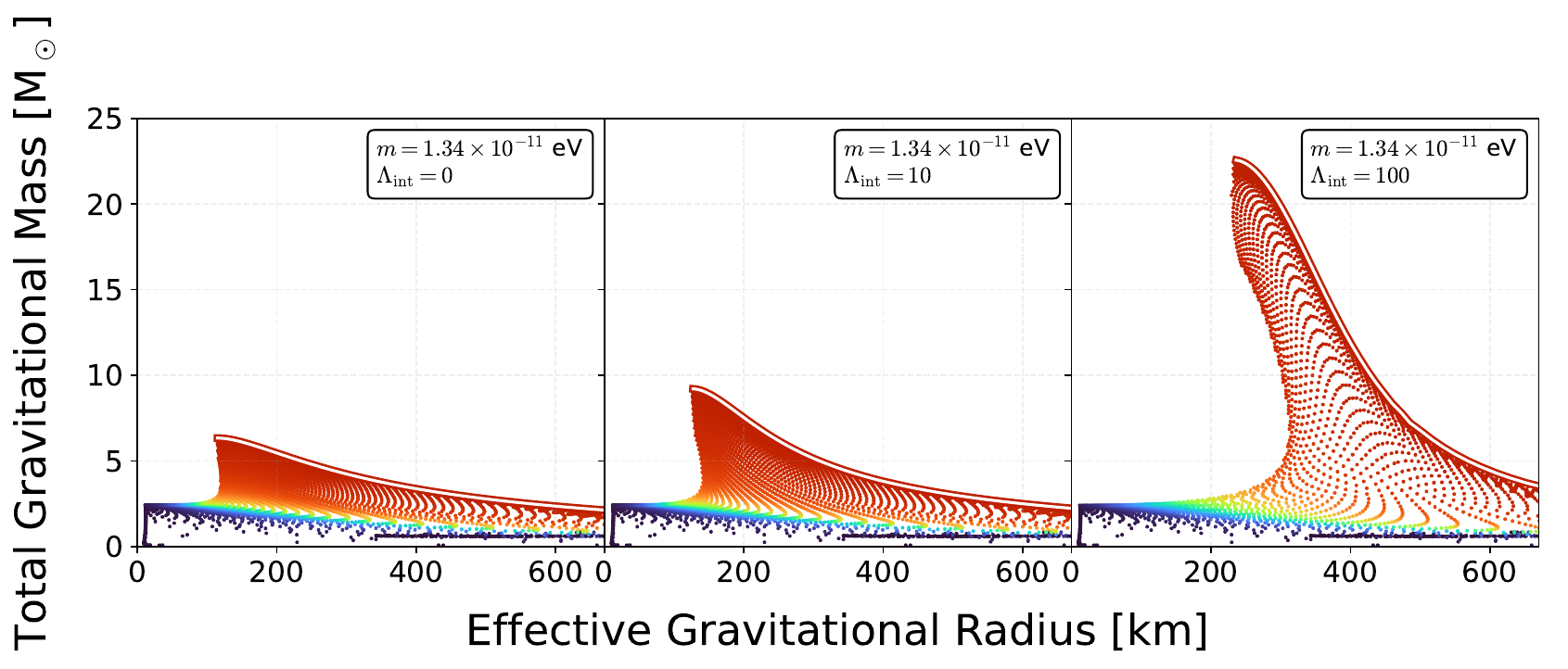}
    \caption{Relation between total gravitational mass $M_\mathrm{tot}$ and effective gravitational radius $R_G$ for different FBS. $R_G$ is the radius where $99\%$ of the restmass is contained. The rows correspond to bosonic masses of $m = \{1, 10, 0.1\}\times 1.34 \e{-10}\,eV$, columns correspond to self-interactions of $\Lambda_\mathrm{int}= \{0, 10, 100\}$ respectively. We use the DD2 EOS for the fermionic part. Notice the different scales of the bottom plots. For pure NS, because the crust has comparatively low density, $R_G$ is significantly smaller than $R_\mathrm{f}$ (compare to \autoref{fig:results:fermion-boson-stars:MR-grid}). For low DM-fractions, the bosonic component tends to form a core and the total compactness of the object increases. For higher DM-fractions, the bosonic component forms a cloud and can significantly decrease the compactness of the object. Figures adapted from \cite{Diedrichs:2023trk}.}
    \label{fig:results:fermion-boson-stars:MRg-grid}
\end{figure}

We show the mass-radius relations of various FBS in \autoref{fig:results:fermion-boson-stars:MR-grid} and \autoref{fig:results:fermion-boson-stars:MRg-grid}. We follow the same steps as described to obtain \autoref{fig:results:fermion-boson-stars:stability-and-MR-curve-example} and only consider the stable FBS from here on. It is important to note that in \autoref{fig:results:fermion-boson-stars:MR-grid} we show the fermionic radius $R_\mathrm{f}$. The bosonic radius $R_\mathrm{b}$, i.e. the radius where $99\%$ of the bosonic restmass-energy is included, can be orders of magnitude smaller or larger than $R_\mathrm{f}$, depending on the mass $m$ and self-interaction parameter $\Lambda_\mathrm{int}$. In \autoref{fig:results:fermion-boson-stars:MRg-grid} we show the effective gravitational radius $R_G$ -- the radius where $99\%$ of the total restmass is contained. Using the total gravitational mass $M_\mathrm{tot}$ and $R_G$, it is possible to infer the compactness $C = M_\mathrm{tot} / R_G$ of the FBS. The compactness can also be understood as a measure of how relativistic a given compact object is. For pure NS, this effective gravitational radius is always smaller than the fermionic one. Which radius ($R_\mathrm{f}$ or $R_G$) is more relevant for a given problem depends on the observation. The fermionic radius is crucial for electromagnetic signatures, such as those observed by the NICER telescope. The effective gravitational radius is more relevant for the inspiral of binary FBS, since it also is a measure of the size of the combined system. It also is related to the tidal deformability through the compactness. An observed effective gravitational radius larger than the fermionic radius would be an indication that dark matter might be present around the NS. \\

We note general trends in the figures. As expected, FBS with small DM-fractions are dominated by the fermionic component and thus show similar behaviour to pure NS in the MR diagram. Likewise, stars with high DM-fraction are dominated by the bosonic component and thus behave similarly to pure boson stars. For masses of $m = \{1, 10\}\times 1.34 \e{-10}\,eV$ in \autoref{fig:results:fermion-boson-stars:MR-grid}, the region of stable configurations extends to lower masses and similar compactness to pure NS. These results are consistent with the lines shown in \cite{DiGiovanni:2021vlu}. Comparing these trends to \autoref{fig:results:fermion-boson-stars:MRg-grid} provides more insight into the structure of these FBS configurations. For $m = 1.34 \e{-9}\,eV$, the bosonic component is concentrated inside of the fermionic one and forms a DM core. For $m = 1.34 \e{-10}\,eV$, the bosonic and fermionic components have roughly similar sizes. For low DM-fraction, the compactness is increased, while for DM dominated configurations, the overall compactness decreases and the DM forms a cloud which envelopes the fermionic component. Increasing the self-interaction strength also leads generally to geometrically larger DM distributions. The general trend is that a lower boson mass $m$ leads to DM cloud-like configurations, while a higher DM mass leads to core-like systems. This is consistent with the behaviour found by \cite{Shakeri:2022dwg} for a different mass range, where increasing the DM mass fraction leads to cloud formation. For DM masses of $m = 1.34 \e{-11}\,eV$, the bosonic component is significantly larger than the fermionic one and forms a DM cloud. This can significantly decrease the compactness of the FBS (notice the different scales on the x-axis). However, the apparent compactness of the fermionic part is seen to increase significantly. Only observing the fermionic radius as in \autoref{fig:results:fermion-boson-stars:MR-grid} would lead to an apparent violation of GR, as the apparent compactness exceeds the Buchdahl limit of $C= 4/9$. The core-cloud behaviour can further be understood when considering the correlation length of the boson. For higher masses, the Compton wavelength is smaller and thus the DM Bose-Einstein condensate is found to be more strongly localized, i.e. extends to smaller radii. The opposite is the case for small DM masses. Observing a Buchdahl-limit violating NS could therefore be an indication for the presence of DM.

\begin{figure}
    \centering
    \includegraphics[width = 0.99\textwidth]{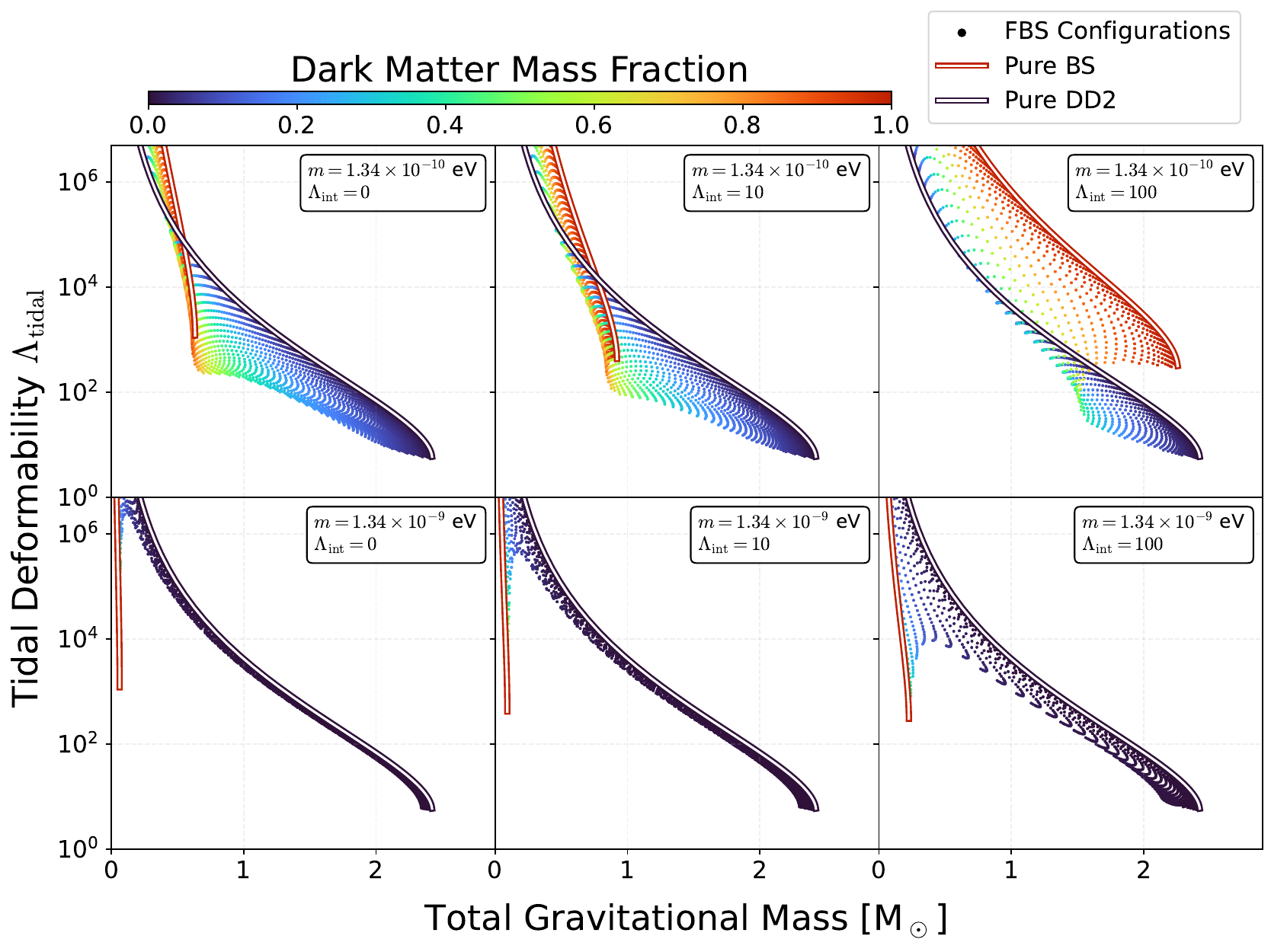}
    \includegraphics[width = 0.99\textwidth]{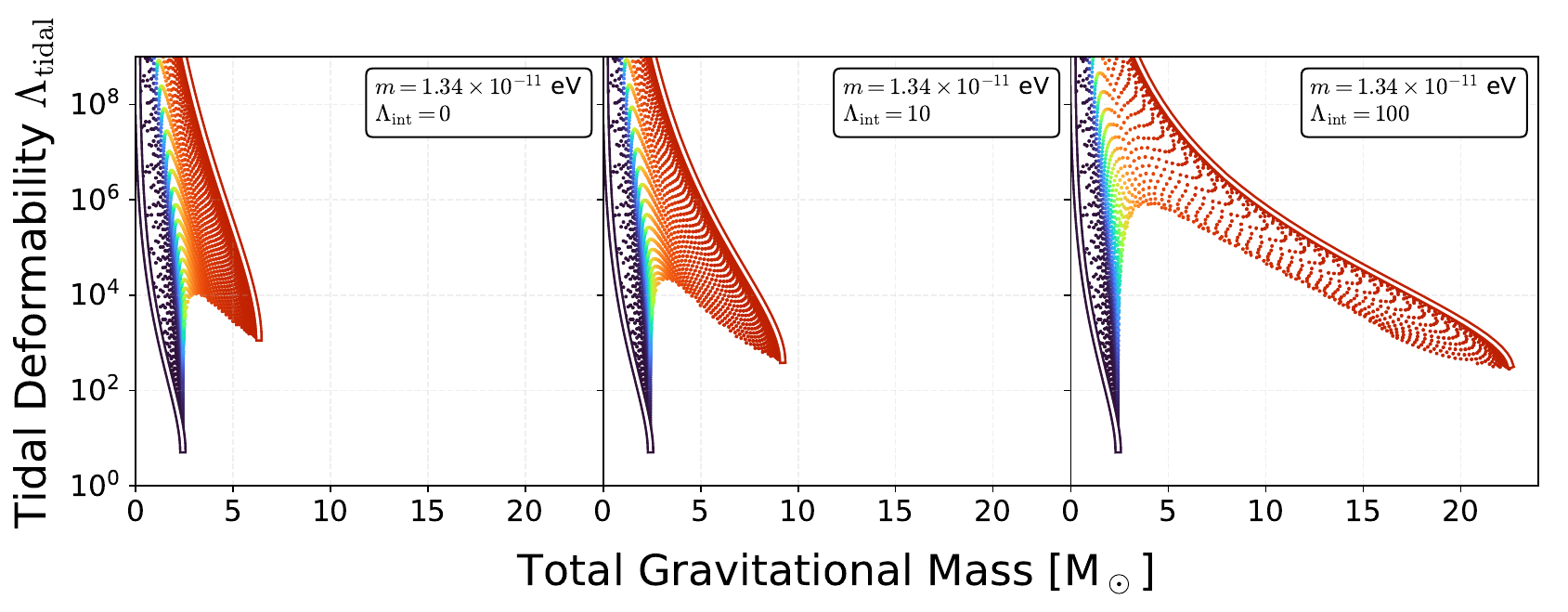}
    \caption{Relation between dimensionless tidal deformability $\Lambda_\mathrm{tidal}=\lambda_\mathrm{tidal}/M_\mathrm{tot}^5$ and total gravitational mass $M_\mathrm{tot}$ for different FBS. The rows correspond to bosonic masses of $m = \{1, 10, 0.1\}\times 1.34 \e{-10}\,eV$, columns correspond to self-interactions of $\Lambda_\mathrm{int}= \{0, 10, 100\}$ respectively. We use the DD2 EOS for the fermionic part. Even small DM-fractions can have a significant influence on the tidal properties of NS. Figures adapted from \cite{Diedrichs:2023trk}.}
    \label{fig:results:fermion-boson-stars:ML-grid}
\end{figure}

We show the relation between the dimensionless tidal deformability $\Lambda_\mathrm{tidal} = \lambda_\mathrm{tidal} / M_\mathrm{tot}^5$ and the total gravitational mass $M_\mathrm{tot}$ in \autoref{fig:results:fermion-boson-stars:ML-grid}. The solution corresponding to pure NS and pure BS are marked by thick black and thick red lines respectively. The lines corresponding to pure boson stars agree with the trend lines shown in \cite{Sennett:2017etc}. For $m = 1.34 \e{-9}\,eV$ and low DM-fractions, where the DM is mostly confined in a DM core, the tidal deformability is only weakly modified compared to a pure NS. For increasing  self-interaction strength $\Lambda_\mathrm{int} \approx 100$, the effect is magnified and the tidal deformability decreases significantly. Only for FBS dominated by DM, the results are close to that of pure boson stars. \\
For smaller boson masses, the effect on the tidal deformability is more pronounced. Since the tidal deformability of boson stars is much higher than that of pure NS, even small amounts of DM can significantly impact the tidal properties of FBS. This is especially apparent when $m = 1.34 \e{-11}\,eV$. Here, for constant central densities $\rho_c$, the tidal deformability increases orders of magnitude as $\varphi_c$ increases. At the point where the bosonic component starts to dominate, there is a turning point where the tidal deformability decreases while the total gravitational mass increases. The tidal deformability converges to the purely bosonic solutions for high DM-fractions. This opens up vast possibilities to probe DM, even for small DM-fractions. While the presence of DM in small quantities would barely be visible in the MR plane, it can significantly change the tidal deformability, even in small quantities, as evidenced in \autoref{fig:results:fermion-boson-stars:comparison-to-measurements}. \\
For masses of $m = 1.34 \e{-11}\,eV$, the tidal behaviour is more dependent on the self-interaction strength $\Lambda_\mathrm{int}$. For weaker self-interactions, the tidal deformability stays roughly in the same order of magnitude for constant central densities $\rho_c$ while gradually converging to the pure bosonic solution when increasing $\varphi_c$. For strong self-interactions, the tidal deformability can be seen to increase as it tends towards the pure boson star solutions. The configurations of high $\Lambda_\mathrm{tidal}$ correspond to DM cloud solutions. This behaviour is consistent with the findings of \cite{Karkevandi:2021ygv}, where an effective EOS was used for modelling the bosonic component. \\
For some parameters for the boson mass and self-interaction, degeneracies in the mass, radius and/or tidal deformabilities arise. That is, there exist FBS configurations with different DM-fractions which produce identical (fermionic) radii and masses. By combining measurements of mass and fermionic radius with measurements of the tidal deformability, it might be possible to break the degeneracy. In this study we used the DD2 EOS to model the fermionic component. We note that using different EOS for the fermionic component might lead to further degeneracies. For example, a NS with a given EOS and no DM might produce an identical total gravitational mass and fermionic radius to an FBS with a non-zero DM-fraction and a different EOS.

\subsubsection{Comparison to Observational Constraints} \label{subsec:results:comparison-to-observational-constraints}

\begin{figure}
    \centering
    \includegraphics[width=0.49\textwidth]{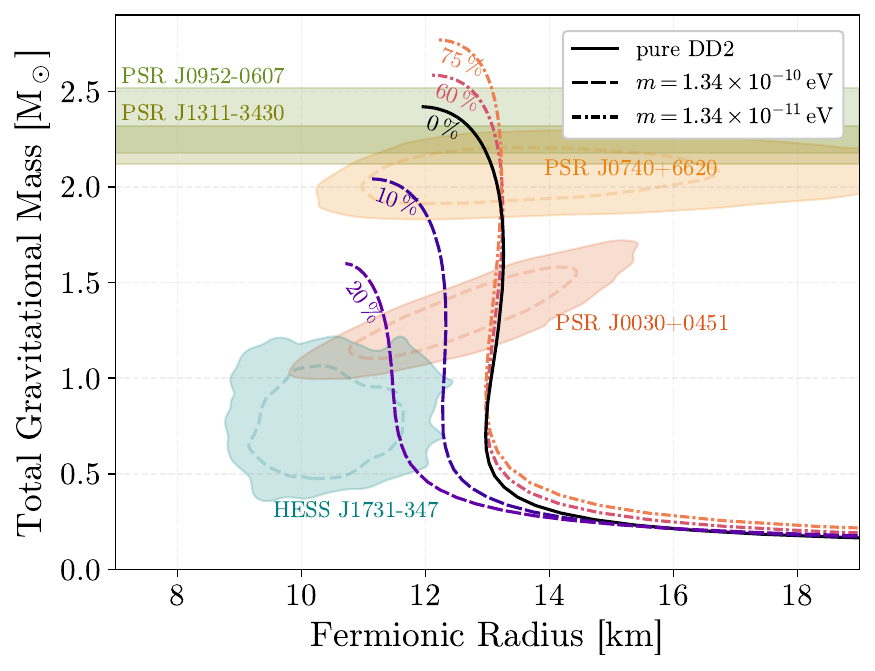}
    \includegraphics[width=0.499\textwidth]{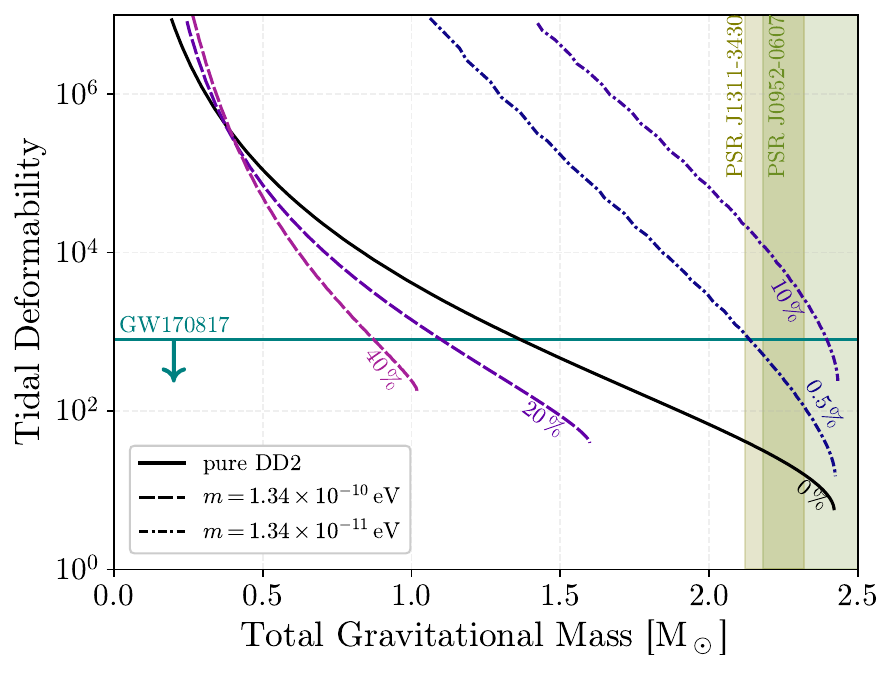}
    \caption{\textbf{Left panel:} Total gravitational mass and fermionic radius of FBS with boson masses of $m = 1.34 \e{-10}\,eV$ and $m = 1.34 \e{-11}\,eV$ and no self-interaction, shown together with observational constraints from PSR J0030+0451 \cite{Riley:2019yda}, PSR J0740+6620 \cite{Riley:2021pdl}, J0952-0607 \cite{Romani:2022jhd}, HESS J1731-347 \cite{Doroshenko2022:aeq} and PSR J1311-3430 \cite{Kandel:2022qor} . The percentage number denotes the DM-fraction. For the NICER and HESS measurements, dashed lines display the $1\sigma$ regions, while the total coloured areas show the $2\sigma$ regions. 
    \textbf{Right panel:} Dimensionless tidal deformability for the same set of parameters shown together with the constraint coming from the GW170817 event \cite{Abbott:2018wiz,LIGOScientific:2017vwq}. Figures taken from \cite{Diedrichs:2023trk}.}
    \label{fig:results:fermion-boson-stars:comparison-to-measurements}  %PSR J0348+0432 \cite{Antoniadis:2013pzd}, PSR J1614-2230 \cite{NANOGrav:2017wvv} and PSR J0952-0607 \cite{Romani:2022jhd}.
\end{figure}

Measurements of the fermionic radius of neutron stars with known masses have been performed by the NICER telescope, which tracks X-ray-emitting hotspots on the NS surface. The measurements include the pulsars PSR J0030+0451 with $M=1.34^{+0.15}_{-0.16}\,M_\odot$ and $R=12.71^{+1.14}_{-1.19}\,km$ \cite{Riley:2019yda} and J0740+6620 with $M=2.072^{+0.067}_{-0.066}\,M_\odot$ and $R=12.39^{+1.30}_{-0.98}\,km$ \cite{Riley:2021pdl}. Even though these measurements only provide two data points in the MR diagram, they can be used to constrain which FBS configurations could exist or which are excluded. These observational constraints also serve to constrain the DM properties. We show the posteriors of these measurements in \autoref{fig:results:fermion-boson-stars:comparison-to-measurements}. The measurements should be compared to \autoref{fig:results:fermion-boson-stars:MR-grid}, where the fermionic radius is shown. FBS solutions with a DM core generally become more compact with increasing DM-fraction. For higher DM-fraction, they are not able to fulfil all observational constraints since they reduce the maximum mass of the FBS compared to pure NS. In contrast, DM cloud solutions can easily reach higher maximum masses due to the FBS being dominated by the bosonic component, which mass scales like $M_\mathrm{max} \approx M^2_p / m$. This is consistent with \cite{Shakeri:2022dwg}, where the FBS is modelled with an effective EOS and also includes the changing photon geodesics due to the DM cloud, and \cite{Rutherford:2022xeb} where the authors performed a Bayesian analysis with the effective EOS. \\

Another constraint comes from the neutron star HESS J1731\ensuremath{-}347 with $M=0.77^{+0.20}_{-0.17}\,M_\odot$ and $R=10.4^{+1.86}_{-0.78}\,km$ \cite{Doroshenko2022:aeq}. This is the lightest known NS, which is difficult to explain using standard stellar evolution, see \cite{Stockinger:2020hse}. The authors of \cite{Doroshenko2022:aeq} propose it to be a strange star (a star made out of hadrons that include strange quarks), but when considering \autoref{fig:results:fermion-boson-stars:MR-grid}, this MR region can also be reached using FBS with DM core solutions. To obtain accurate constraints specifically for FBS, one would have to repeat their analysis with an actual bosonic component, which we leave for future work. But as it stands, the measurement of HESS J1731\ensuremath{-}347 favours DM core solutions and disfavours DM cloud solutions. \\

There is also the observation of the binary neutron star merger GW170817 \cite{Abbott:2018wiz}. The authors derived constraints on the tidal deformability with minimal assumptions on the nature of the involved compact objects. They use a mass-weighted linear combination of the individual tidal deformabilities and obtained a limit of $\Lambda_\mathrm{tidal} < 630$. Alternatively, assuming that all neutron stars have the same EOS, \cite{Abbott:2018exr} derived constraints on the tidal deformability with the help of universal relations (see \cite{Yagi:2013bca, Yagi:2013awa}). These constraints are not perfectly applicable to FBS, as the I-Love-Q relations are not necessarily applicable. The use of universal relations is further complicated by the fact that two FBS might have the same EOS but different DM-fractions. Some authors \cite{Maselli:2017vfi} have argued that the relations might be applicable, but we leave this for future work. However, the present measurements should be enough to get a rough understanding of which types of FBS might be favoured or disfavoured. It can be seen in \autoref{fig:results:fermion-boson-stars:comparison-to-measurements} that generally low tidal deformabilities are favoured. We leave a more thorough quantitative analysis of constraints on these FBS models for future work. \\

Previous studies using an effective EOS for the bosonic component reach similar conclusions and have placed initial constraints on different DM mass ranges, such as \cite{Giangrandi:2022wht,Karkevandi:2021ygv,Sagun:2022ezx}.  Overall, measurements of mass, radius and tidal deformability seem complementary and combining them in a quantitative analysis might be able to significantly constrain the parameter space. FBS with DM cloud solutions can have large tidal deformabilities, even for small DM-fractions. These large tidal deformabilities would most likely be observable in GW measurements of inspiralling binary systems.  The effects on $\Lambda_\mathrm{tidal}$ of DM cores inside of NS on the other hand can be obscured since they tend to only slightly change the global tidal properties, even for large DM-fractions. This is especially true for large DM masses. DM core solutions are however able to explain the HESS measurement, and assuming different DM-fractions for different FBS, these DM masses are not ruled out by the constraint on the maximum total gravitational mass. As noted before, these effects are somewhat degenerate with the neutron star EOS, since different EOS with different DM-fractions might lead to very similar observables. These degeneracies could be broken for example by looking at correlations between the DM-fraction inside FBS and the local DM density distribution in the galactic disc \cite{Giangrandi:2022wht,Sagun:2022ezx}.

\subsubsection{Comparison with an effective EOS} \label{subsec:results:comparison-to-effecive-eos}

Due to the significant numerical effort needed to solve the full system of equations for FBS \eqref{eq:fermion-boson-stars:scalar-fermion-boson-stars:TOV-equations-grr}-\eqref{eq:fermion-boson-stars:scalar-fermion-boson-stars:TOV-equations-P} self-consistently, earlier studies \cite{Leung:2022wcf,Colpi:1986ye} have used an effective EOS $P(e)$ for the scalar field, which treats it like a perfect fluid with pressure $P$ and a total energy density $e$. The effective EOS was originally derived in \cite{Colpi:1986ye} for the cases where the self-interaction strength is large $\Lambda_{\mathrm{int}} = \lambda / 8 \pi m^2 \gg 0$ . The effective EOS models exclusively the ground state of the scalar field and assumes that the energy-momentum tensor of the scalar field \eqref{eq:fermion-boson-stars:scalar-fermion-boson-stars:energy-momentum-tensor-scalar-field} is isotropic (which is only valid in the given limit). The effective EOS has the advantage that the scalar field must not be solved for directly, and the evolution equations simplify to the default TOV equations. For a scalar field with self-interaction potential \eqref{eq:results:fermion-boson-stars:quartic-self-interaction-potential}, the effective EOS is given by
\begin{align}
    P = \frac{4}{9} \rho_0 \left[ \left( 1 + \frac{3}{4} \frac{e}{\rho_0} \right)^{1/2} - 1 \right]^2  \: , \label{eq:results:fermion-boson-stars:effective-bosonic-eos}
\end{align}

where $\rho_0= m^4/2 \lambda$. Note that our expressions for $\rho_0$ and $\Lambda_{\mathrm{int}}$ differ from \cite{Leung:2022wcf,Colpi:1986ye} by a factor of two due to the different normalization of the scalar field $\varphi$ and the self-interaction parameter $\lambda$ in the potential \eqref{eq:results:fermion-boson-stars:quartic-self-interaction-potential}. The authors of \cite{Leung:2022wcf} used the effective EOS in a two-fluid system of two minimally coupled perfect fluids (i.e. which interact only gravitationally) to compute the tidal deformability $\Lambda_\mathrm{tidal}$ of FBS. One of the fluids was taken to be that of nuclear matter, just as was done in this work for the fermionic component. The other fluid describes the scalar field using the effective EOS. Hereafter, we compare the results obtained from integrating the two-fluid model (see \cite{Leung:2022wcf} for details) and from solving the full system of equations for the FBS \eqref{eq:fermion-boson-stars:scalar-fermion-boson-stars:TOV-equations-grr}-\eqref{eq:fermion-boson-stars:scalar-fermion-boson-stars:TOV-equations-P}. In addition, we compute the tidal deformability in two ways. First, as one would for a single-fluid system (details in \cite{Leung:2022wcf}) for the two-fluid model, and second, as was described in Chapter \ref{subsec:fermion-boson-stars:scalar-fermion-boson-stars} for the full system. We further note that the effective EOS is only valid for $\lambda > 0$ because otherwise, the pressure \eqref{eq:results:fermion-boson-stars:effective-bosonic-eos} would become negative.

\begin{figure}[h]
 	\centering
    \includegraphics[width=0.6\textwidth]{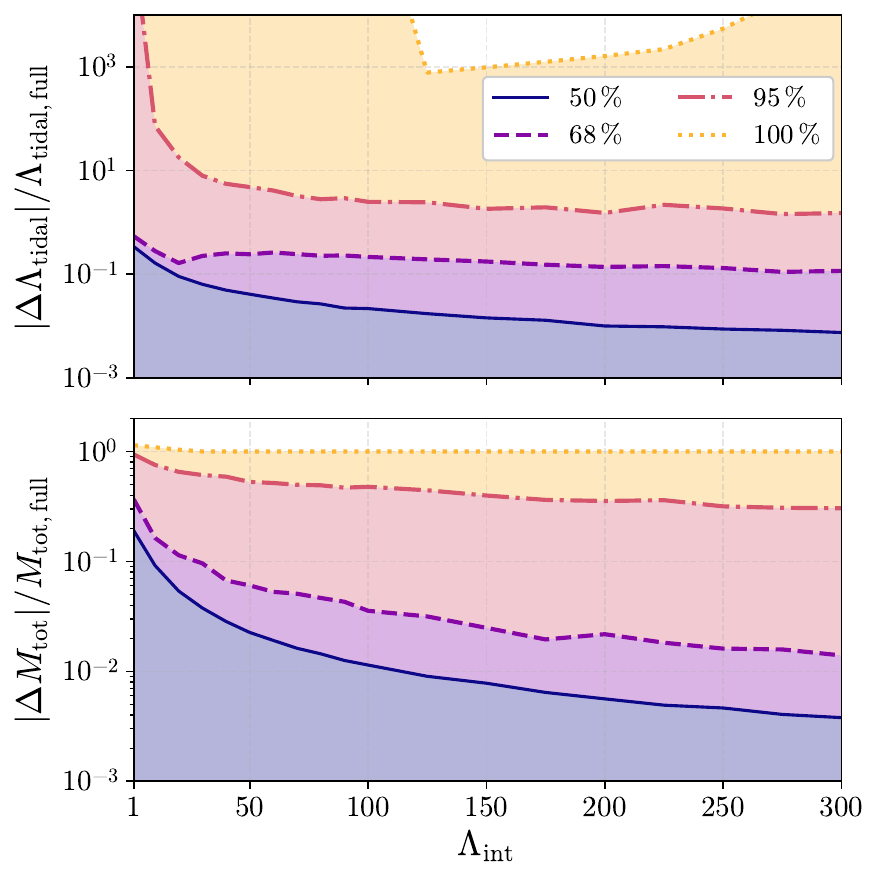}
    \caption{Distribution of the relative error (e.g. $|\Lambda_\mathrm{tidal, full} - \Lambda_\mathrm{tidal, eff}| / \Lambda_\mathrm{tidal, full}$) of the observables obtained using the full system and the effective EOS. Shown are values for the dimensionless tidal deformability $\Lambda_\mathrm{tidal}$ (upper panel) and total gravitational mass $M_{\mathrm{tot}}$ (lower panel) as a function of the interaction strength $\Lambda_\mathrm{int}$. The continuous line shows the boundary, below which half of the FBS configurations lie, meaning that half of them have a relative error of less than the shown value for a given $\Lambda_\mathrm{int}$. Similarly the dashed/dash-dotted/dotted lines show the line below which $68\,\%$/$95\,\%$/$100\,\%$ of configurations lie. Only stable FBS configurations were considered for the relative error at a given $\Lambda_\mathrm{int}$ and all computations were performed for $m=6.7 \e{-11}\,eV$. The agreement between the full system and the effective EOS becomes generally better for large $\Lambda_\mathrm{int}$, however, at some point, numerical inaccuracies in the full system dominate the relative error, which starts to be problematic for $\Lambda_\mathrm{int} \gtrsim 400$. Figure taken from \cite{Diedrichs:2023trk}.}
    \label{fig:results:fermion-boson-stars:error-comparison-effectiveEOS-fullsys}
\end{figure}

Regarding the initial conditions of the two-fluid model, we choose the same conditions as were used in \cite{Leung:2022wcf}. For better comparability between the full system and the effective EOS, we first want to find an expression that relates the scalar field $\varphi(r)$ to the energy density $e_{\textrm{eff}}$ of the effective fluid. To derive this relation, we set the $T_{tt}$-component of \eqref{eq:fermion-boson-stars:scalar-fermion-boson-stars:energy-momentum-tensor-scalar-field} equal to the $T_{tt}$-component of a perfect fluid. We therefore take  $T^{(\varphi)}_{tt} \stackrel{!}{=} e_{\textrm{eff}} \cdot \alpha^2$ and then use the approximations used in \cite{Colpi:1986ye} (i.e. local flatness and neglecting spatial derivatives, which is valid in the limit of strong self-interactions). We obtain an expression that depends only on the scalar field value $\varphi(r)$ (see \eqref{eq:fermion-boson-stars:scalar-fermion-boson-stars:scalar-field-harmonic-ansatz}), the scalar field mass $m$ and the self-interaction parameter $\lambda$
\begin{equation}
    e_{\textrm{eff}} (\varphi) = 2 m^2 \varphi^2 + \frac{3}{2} \lambda \varphi^4 \: , \label{eq:results:fermion-boson-stars:phi-to-energy-density}
\end{equation}
where $\omega^2 / \alpha^2  = m^2 + \lambda \varphi^2$ was substituted using the Klein-Gordon equation \eqref{eq:fermion-boson-stars:scalar-fermion-boson-stars:TOV-equations-phi}. Equation \eqref{eq:results:fermion-boson-stars:phi-to-energy-density} is a generic result and holds for all radii (under the approximations stated above). To obtain the initial conditions for the central energy density of the scalar field $e_{\textrm{eff}, c}$, one simply plugs in the corresponding central value of the scalar field $\varphi_c$. \\

In \autoref{fig:results:fermion-boson-stars:error-comparison-effectiveEOS-fullsys}, we show the relative error for the total gravitational mass $M_{\mathrm {tot}}$ and the tidal deformability $\Lambda_{\mathrm{tidal}}$, computed using the full system and the effective two-fluid system, with respect to the self-interaction strength $\Lambda_{\mathrm{int}}$.  We observe that the errors (the shaded regions) generally decrease when the self-interaction strength $\Lambda_{\mathrm{int}}$ increases. This is consistent with the assumption that the effective EOS \eqref{eq:results:fermion-boson-stars:effective-bosonic-eos} becomes exact only in the limit of strong self-interactions. For small $\Lambda_{\mathrm{int}}$, the relative error reaches $100\%$ for the total mass and diverges for the tidal deformability. This is to be expected since the total mass converges to zero for pure boson stars when using the effective EOS in the limit $\Lambda_{\mathrm{int}} \rightarrow 0$ (see Fig. 2 in \cite{Colpi:1986ye}), whereas it reaches a constant value when computing the mass using the full system. Likewise, due to the definition of the dimensionless tidal deformability $\Lambda_\mathrm{tidal} = \lambda_\mathrm{tidal} / M_\mathrm{tot}^5$, a diverging error is to be expected. For $\Lambda_{\mathrm{int}} \approx 100$ the maximal error of the total mass (tidal deformability) is on the order of $88\,\%$ ($>10^4\,\%$), whereas the lower 95-th percentiles of errors are noticeably smaller at around $< 47\,\%$ ($< 240\,\%$). This means that only $5\%$ of the computed configurations have relative errors higher than $47\,\%$ ($250\,\%$). The median error denoted by the solid blue line is around $1\,\%$ ($2\,\%$). At $\Lambda_{\mathrm{int}}=300$ the maximal error reaches $85\,\%$ ($>10^4\,\%$) and the median error reaches $0.4\,\%$ ($0.8\,\%$). Asymptotically, the error is constrained by floating-point precision and the inherent non-exactness of the effective EOS as compared to the full system.

\begin{figure}
    \centering
    \includegraphics[width=0.499\textwidth]{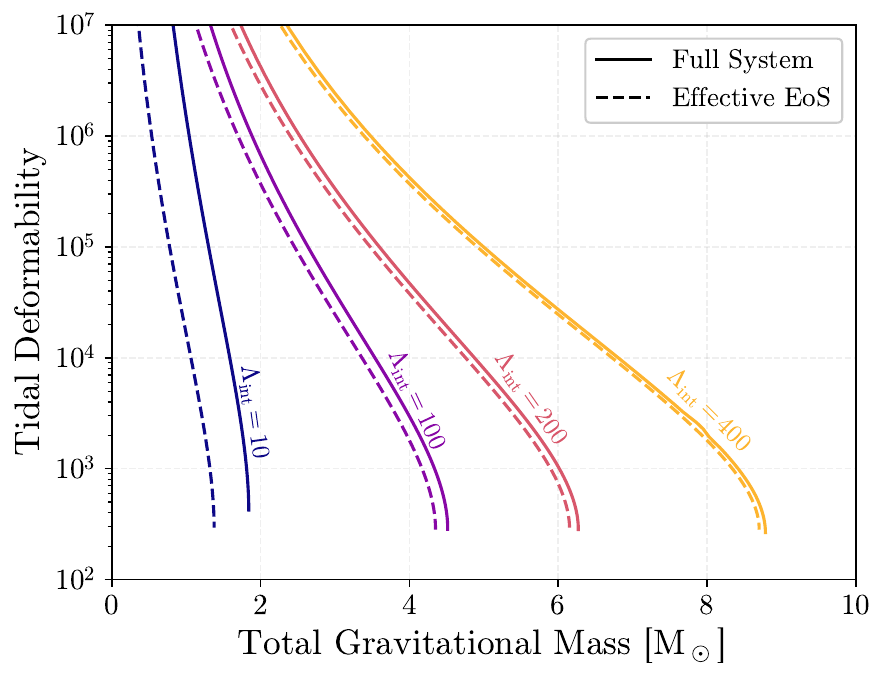}
    \includegraphics[width=0.49\textwidth]{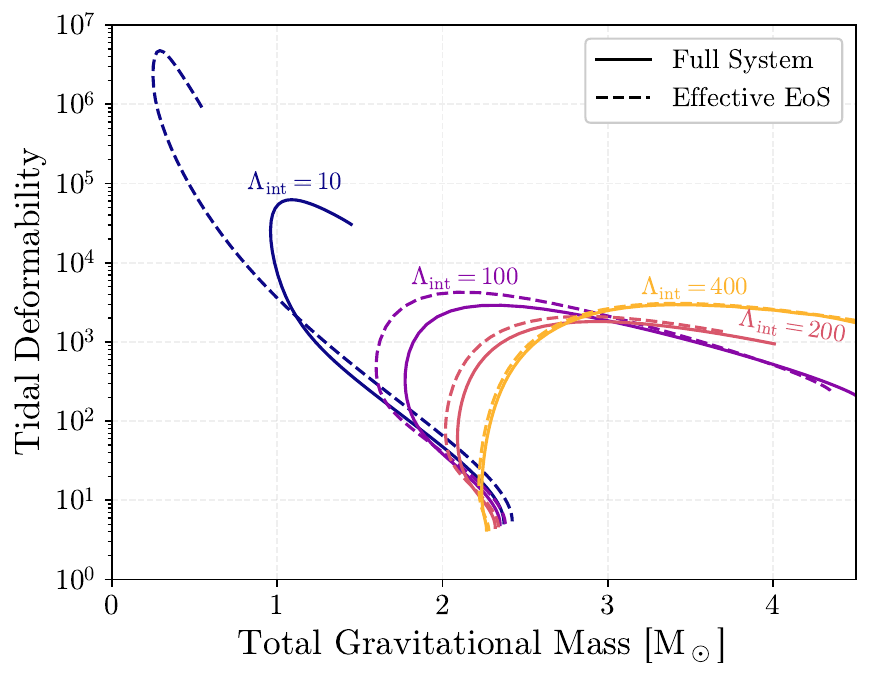}
    \caption{\textbf{Left panel:} Tidal deformability $\Lambda_\mathrm{tidal}$ with respect to the total gravitational mass $M_\mathrm{tot}$ for pure boson stars with various self-interaction strengths $\Lambda_{\mathrm{int}}$. For all cases, the boson mass is $m=6.7 \e{-11}\,eV$. The solid lines are the values obtained using the full system \eqref{eq:fermion-boson-stars:scalar-fermion-boson-stars:TOV-equations-grr}-\eqref{eq:fermion-boson-stars:scalar-fermion-boson-stars:TOV-equations-P} and the dashed lines correspond to the solutions using the effective bosonic EOS \eqref{eq:results:fermion-boson-stars:effective-bosonic-eos}. 
    \textbf{Right panel:} Tidal deformability $\Lambda_\mathrm{tidal}$ with respect to the total gravitational mass $M_\mathrm{tot}$ of different FBS for different self-interaction strengths $\Lambda_{\mathrm{int}}$. For all cases, the boson mass is $m=6.7 \e{-11}\,eV$. All lines have a constant central value of the scalar field $\varphi_c = 0.02$, but different central densities $\rho_c$. Only stable FBS are shown. The solid lines are the values obtained using the full system \eqref{eq:fermion-boson-stars:scalar-fermion-boson-stars:TOV-equations-grr}-\eqref{eq:fermion-boson-stars:scalar-fermion-boson-stars:TOV-equations-P} and the dashed lines correspond to the solutions using the effective bosonic EOS \eqref{eq:results:fermion-boson-stars:effective-bosonic-eos}. Figures taken from \cite{Diedrichs:2023trk}.}
    \label{fig:results:fermion-boson-stars:TLN-full-eff-comparison-Lambda_int}
\end{figure}

To gain a better understanding of how the effective EOS and the full system compare, we compute the tidal deformability $\Lambda_{\mathrm{tidal}}$ using both systems. In the left panel of \autoref{fig:results:fermion-boson-stars:TLN-full-eff-comparison-Lambda_int}, we show the tidal deformability of pure boson stars computed for different self-interaction strengths of $\Lambda_{\mathrm{int}} = \{10, 100, 200, 400 \}$. The solid lines show the solutions using the full system and the dashed lines are the values obtained using the effective EOS. The effective EOS is able to reproduce the solution of the full system qualitatively, even for a small self-interaction strength $\Lambda_{\mathrm{int}}$. With increasing $\Lambda_{\mathrm{int}}$, the agreement between the full and the effective system becomes better. At around $\Lambda_{\mathrm{int}} = 400$, the quantitative agreement reaches a few percent of relative difference. \\

Next, we consider mixed configurations where both the scalar field and central density are non-zero. The right panel of \autoref{fig:results:fermion-boson-stars:TLN-full-eff-comparison-Lambda_int} shows the tidal deformability with respect to the total gravitational mass of the FBS. We compute several curves of constant central scalar field $\varphi_c$ at different $\Lambda_{\mathrm{int}} = \{10, 100, 200, 400 \}$. The choice of a constant $\varphi_c$ is per se arbitrary but was made to make our work more easily reproducible in future works. The solid lines show the solutions obtained using the full system and the dashed lines were computed using the effective EOS (all other values being equal). With increasing self-interaction strength $\Lambda_{\mathrm{int}}$, the solutions using the effective EOS converge to the solutions with the full system with increasing accuracy. Even though at lower $\Lambda_{\mathrm{int}} < 200$ the deviations are quite large, the qualitative trend is correctly reproduced by the effective EOS. At $\Lambda_{\mathrm{int}}=400$, both systems produce reasonably similar results (within a few percent of relative difference). This justifies the usage of the effective EOS for large $\Lambda_{\mathrm{int}} \gtrsim 400$ also to compute the tidal deformability $\Lambda_{\mathrm{tidal}}$. \\

We close the analysis by providing a few notes on the usefulness of the effective EOS \eqref{eq:results:fermion-boson-stars:effective-bosonic-eos} and the two-fluid system. For most FBS configurations, we were able to verify the general notion that the effective EOS becomes asymptotically more accurate and is able to reproduce the behaviour of the full system. However, we find that a significant percentage of FBS configurations with high relative errors remain, especially when considering the tidal deformability, where the relative error surpasses 200\% for roughly five percent of all configurations. This can be explained by the different low-mass limits of the full and the effective system while considering the definition of the dimensionless tidal deformability $\Lambda_\mathrm{tidal} = \lambda_\mathrm{tidal} / M_\mathrm{tot}^5$. Nevertheless, we conclude that the usage of the effective EOS is justified in the cases where the self-interactions are strong (roughly $\Lambda_{\mathrm{int}} \gtrsim 400$), as the errors are acceptable for most configurations with high total gravitational mass. Of course, solving the full system \eqref{eq:fermion-boson-stars:scalar-fermion-boson-stars:TOV-equations-grr}-\eqref{eq:fermion-boson-stars:scalar-fermion-boson-stars:TOV-equations-P} will always yield the exact results in theory. In practice, it can be numerically difficult to integrate the full system for strong self-interactions $\Lambda_{\mathrm{int}} \gtrsim 400$ because 
\begin{itemize}
\itemsep0em
\item[$1)$] the frequency $\omega$ must be tuned up to higher accuracy than what is possible using 64-bit floating-point numbers and
\item[$2)$] increasingly small step-sizes are needed, to solve the equations correctly.
\end{itemize}

During our tests, we could determine that the more relevant constraining factor is the needed accuracy for $\omega$, rather than the step-size. Smaller initial values for the scalar field $\varphi_c$ lead to larger bosonic radii $\gg 10\,km$, for which the numerical integration becomes problematic. This concerns roughly $5\,\%$ of the considered configurations. In contrast, when the two-fluid system is used together with the effective bosonic EOS, the solution is numerically robust and does not require numerical root-finding for $\omega$, and can manage well with larger numerical step-sizes. With equal step-sizes and initial conditions, we found that the two-fluid system takes around two orders of magnitude less computation time than solving the full system. The speedup can be increased further when considering that the two-fluid system also tolerates larger step-sizes while staying numerically accurate.

\subsection{Fermion Proca Stars} \label{subsec:results:fermion-proca-stars}

We present our results regarding fermion Proca stars (FPS), mixed systems of neutron stars and vector dark matter, modelled by a complex vector field. We consider a quartic self-interaction potential with a mass term of the form
\begin{align}
  V(A_\mu \bar{A}^\mu ) = m^2 A_\mu \bar{A}^\mu + \frac{\lambda}{2} ( A_\mu \bar{A}^\mu )^2 \: , \label{eq:results:fermion-proca-stars:quartic-self-interaction-potential}
\end{align}

where $m$ is the vector boson mass and $\lambda$ is the self-interaction parameter. Similar to the case with the scalar field, we define the effective self-interaction parameter $\Lambda_\mathrm{int} = \lambda / 8\pi m^2$. As is the case for the scalar boson case, the parameter $\Lambda_\mathrm{int}$ is a useful measure for the self-interaction strength and parametrizes scaling relations for the total gravitational mass $M_\mathrm{max} \approx  1.058 M^2_p / m$ (for small $\Lambda_\mathrm{int}$) and $M_\mathrm{max} \approx \sqrt{\Lambda_\mathrm{int}} \ln(\Lambda_\mathrm{int})\, M^2_p / m$ (for large $\Lambda_\mathrm{int}$), as found by \cite{Brito:2015pxa} and \cite{Minamitsuji:2018kof} respectively. We further note that the self-interaction parameter $\Lambda_\mathrm{int}$ in our work differs from the one used in \cite{Minamitsuji:2018kof} by a factor of two, even though they are defined in the same way. This discrepancy is due to the different normalization used for the vector field. \\

We hereafter consider a variety of different cases of FPS, starting with radial profiles in Chapter \ref{subsec:results:radial-profiles}. In Chapter \ref{subsec:results:stability-and-MR-relations}, we analyse the stability of FPS and show the mass-radius relations of stable configurations. For that, we use the same parameter range -- namely $m = \{0.1, 1, 10\}\times 1.34 \e{-10}\,eV$ and $\Lambda_\mathrm{int} = \{0, 10, 100 \}$, see Chapter \ref{subsec:results:fermion-boson-stars} -- as was used for FBS to enable simpler comparison between the scalar field and vector field cases. Arguments about the mass-range and self-interaction ranges chosen apply here in the same way. We further use the DD2 EOS in our analysis of FPS. In Chapter \ref{subsec:results:study-of-higher-modes-and-different-eos}, we briefly study higher modes of FPS as well as FPS with different EOS for the neutron star component. This is mainly done to solidify the analysis of the FPS properties and decrease the effect of the choice of the EOS on our conclusions.

\subsubsection{Radial Profiles} \label{subsec:results:radial-profiles}

\begin{figure}[h]
\centering
    \includegraphics[width=0.49\textwidth]{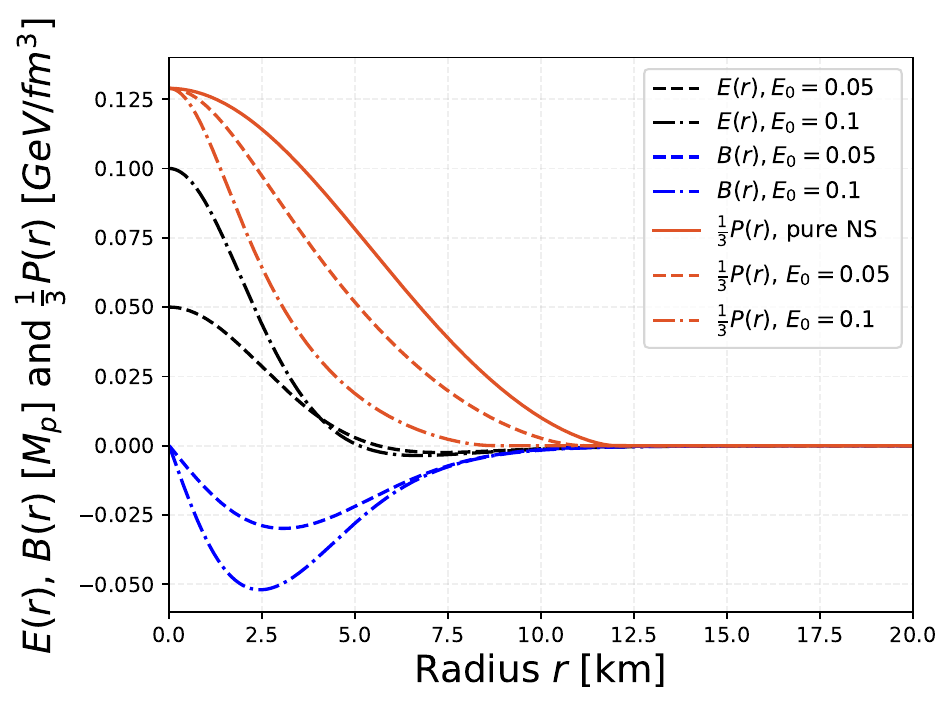}
    \includegraphics[width=0.49\textwidth]{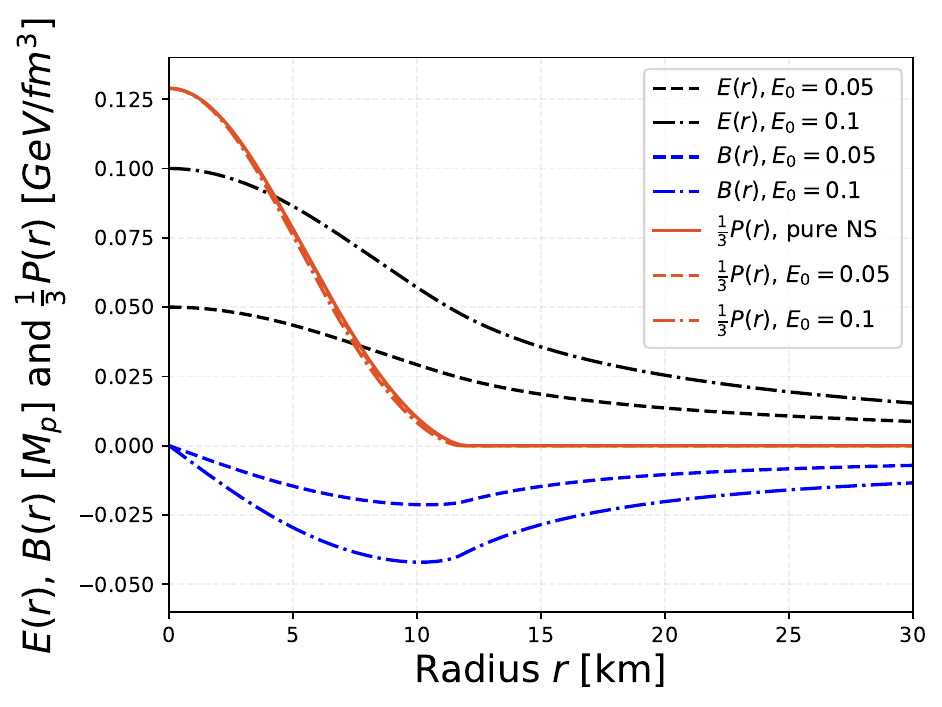}
    \caption{\textbf{Left panel:} Radial profiles of the pressure $P(r)$ (orange) and the vector field components $E(r)$ (black), $B(r)$ (blue) of the zeroth mode of different FPS with potential \eqref{eq:results:fermion-proca-stars:quartic-self-interaction-potential}. The mass is $m=1.34 \e{-10}\,eV$ and $\Lambda_\mathrm{int} = 0$. The FPS have a central density of $\rho_c = 5 \rho_{\mathrm{sat}}$ and differing central vector field amplitudes $E_0$. The pressure has been re-scaled by a factor of $3$ for convenience. The DM forms a core and compactifies the fermionic component.
    \textbf{Right panel:} Same as in the left panel, but this time the vector boson mass is set to $m=1.34 \e{-11}\,eV$. The DM forms a cloud around the fermionic component. The radius of the fermionic component is barely affected by the field. A kink can be seen in the profile for $B(r)$ at roughly $11.5\,km$, corresponding to the point where the fermionic radius is located. This illustrates the gravitational back-reaction between the vector field and NS matter.}
    \label{fig:results:fermion-proca-stars:radial-profiles-1}
\end{figure}

\begin{figure}[h]
\centering
    \includegraphics[width=0.49\textwidth]{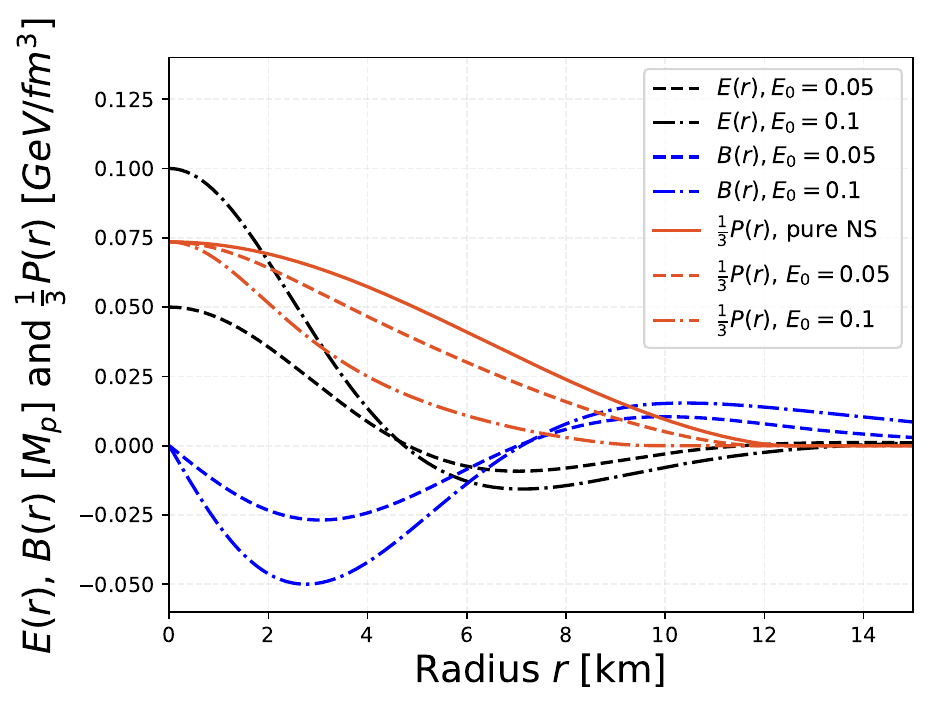}
    \includegraphics[width=0.49\textwidth]{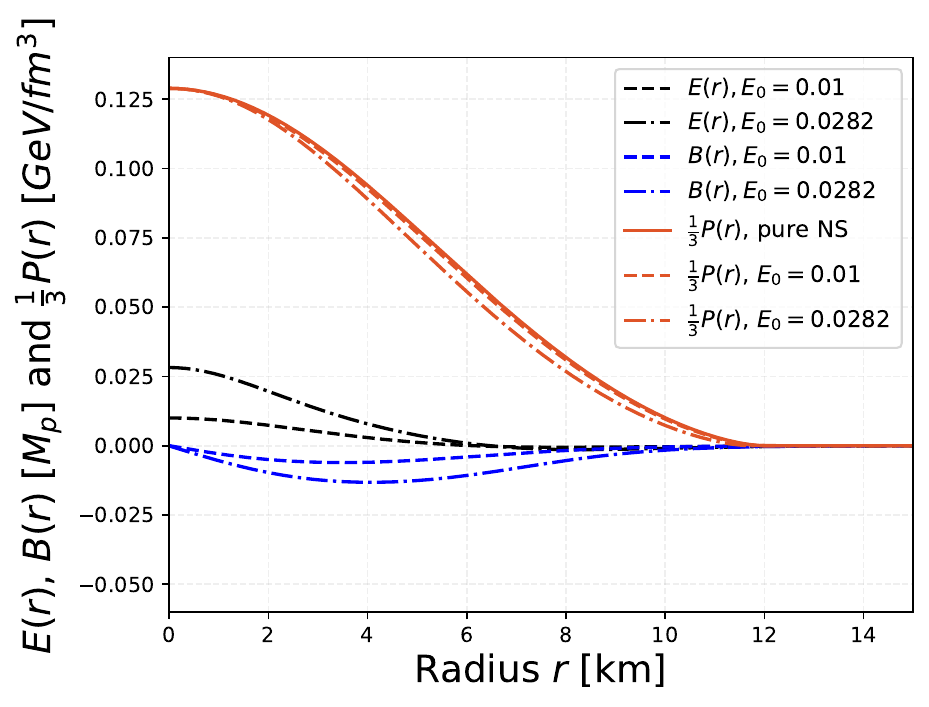}
    \caption{\textbf{Left panel:} Radial profiles of the pressure $P(r)$ (orange) and the vector field components $E(r)$ (black), $B(r)$ (blue) of the first mode of different FPS with potential \eqref{eq:results:fermion-proca-stars:quartic-self-interaction-potential}. The mass is $m=1.005 \e{-10}\,eV$ and $\Lambda_\mathrm{int} = 0$. The FPS has a central density of $\rho_c = 4 \rho_{\mathrm{sat}}$ and differing central vector field amplitudes $E_0$. The pressure has been re-scaled by a factor of $3$ for convenience.
    \textbf{Right panel:} Radial profiles of the pressure $P(r)$ (orange) and the vector field components $E(r)$ (black), $B(r)$ (blue) of an FPS in the zeroth mode with potential \eqref{eq:results:fermion-proca-stars:quartic-self-interaction-potential}. The mass is $m=1.34 \e{-10}\,eV$ and the self-interaction strength is $\Lambda_\mathrm{int} = 50$. The FPS has a central density of $\rho_c = 5 \rho_{\mathrm{sat}}$ and differing central vector field amplitudes $E_0$. The pressure has been re-scaled by a factor of $3$ for convenience. Due to the analytical bound on $E_0$ \eqref{eq:fermion-boson-stars:fermion-proca-stars:analytical-bound-amplitude-rewritten}, the maximal amplitude is roughly $E_{0,\text{crit}} \approx 0.0282$. The limited field amplitude strongly limits the effect on the fermionic component.}
    \label{fig:results:fermion-proca-stars:radial-profiles-2}
\end{figure}

One of the first steps when analysing self-gravitating objects is to consider radial profiles of the matter-fields constituting the object. In the context of FPS, we consider three main relevant physical quantities: the radial dependence of the pressure $P(r)$ and the vector field components $E(r)$, $B(r)$. Even though the radial distribution of physical quantities is not yet possible to be observed directly (although one could infer the DM-field distribution using the geodesic motion of light \cite{Shakeri:2022dwg}), a good understanding of the internal structure of FPS can be used to deduce their global quantities and vice-versa. Knowledge about the internal distribution is also relevant for numerical applications. We also include the radial profiles here to facilitate reproducibility of this work and for the sake of code-validation for future works. \\
Radial profiles of pure Proca stars have already been discussed by \cite{Brito:2015pxa} and for the case of a quartic self-interaction potential like \eqref{eq:results:fermion-proca-stars:quartic-self-interaction-potential} by \cite{Minamitsuji:2018kof}. We used the results of \cite{Minamitsuji:2018kof} in particular to verify that our code \cite{Diedrichs-Becker-Jockel} reproduces the results correctly and consistently. \\

In \autoref{fig:results:fermion-proca-stars:radial-profiles-1}, we show radial profiles of the pressure $P(r)$ (orange) and the vector field components $E(r)$ (black), $B(r)$ (blue) of the zeroth mode of different FPS with potential \eqref{eq:results:fermion-proca-stars:quartic-self-interaction-potential}. In the left panel, we take a boson mass of $m=1.34 \e{-10}\,eV$ and an interaction strength of $\Lambda_\mathrm{int} = 0$. The FPS have central densities of $\rho_c = 5 \rho_{\mathrm{sat}}$ and differing central vector field amplitudes $E_0$. The radial profile of a pure NS is shown with the orange continuous line and has no corresponding vector field (because it would be zero everywhere). The presence of the DM can be seen to compactify the NS component with increasing central field amplitude $E_0$. The field forms a DM core configuration. In the right panel of \autoref{fig:results:fermion-proca-stars:radial-profiles-1}, all parameters are left equal except for the vector boson mass, which is set to $m=1.34 \e{-11}\,eV$. Due to the low mass, the correlation length increases, which increases the size of the vector field component and forms a DM cloud configuration. Since the amount of energy density of the vector field is distributed inside and outside of the NS component, the effect on the radius is small. At around $r=11.5\,km$, a kink can be seen in the radial profile of the field component $B(r)$. This point coincides with the point where the fermionic radius of the FBS is located. This illustrates the gravitational back-reaction between the vector field and the NS component of the FBS. \\
In \autoref{fig:results:fermion-proca-stars:radial-profiles-2}, radial profiles of the pressure $P(r)$ (orange) and the vector field components $E(r)$ (black), $B(r)$ (blue) of an FPS are shown. In the left panel, we show an FPS in the first mode, which can be identified by the fact that the $E(r)$ component crosses the x-axis twice and $B(r)$ crosses it once. The boson mass is $m=1.005 \e{-10}\,eV$ and $\Lambda_\mathrm{int} = 0$. This time, the central density is taken to be $\rho_c = 4 \rho_{\mathrm{sat}}$ and the central vector field amplitudes vary. The right panel of \autoref{fig:results:fermion-proca-stars:radial-profiles-2} shows an FBS in the zeroth mode with a vector boson mass of $m=1.34 \e{-10}\,eV$ and the self-interaction strength is $\Lambda_\mathrm{int} = 50$. Due to the analytical bound on $E_0$ \eqref{eq:fermion-boson-stars:fermion-proca-stars:analytical-bound-amplitude-rewritten}, the maximal amplitude is roughly $E_{0,\text{crit}} \approx 0.0282$. The limited field amplitude strongly limits the possible effect on the fermionic component and thus on the fermionic radius, especially in the limit of large $\Lambda_\mathrm{int}$ . It may therefore be difficult to detect strongly self-interacting vector DM within a NS if one only considers measurements of the fermionic radius. It is also conceivable that the maximum amplitude $E_{0,\text{crit}}$ implies a maximum amount of possible accretion of vector DM, which could be used to set bounds on the DM self-interaction strength. We leave the analysis of this aspect for a future work.

\subsubsection{Stability and Mass-Radius Relations} \label{subsec:results:stability-and-MR-relations}

\begin{figure}[h]
\centering
    \includegraphics[width=0.499\textwidth]{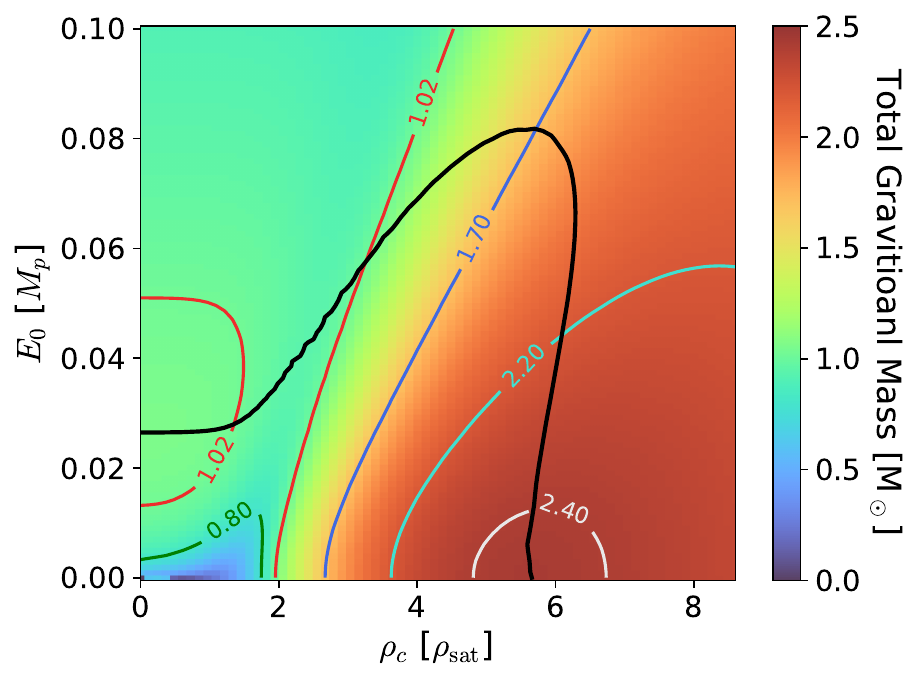}
    \includegraphics[width=0.49\textwidth]{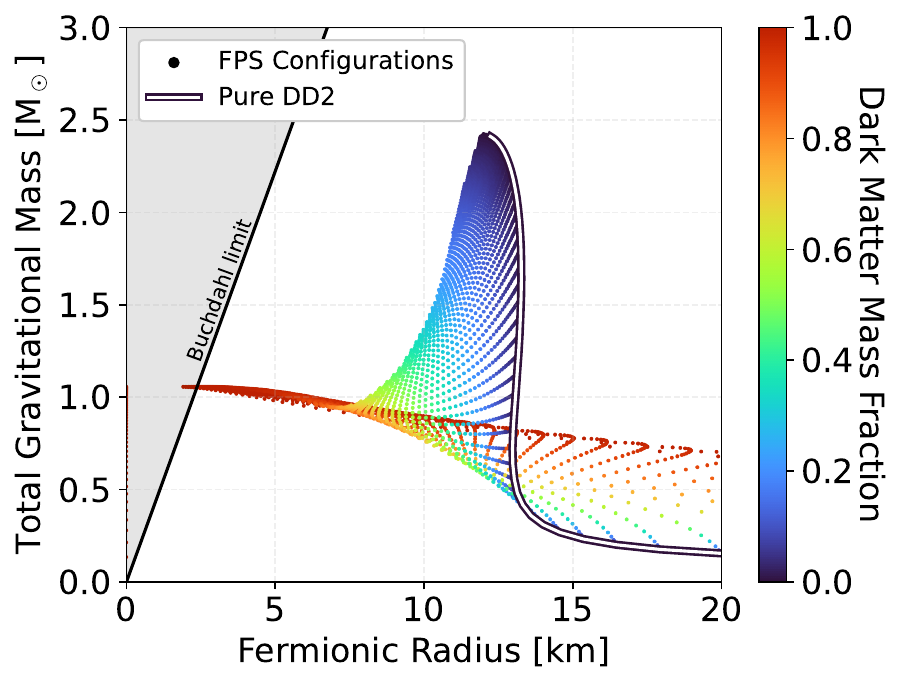}
    \caption{\textbf{Left panel:} Total gravitational mass of different FPS as a function of the restmass density $\rho_c$ and central vector field amplitude $E_0$. The black line corresponds to the stability curve, which separates stable solutions (in the lower left region) from unstable solutions (everywhere else). The qualitative behaviour of the stability curve of FPS is similar to the case with FBS (see \autoref{fig:results:fermion-boson-stars:stability-and-MR-curve-example})
    \textbf{Right panel:} Mass-radius diagram displaying the fermionic radius vs the total gravitational mass for FPS configurations that are within the stability region displayed in the left panel. Each point corresponds to a single configuration and is colour-coded according to the DM-fraction $N_\mathrm{b}/(N_\mathrm{b} + N_\mathrm{f})$. The solid black-white line shows the mass-radius curve for pure fermionic matter. In both cases, a vector field with mass of $m=1.34 \e{-10}\,eV$ and no self-interactions was considered in addition to the DD2 EOS for the fermionic part.}
    \label{fig:results:fermion-proca-stars:stability-and-MR-curve-lamda0}
\end{figure}

\begin{figure}[h]
\centering
    \includegraphics[width=0.499\textwidth]{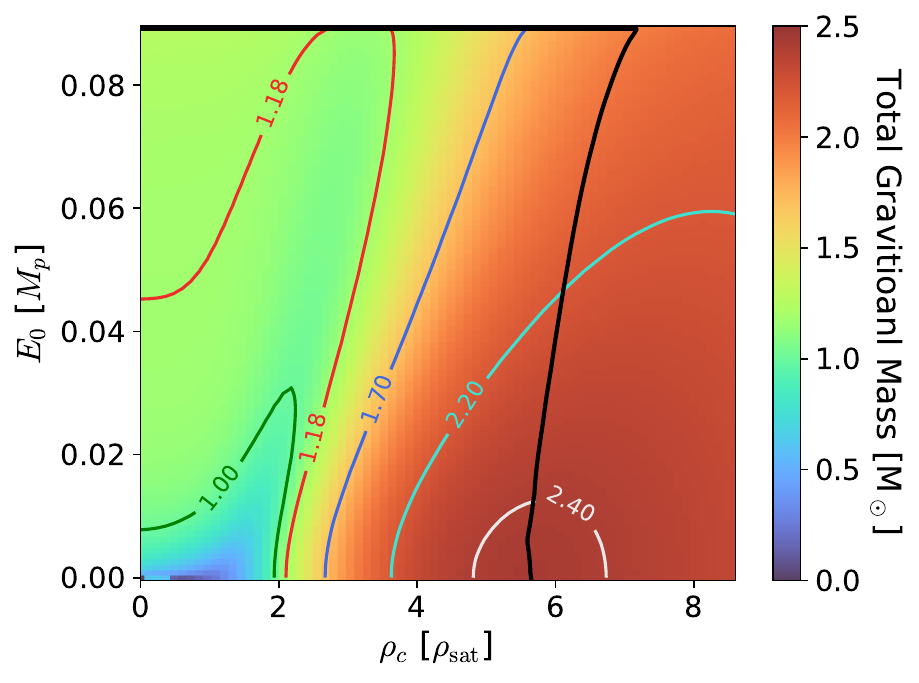}
    \includegraphics[width=0.49\textwidth]{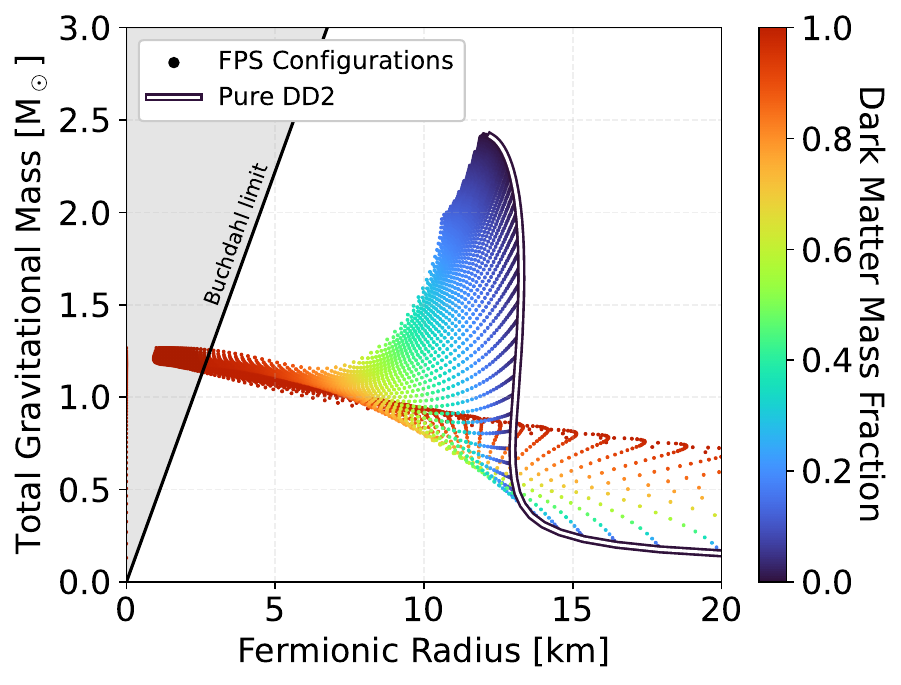}
    \caption{\textbf{Left panel:} Total gravitational mass of different FPS as a function of the restmass density $\rho_c$ and central vector field amplitude $E_0$, with $m=1.34 \e{-10}\,eV$ and $\Lambda_\mathrm{int}=5$. The black line corresponds to the stability curve, which separates stable solutions (in the lower left region) from unstable solutions (everywhere else). The stability curve reaches configurations with the maximum possible vector field amplitude $E_{0,\text{crit}} \approx 0.089$. This is a feature unique to FPS.
    \textbf{Right panel:} Mass-radius diagram displaying the fermionic radius vs the total gravitational mass for FPS configurations that are within the stability region displayed in the left panel. Each point corresponds to a single configuration and is colour-coded according to the DM-fraction $N_\mathrm{b}/(N_\mathrm{b} + N_\mathrm{f})$. The solid black-white line shows the mass-radius curve for pure fermionic matter. A vector field with mass of $m=1.34 \e{-10}\,eV$ and $\Lambda_\mathrm{int}=5$ was considered in addition to the DD2 EOS for the fermionic part.}
    \label{fig:results:fermion-proca-stars:stability-and-MR-curve-lamda5}
\end{figure}

We compute a grid of FPS with different central densities $\rho_c$ and central vector field amplitudes $E_0$. This can be seen in the left panel of \autoref{fig:results:fermion-proca-stars:stability-and-MR-curve-lamda0}, where we compute FPS with a quartic self-interaction potential \eqref{eq:results:fermion-proca-stars:quartic-self-interaction-potential} with $m=1.34 \e{-10}\,eV$ and $\Lambda_\mathrm{int}=0$. We additionally compute the stability curve, which defines the boundary between configurations stable under linear radial perturbations and configurations which are unstable. We use the same methods as were used for the FBS, because the stability criterion \eqref{eq:fermion-boson-stars:scalar-fermion-boson-stars:FBS-stability-criterion-rewritten} generalizes for systems of two gravitationally interacting fluids. The shape of the stability curve for FPS is qualitatively very similar to the FBS case. For pure neutrons stars and Proca stars respectively, the curve converges on the $\rho_c$- and $E_0$-axis at the point, where the non-mixed configurations have their maximum gravitational masses. Taking only the FPS inside of the stability region, enclosed by the stability curve, and plotting them in an MR diagram leads to the graph in the right panel of \autoref{fig:results:fermion-proca-stars:stability-and-MR-curve-lamda0}. As with the scalar case, stable FPS configurations now form an MR region instead of an MR curve (in the case of single-fluid systems). The stable configurations can be seen to form core or cloud solutions, depending on their DM-fraction $N_\mathrm{b}/(N_\mathrm{b} + N_\mathrm{f})$. The FPS with high DM-fractions have masses of roughly $1\,M_\odot$, which is higher than in the scalar field case with equal boson mass $m$. This can be explained through the differing scaling relations for pure Proca stars and boson stars. Another point where FPS differ from FBS is the existence of a maximal amplitude $E_{0,\text{crit}}$ \eqref{eq:fermion-boson-stars:fermion-proca-stars:analytical-bound-amplitude} for the bosonic vector field. When increasing the self-interaction strength $\Lambda_\mathrm{int}$, the maximal possible vector field amplitude shrinks, which affects the shape of the stability curve as well. In \autoref{fig:results:fermion-proca-stars:stability-and-MR-curve-lamda5} (left panel), we show such a case where the self-interaction strength is $\Lambda_\mathrm{int} = 5$. The stability curve does not reach the $E_0$-axis any more and instead rises vertically from the pure NS configurations until it reaches the FPS with maximal central vector field amplitude $E_{0,\text{crit}} \approx 0.089$. We have manually extended the stability curve so that it proceeds horizontally until it reaches the $E_0$-axis. It is noteworthy that this behaviour starts at surprisingly small self-interaction strengths and persists up to higher $\Lambda_\mathrm{int}$. In principle, also a third behaviour of the stability curve of FPS is conceivable. For some specific $\Lambda_\mathrm{int}$, it should be possible that the stability curve does not admit one continuous shape like in \autoref{fig:results:fermion-proca-stars:stability-and-MR-curve-lamda0} or \autoref{fig:results:fermion-proca-stars:stability-and-MR-curve-lamda5}, but that the stability curve is cut into two parts. Namely one part which starts at the $E_0$-axis and then rises to reach the edge where $E_{0,\text{crit}}$ is located, and another part which starts at the $\rho_c$-axis and then rises roughly vertically until it too reaches the analytical bound for the vector field amplitude $E_{0,\text{crit}}$ (think of a horizontal line cutting through the stability curve in \autoref{fig:results:fermion-proca-stars:stability-and-MR-curve-lamda0} at e.g. $E_0 = 0.06$). During our testing, we did not find any case where the stability curve follows this behaviour. However, there is also no reason, that we are aware of, why such a behaviour of the stability curve should be forbidden, which is why we presume that such a case might exist. \\

\begin{figure}
    \centering
    \includegraphics[width = 0.99\textwidth]{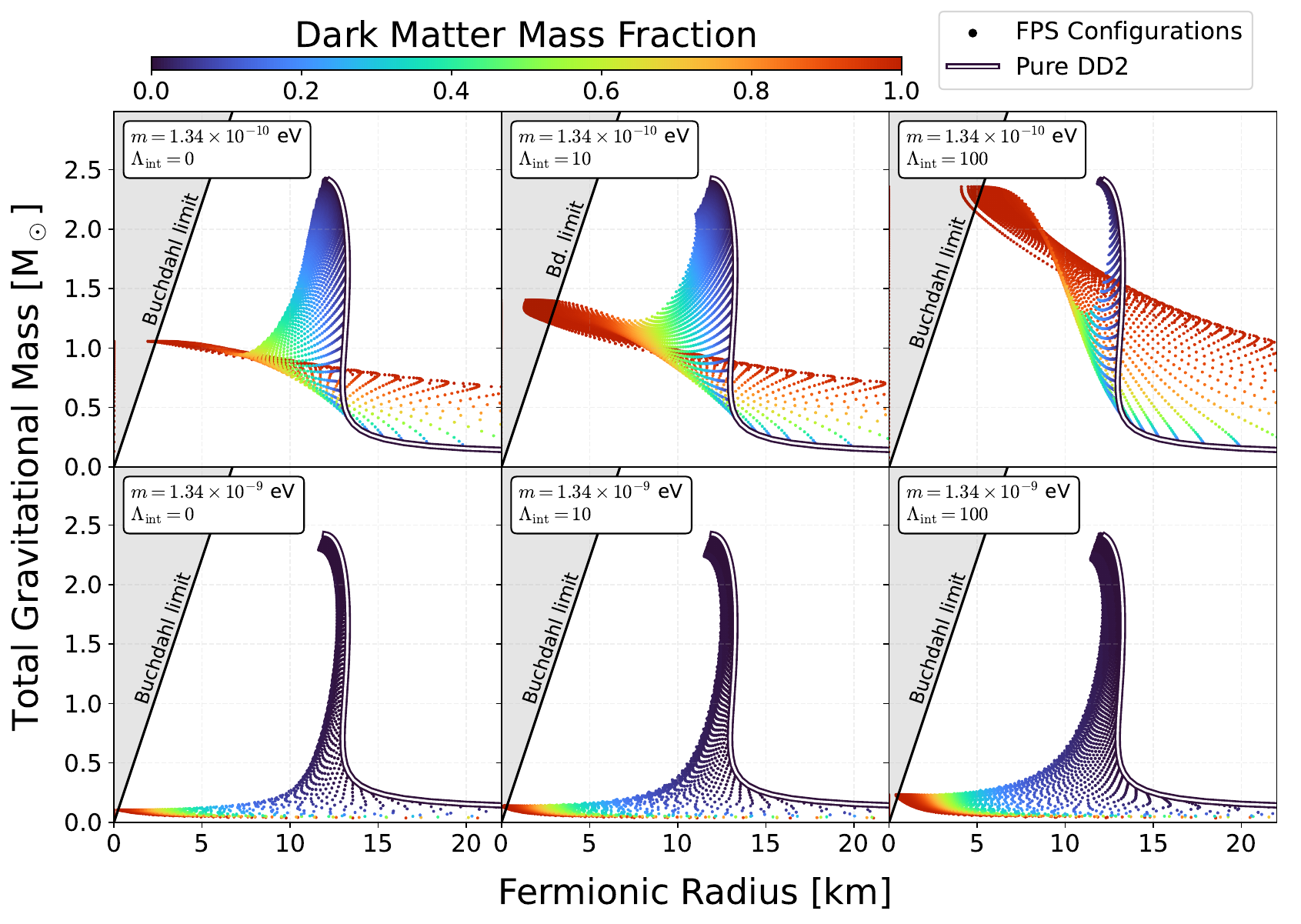}
    \includegraphics[width = 0.99\textwidth]{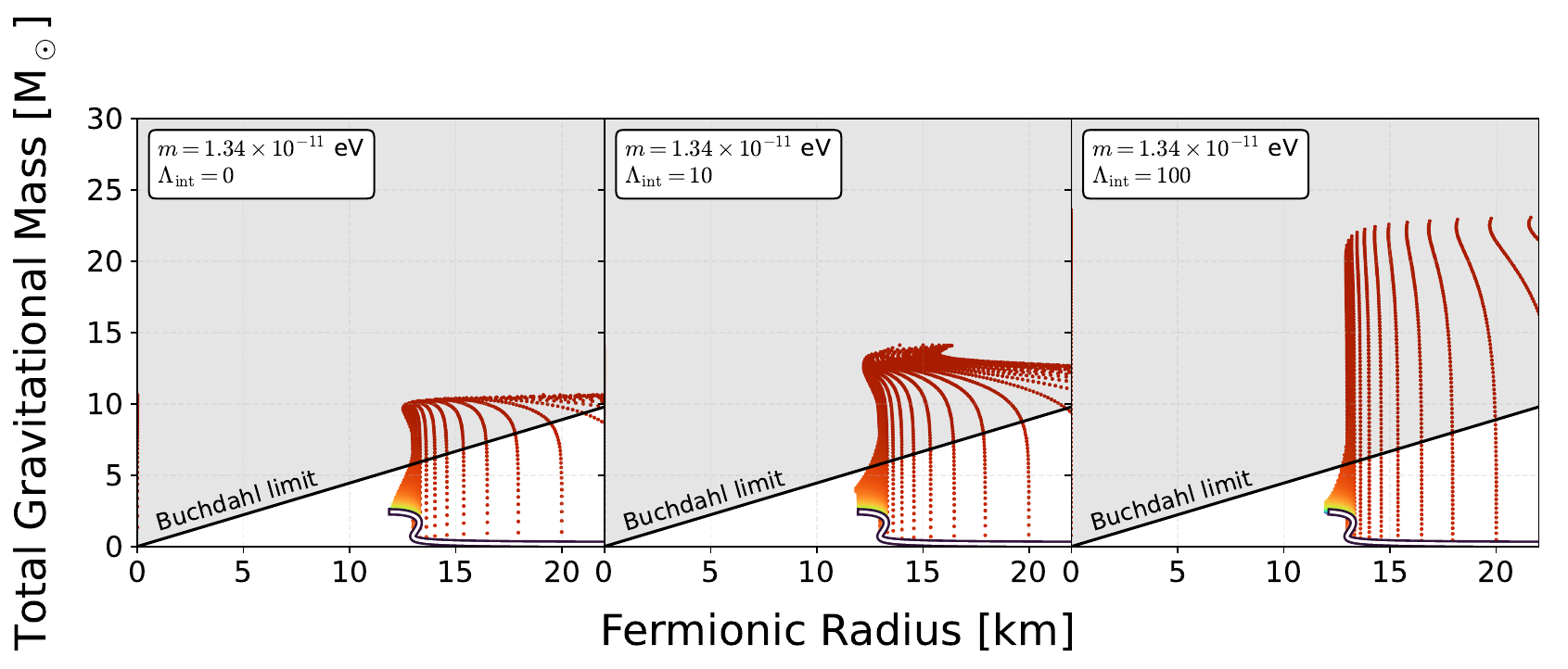}
    \caption{Relation between total gravitational mass $M_\mathrm{tot}$ and fermionic radius $R_\mathrm{f}$ for different FPS. The rows correspond to bosonic masses of $m = \{1, 10, 0.1\}\times 1.34 \e{-10}\,eV$, columns correspond to self-interactions of $\Lambda_\mathrm{int}= \{0, 10, 100\}$ respectively. We use the DD2 EOS for the fermionic part. Notice the different scale of the bottom plots. The grey region marks the Buchdahl limit, where no stable NS can exist. Observing only $R_\mathrm{f}$ of these systems would appear to violate the Buchdahl limit, even though the FPS as a whole does not.}
    \label{fig:results:fermion-proca-stars:MR-grid}
\end{figure}

\begin{figure}
    \centering
    \includegraphics[width = 0.95\textwidth]{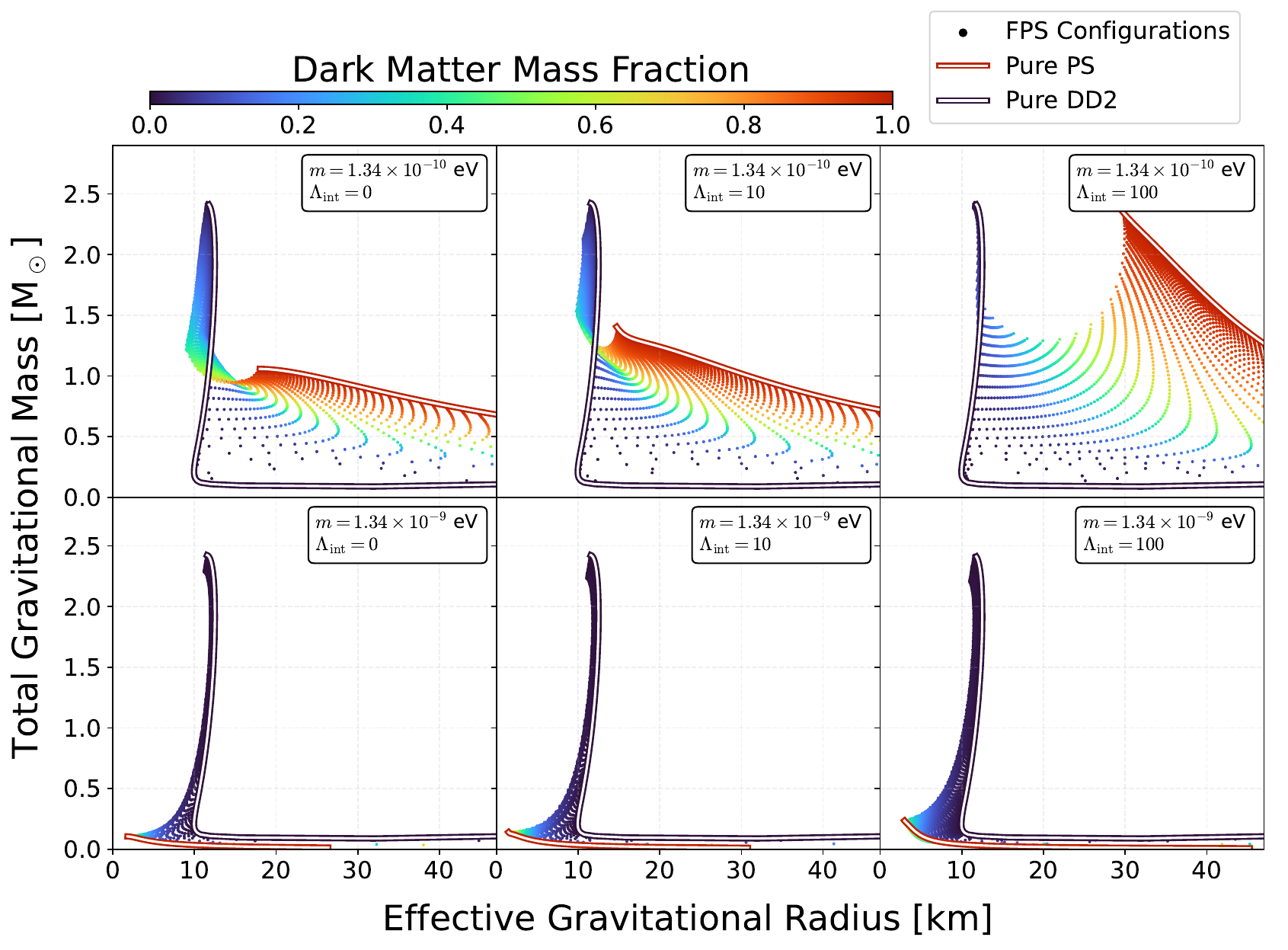}
    \includegraphics[width = 0.95\textwidth]{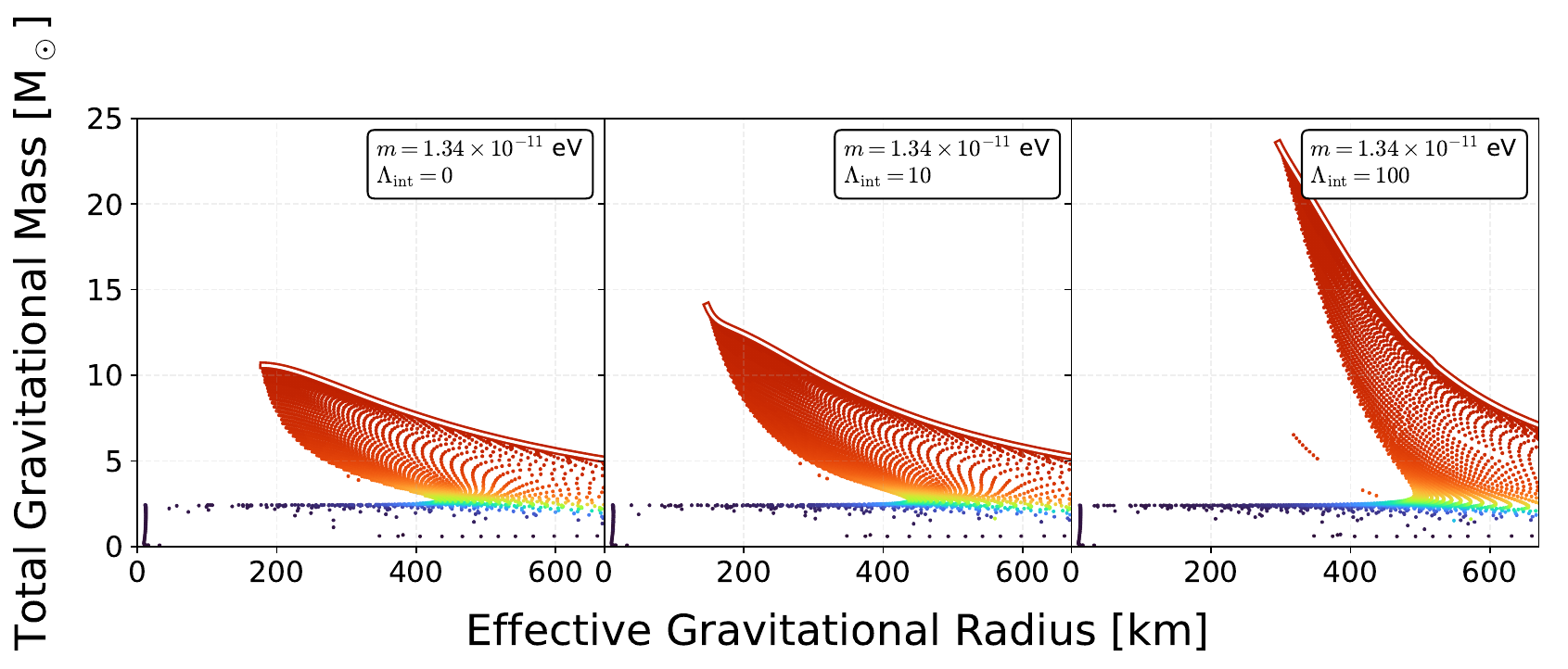}
    \caption{Relation between total gravitational mass $M_\mathrm{tot}$ and effective gravitational radius $R_G$ for different FPS. $R_G$ is the radius where $99\%$ of the restmass is contained. The rows correspond to bosonic masses of $m = \{1, 10, 0.1\}\times 1.34 \e{-10}\,eV$, columns correspond to self-interactions of $\Lambda_\mathrm{int}= \{0, 10, 100\}$ respectively. We use the DD2 EOS for the fermionic part. Notice the different scales of the bottom plots. For pure NS, because the crust has comparatively low density, $R_G$ is significantly smaller than $R_\mathrm{f}$ (compare to \autoref{fig:results:fermion-proca-stars:MR-grid}). $R_G$ tends to be higher as compared to FBS (see \autoref{fig:results:fermion-boson-stars:MRg-grid}).}
    \label{fig:results:fermion-proca-stars:MRg-grid}
\end{figure}

We compute various FPS with different values of the vector boson masses $m = \{1, 10, 0.1\}\times 1.34 \e{-10}\,eV$ and self-interaction strengths $\Lambda_\mathrm{int}= \{0, 10, 100\}$. In \autoref{fig:results:fermion-proca-stars:MR-grid}, we show the mass and fermionic radii of all stable FPS configurations in an MR diagram. In \autoref{fig:results:fermion-proca-stars:MRg-grid}, we show the mass plotted against the effective gravitational radius. The stable solutions have been obtained using the stability curve, in the same way as was done previously for FBS. We hereafter discuss some general trends and compare the results to the one obtained for FBS. We find that many of the general conclusions regarding FBS can also be applied to FPS. FPS with small DM-fractions are dominated by the fermionic component, leading to only small changes in the fermionic radius. In the case of DM dominated FPS, the solutions behave similar to pure Proca stars. This leads to higher masses as compared to FBS, where the total gravitational mass of pure boson stars will be roughly half that of a Proca star with the same boson mass, as can be seen well for the cases where $m = \{1, 0.1\}\times 1.34 \e{-10}\,eV$. FPS can thus reach higher total gravitational masses as compared to FBS with the same DM mass and self-interaction strength. For $m = 1.34 \e{-9}\,eV$, the bosonic component is concentrated inside of the fermionic one and forms a DM core. Even small amounts of DM can have a significant impact on the fermionic radius, since the whole vector field is concentrated entirely inside of the NS component. More massive DM particles can thus have larger effects on the fermionic radius compared to low-mass DM at similar DM-fractions. This is due to the cloud-like structure of low-mass DM. For small DM  masses, the majority of the DM will be concentrated outside of the NS part -- due to its larger correlation length -- and will thus have smaller effects on the fermionic radius. The general behaviour of FBS hence also holds for FPS. The smaller the mass and the larger the self-interaction strength, the more likely the formation of a DM cloud is. The opposite is true for DM core solutions. Fermion Proca stars however tend to produce configurations with larger total masses and their halos also extend to larger radii, as can be seen from the gravitational radius in \autoref{fig:results:fermion-proca-stars:MRg-grid}. \\
In general, the gravitational radius of FPS is larger in size as compared to scalar FBS (note the different scale of the x-axis of the cases where $m = \{1, 10\}\times 1.34 \e{-10}\,eV$ in \autoref{fig:results:fermion-boson-stars:MRg-grid}). The larger gravitational radius suggests that FPS have larger tidal deformabilities, compared to their scalar field counterparts (FBS) with equal $m$ and $\Lambda_\mathrm{int}$, since objects with larger radii are generally easier to tidally disrupt. This could favour higher vector boson masses compared to the corresponding scalar boson mass in the case of FBS. Without an explicit quantitative analysis of the tidal deformability of FPS however, it is not possible to definitively verify this hypothesis. When considering the gravitational radius of FPS with small boson masses of $m = 1.34 \e{-11}\,eV$ (see bottom row of \autoref{fig:results:fermion-proca-stars:MRg-grid}), the transition between DM-dominated and NS-dominated configurations appears more abrupt than in the FBS case (compare to \autoref{fig:results:fermion-boson-stars:MRg-grid}). For example, when starting with a system with a DM-fraction of roughly $0\%$ or $80\%$, increasing the DM-fraction by small amounts can massively impact the total mass and gravitational radius of the combined system. Finally, note the outlier points in \autoref{fig:results:fermion-proca-stars:MRg-grid} for $m = 1.34 \e{-11}\,eV$ and $\Lambda_\mathrm{int}=100$ at roughly $R_G=350\,km$. These are likely to be numerical artefacts and should thus not be regarded as physical. This is to be expected since for small DM masses and large self-interactions, the numerical solution gets increasingly difficult. This problem could be avoided by using smaller step-sizes and higher numerical precision, but would also warrant longer run-times of the code. \\

\begin{figure}[h]
\centering
    \includegraphics[width=0.49\textwidth]{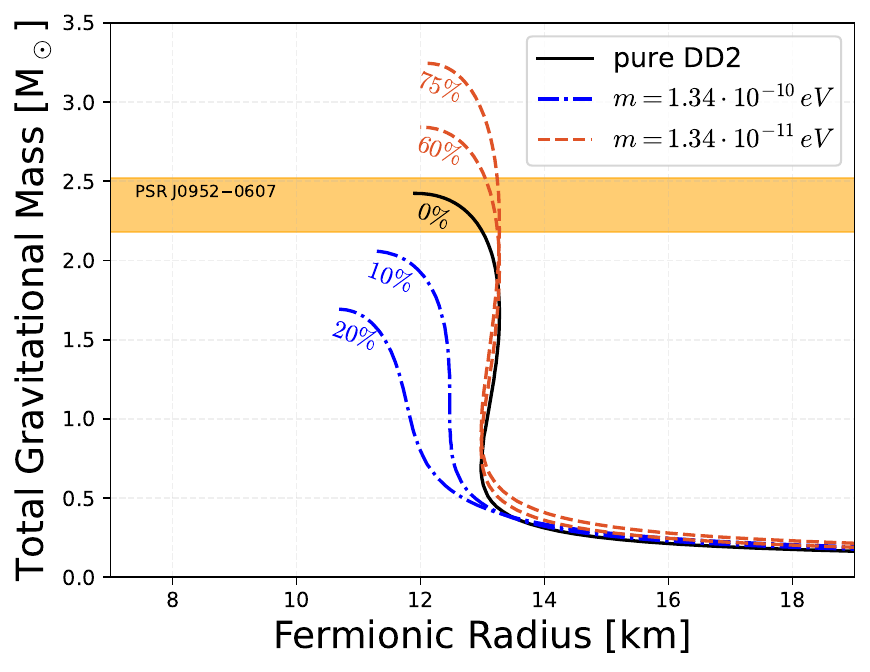}
    \includegraphics[width=0.49\textwidth]{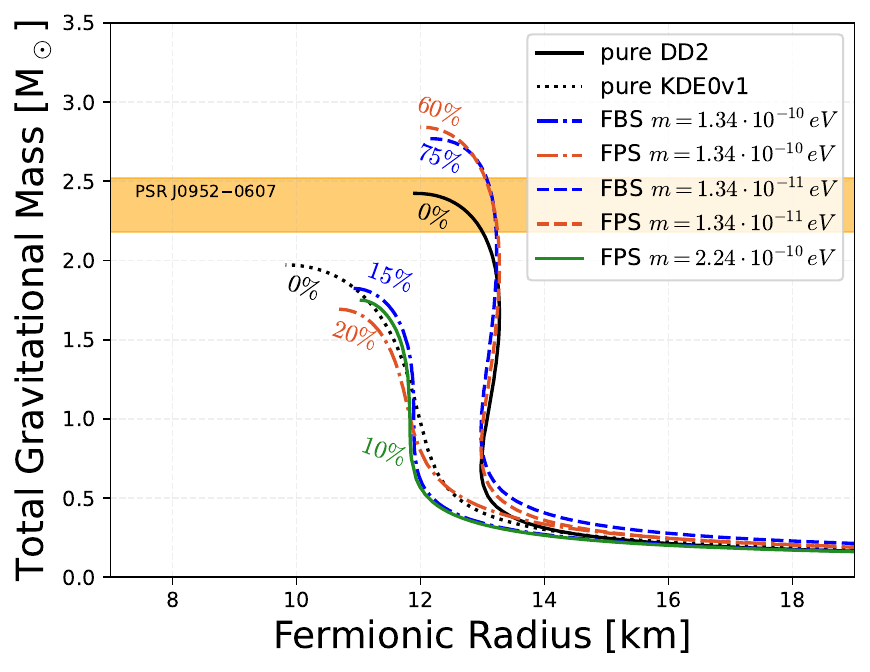}
    \caption{\textbf{Left panel:} Mass-radius relations of FPS with the DDS EOS \cite{Hempel:2009mc} for vector boson masses $m = \{1, 0.1\}\times 1.34 \e{-10}\,eV$, no self-interactions and constant DM-fractions $N_\mathrm{b}/(N_\mathrm{b} + N_\mathrm{f})$. This figure should be compared to \autoref{fig:results:fermion-boson-stars:comparison-to-measurements} (left panel) as the same masses and DM-fractions were chosen. In both figures, the orange band marks the observational constraint of J0952-0607 \cite{Romani:2022jhd} and the percentage numbers denote the respective DM-fractions.
    \textbf{Right panel:} Mass-radius relations of FPS (orange and green lines) and FBS (blue lines) with the DD2 EOS for different boson masses, no self-interactions and different DM-fractions. The black lines correspond to the pure NS with the DD2 EOS and KDE0v1 EOS \cite{Schneider:2017tfi} respectively. FPS and FBS solutions with different masses and DM-fractions can both be degenerate with each other or also degenerate with pure NS with a different EOS.}
    \label{fig:results:fermion-proca-stars:MR-curves-Nbfrac}
\end{figure}

\begin{figure}[h]
\centering
    \includegraphics[width=0.495\textwidth]{FBS-MR_scatter.pdf}
    \includegraphics[width=0.495\textwidth]{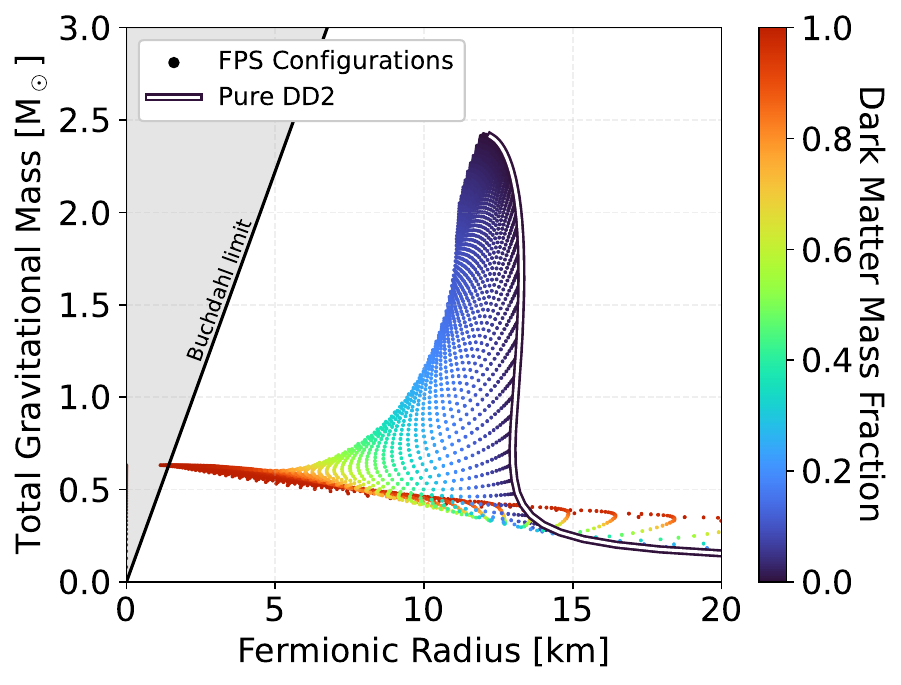}
    \caption{\textbf{Left panel:} Mass-radius diagram displaying the fermionic radius vs the total gravitational mass for stable FBS configurations with scalar boson mass of $m = 1.34 \e{-10}\,eV$ and no self-interaction. Each point corresponds to a single configuration and is colour-coded according to the DM-fraction $N_\mathrm{b}/(N_\mathrm{b} + N_\mathrm{f})$. The solid black-white line shows the mass-radius curve for pure fermionic matter.
    \textbf{Right panel:} Mass-radius diagram displaying the fermionic radius vs the total gravitational mass for stable FPS configurations with vector boson mass of $m = 2.24 \e{-10}\,eV$ and no self-interaction. Each point corresponds to a single configuration and is colour-coded according to the DM-fraction $N_\mathrm{b}/(N_\mathrm{b} + N_\mathrm{f})$. The solid black-white line shows the mass-radius curve for pure fermionic matter. The vector boson mass has been chosen so that in the limit of pure BS/PS, the same total gravitational mass is produced. Both diagrams show only marginal differences.}
    \label{fig:results:fermion-proca-stars:MR-curves-FBS-FPS-comparison}
\end{figure}

Next, we show mass-radius relations of fermion Proca stars with a fixed DM-fraction. In the left panel of \autoref{fig:results:fermion-proca-stars:MR-curves-Nbfrac}, we show different FPS with the DDS EOS \cite{Hempel:2009mc} for vector boson masses $m = \{1, 0.1\}\times 1.34 \e{-10}\,eV$, no self-interactions and constant DM-fractions $N_\mathrm{b}/(N_\mathrm{b} + N_\mathrm{f})$. This figure should be explicitly compared to \autoref{fig:results:fermion-boson-stars:comparison-to-measurements} (left panel) as the same masses and DM-fractions were chosen. The MR curve of a pure NS with the DD2 EOS is shown as a reference. Just like FBS, FPS can either increase or decrease the maximum total gravitational mass of the combined system, depending on the boson mass. Overall, FPS tend to produce configurations with larger gravitational masses compared to FBS with equal parameters (mass, self-interaction and DM-fraction). This is not surprising, when considering the scaling relations of pure BS and PS respectively. The presence of light bosonic DM can help to increase the total gravitational mass of a NS, making EOS which do not fulfil the observational constraints for the maximum NS mass possible again. Vector dark matter has a larger effect on the gravitational mass than scalar DM and thus, smaller amounts of vector DM are needed to produce an equal increase in the total gravitational mass. In the right panel of \autoref{fig:results:fermion-proca-stars:MR-curves-Nbfrac}, we now show different FPS (orange and green lines) and FBS (blue lines) with the DDS EOS \cite{Hempel:2009mc} for different boson masses, no self-interactions and constant DM-fractions $N_\mathrm{b}/(N_\mathrm{b} + N_\mathrm{f})$. The parameters were chosen in a way to illustrate the degeneracies that can arise from different DM models or EOS for the NS component. For example, FPS and FBS with boson masses of $m = 1.34 \e{-11}\,eV$ (dashed lines) produce virtually indistinguishable mass-radius relations, when the FPS and the FBS have a DM-fraction of $60\%$ and $75\%$ respectively. A similar behaviour can be seen for the cases where the boson mass is $m = 1.34 \e{-10}\,eV$ (dot-dashed lines). Here, FBS with $15\%$ DM-fraction produce very similar MR curves to FPS with $20\%$ DM-fraction. In addition, the resulting MR curves are comparable to the curve corresponding to a pure NS with the KDE0v1 EOS \cite{Schneider:2017tfi} and also to the curve corresponding to an FPS with  $10\%$ DM-fraction and a different vector boson mass of $m = 2.24 \e{-10}\,eV$ (green line). In conclusion, FPS can produce degenerate results in the MR plane with both FBS and pure NS, given that different DM-fractions and EOS are allowed. Additional observables, such as the tidal deformability, are needed to break the degeneracy. However it seems difficult to prevent degenerate solutions from existing in general, since FPS themselves can be degenerate with other FPS-solutions that have different boson masses and DM-fractions. \\

We further explore the notion of degeneracy between FPS and FBS solutions. In \autoref{fig:results:fermion-proca-stars:MR-curves-FBS-FPS-comparison}, we show the stable FBS/FPS solutions in an MR diagram of total mass vs fermionic radius. We made use of the scaling relations of the maximum mass for pure boson stars ($M_\mathrm{max} \approx 0.633 M^2_p / m$) and pure Proca stars ($M_\mathrm{max} \approx  1.058 M^2_p / m$), to match the masses in a way that both FPS and FBS will have the same mass in the pure BS/PS limit. To guarantee matching solutions in this limit, we chose a scalar boson mass of $m = 1.34 \e{-10}\,eV$ and we chose a mass of $1.058 \div 0.633 \approx 1.671$ times the mass of the scalar boson -- i.e. $m = 2.24 \e{-10}\,eV$ -- for the vector boson. We find a high degree of similarity between the MR region of FBS and FPS with the scaled masses, making both solutions almost indistinguishable. The small differences present between the left and right panel of \autoref{fig:results:fermion-proca-stars:MR-curves-FBS-FPS-comparison} can be attributed to a slightly different grid-spacing used for the initial conditions $\rho_c$, $\varphi_c$ (and $\rho_c$, $E_0$) -- see the bottom parts of the MR regions at small total gravitational masses $M_\mathrm{tot} <0.5\,M_\odot$ and also at radii $R_\mathrm{f} > 15\,km$. Apart from that, the colour shading reveals a different distribution of DM-fractions for a given $M_\mathrm{tot}$ and $R_\mathrm{f}$, even though the difference is slight. \\
We expect a similar behaviour to hold when considering different scalar and vector boson masses (with zero-self-interaction), given that they differ by the same factor of $\approx 1.671$. This adds further confidence to the observation that FBS and FPS might be difficult to distinguish, since a given solution might be another system but with different boson mass (or DM-fraction). Because similar scaling relations also exist for BS and PS in the large $\Lambda_\mathrm{int}$ limit, a similar procedure might be possible when also matching the self-interaction strength appropriately. An independent measurement of the DM particle mass will make it possible to break this degeneracy to a certain degree, but it would also be necessary to constrain the self-interaction strength and the DM-fraction through other means, such as correlations of the DM abundance in the galactic disk (see \cite{Giangrandi:2022wht,Sagun:2022ezx}) or using the bound on the maximal vector field amplitude.

\subsubsection{Study of Higher Modes and Different EOS} \label{subsec:results:study-of-higher-modes-and-different-eos}

\begin{figure}[h]
\centering
    \includegraphics[width=0.495\textwidth]{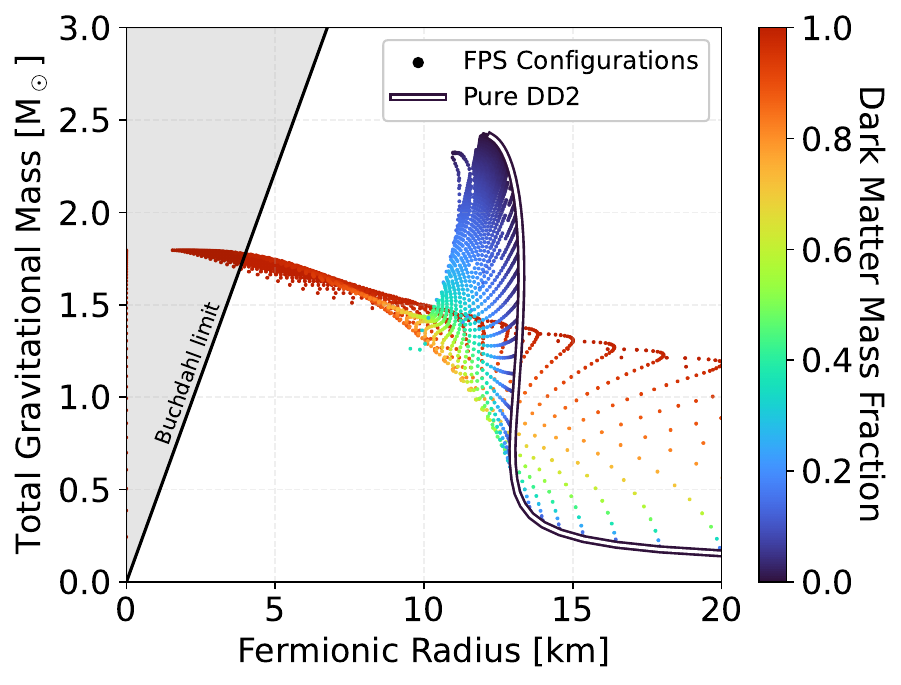}
    \includegraphics[width=0.495\textwidth]{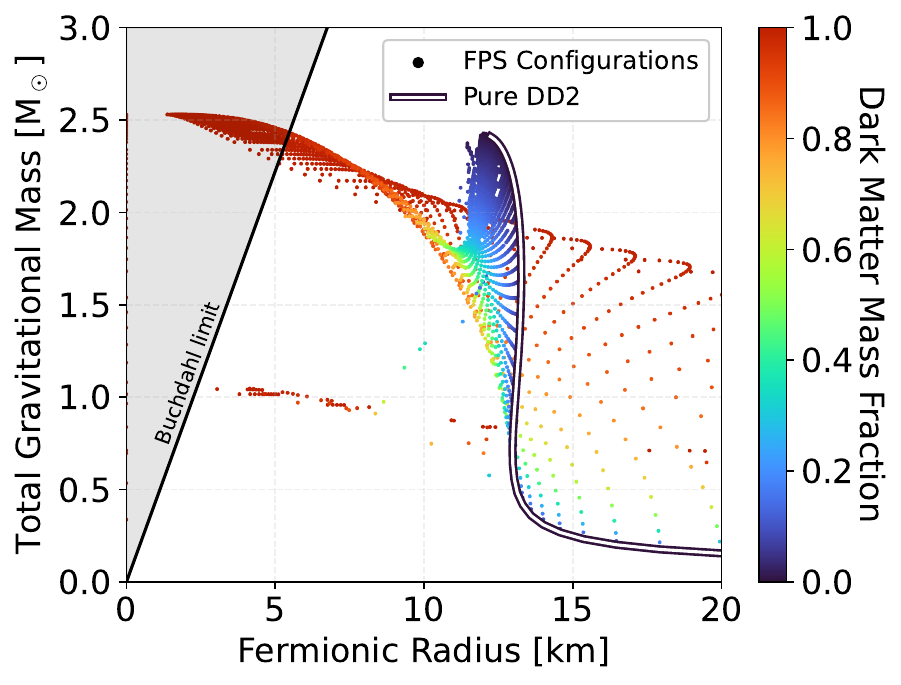}
    \caption{\textbf{Left panel:} Mass-radius diagram displaying the fermionic radius vs the total gravitational mass for stable FPS configurations in the first mode with vector boson mass of $m = 1.34 \e{-10}\,eV$ and no self-interaction. Each point corresponds to a single configuration and is colour-coded according to the DM-fraction $N_\mathrm{b}/(N_\mathrm{b} + N_\mathrm{f})$. The solid black-white line shows the mass-radius curve for pure fermionic matter.
    \textbf{Right panel:} Mass-radius diagram displaying the fermionic radius vs the total gravitational mass for stable FPS configurations in the second mode with vector boson mass of $m = 2.24 \e{-10}\,eV$ and no self-interaction. Each point corresponds to a single configuration and is colour-coded according to the DM-fraction $N_\mathrm{b}/(N_\mathrm{b} + N_\mathrm{f})$. The solid black-white line shows the mass-radius curve for pure fermionic matter.}
    \label{fig:results:fermion-proca-stars:MR-curves-mode1-mode2-comparison}
\end{figure}

We close our analysis of fermion Proca stars by broadening our analysis to FPS with different EOS for the fermionic component and to FPS, where the bosonic component exists in a higher mode. Higher modes are usually assumed to be unstable, but as numerical simulations of scalar boson stars have shown \cite{DiGiovanni:2021vlu,Bernal:2009zy}, higher modes might be dynamically stable, when gravitationally interacting in a multi-component system. We therefore start by considering FPS in the first and second mode in \autoref{fig:results:fermion-proca-stars:MR-curves-mode1-mode2-comparison}. \\

In the left panel of figure \autoref{fig:results:fermion-proca-stars:MR-curves-mode1-mode2-comparison}, we show the total gravitational mass and the fermionic radius of stable FPS configurations, where the bosonic component is in the first mode (as opposed to the ground mode, which is the zeroth mode). The vector boson mass is $m = 1.34 \e{-10}\,eV$  and the self-interaction is set to zero. We first note the fact, that stable solutions under linear radial perturbations, according to the stability criterion \eqref{eq:fermion-boson-stars:scalar-fermion-boson-stars:FBS-stability-criterion}, exist at all. This is a non-trivial statement as higher modes of PS (and also of boson stars) are usually believed to be unstable. Note that our stability analysis does not consider the dynamical stability of the higher modes, so they might still be unstable in non-static scenarios. It is however possible that the higher modes of the bosonic part might be stabilised through the gravitational interaction with the fermionic part of the FPS. Generally, the FPS in the first mode exhibits higher gravitational masses in the configurations dominated by the bosonic component, compared to FPS in the zeroth mode (compare to \autoref{fig:results:fermion-proca-stars:stability-and-MR-curve-lamda0}). The numerical value of the frequency $\omega$ in the higher mode is also larger than the frequency in lower modes. This behaviour is consistent with earlier works, which studied pure Proca stars analytically \cite{Minamitsuji:2018kof} and numerically \cite{Sanchis-Gual:2017bhw}, and also observed that higher frequencies lead to larger total gravitational mass. The left panel of figure \autoref{fig:results:fermion-proca-stars:MR-curves-mode1-mode2-comparison} shows a number of outlier points at around $11\,km$ and $2.3\,M_\odot$. These are likely to be numerical artefacts due to the increased difficulty of finding accurate numerical solutions for higher modes. The right panel of \autoref{fig:results:fermion-proca-stars:MR-curves-mode1-mode2-comparison} shows stable FPS configurations in the second mode. The vector boson mass is $m = 1.34 \e{-10}\,eV$  and the self-interaction is set to zero. Here also, the existence of stable solutions is to be acknowledged. In the limit of high DM-fractions, the FPS converge to the solution of pure PS and reach total gravitational masses of $2.5$ times that of Proca stars in the zeroth mode (compare to \autoref{fig:results:fermion-proca-stars:stability-and-MR-curve-lamda0}). In comparison to the case in the first mode, the quality of the overall solution can be seen to deteriorate further. We believe the outlier points at roughly $<13\,km$ and $1\,M_\odot$ to be non-physical numerical artefacts. Solutions of FPS in even higher modes should be considered with great care, given that the difficulty of obtaining accurate numerical solutions increases further. The quality of the solution is however sufficient to gain a qualitative understanding of FPS in higher modes. In conclusion, higher modes are stable under linear radial perturbations and increase the total gravitational mass of FPS by substantial amounts. \\

\begin{figure}[h]
\centering
    \includegraphics[width=0.499\textwidth]{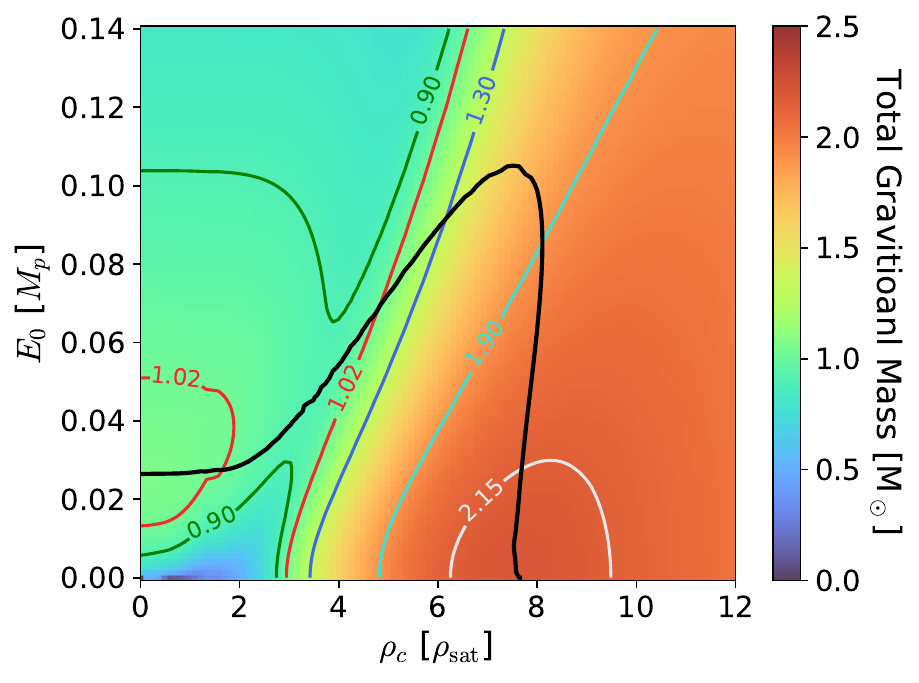}
    \includegraphics[width=0.49\textwidth]{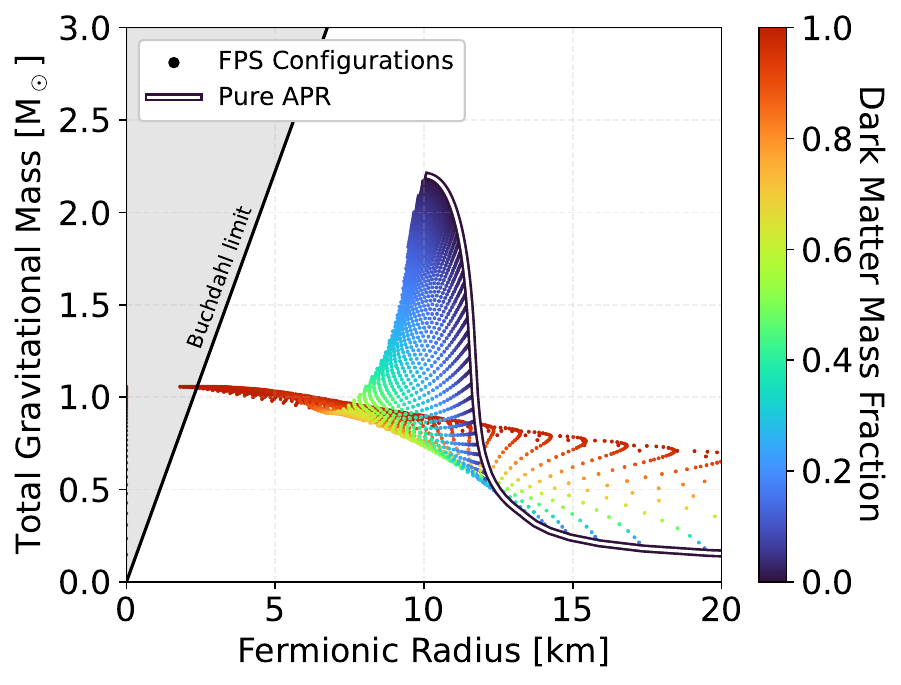}
    \caption{\textbf{Left panel:} Total gravitational mass of different FPS as a function of the restmass density $\rho_c$ and central vector field amplitude $E_0$. The black line corresponds to the stability curve, which separates stable solutions (in the lower left region) from unstable solutions (everywhere else). The qualitative behaviour of the stability curve of FPS is similar to the case with FBS (see \autoref{fig:results:fermion-boson-stars:stability-and-MR-curve-example})
    \textbf{Right panel:} Mass-radius diagram displaying the fermionic radius vs the total gravitational mass for FPS configurations that are within the stability region displayed in the left panel. Each point corresponds to a single configuration and is colour-coded according to the DM-fraction $N_\mathrm{b}/(N_\mathrm{b} + N_\mathrm{f})$. The solid black-white line shows the mass-radius curve for pure fermionic matter. In both cases, a vector field with a mass of $m=1.34 \e{-10}\,eV$ and no self-interactions was considered in addition to the APR EOS \cite{Schneider:2019vdm} for the fermionic part.}
    \label{fig:results:fermion-proca-stars:stability-and-MR-curve-APR}
\end{figure}

\begin{figure}[h]
\centering
    \includegraphics[width=0.507\textwidth]{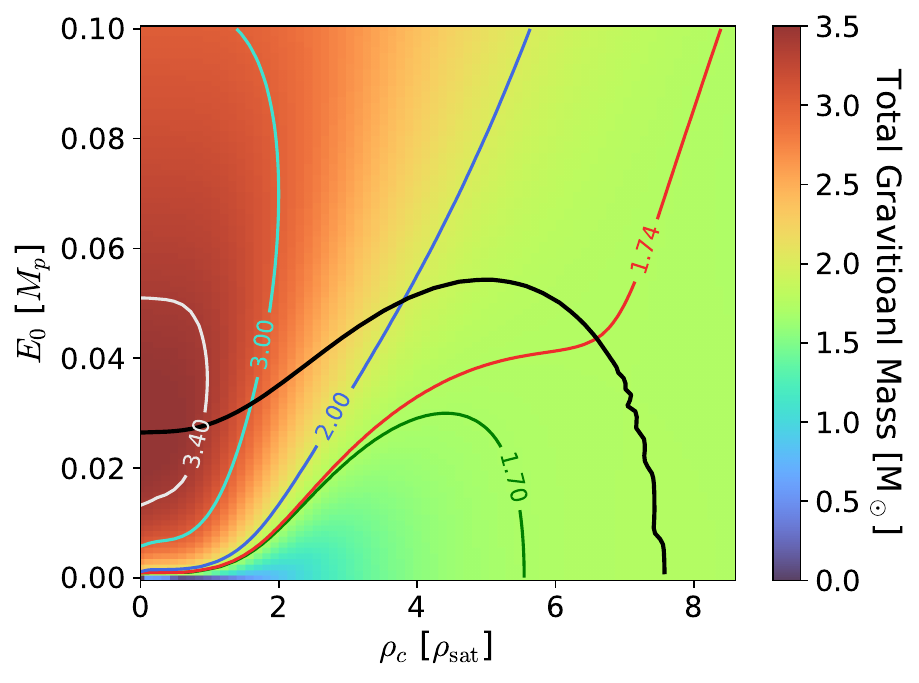}
    \includegraphics[width=0.48\textwidth]{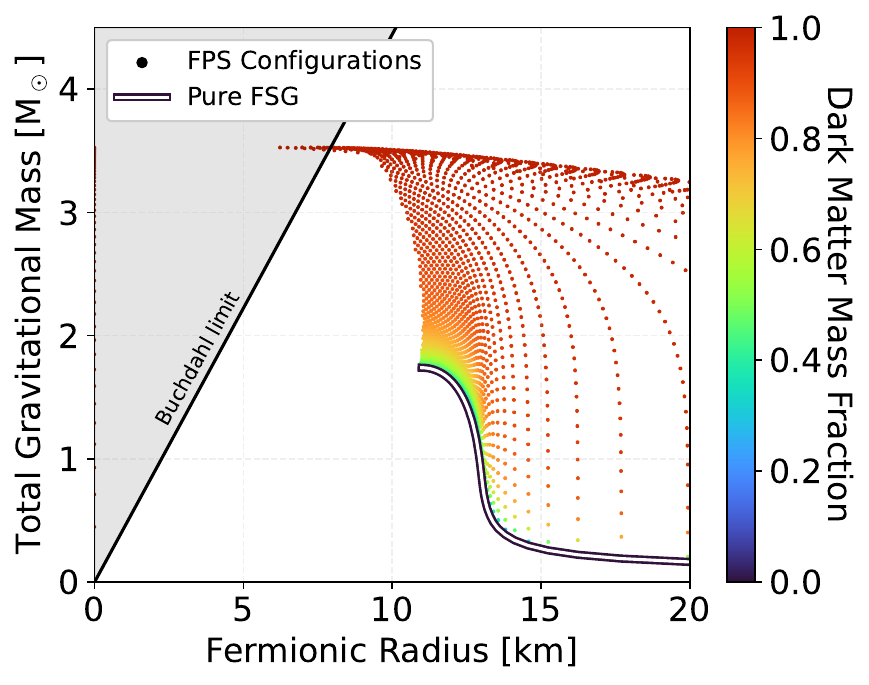}
    \caption{\textbf{Left panel:} Total gravitational mass of different FPS as a function of the restmass density $\rho_c$ and central vector field amplitude $E_0$. The black line corresponds to the stability curve, which separates stable solutions (in the lower left region) from unstable solutions (everywhere else). The qualitative behaviour of the stability curve of FPS is similar to the case with FBS (see \autoref{fig:results:fermion-boson-stars:stability-and-MR-curve-example})
    \textbf{Right panel:} Mass-radius diagram displaying the fermionic radius vs the total gravitational mass for FPS configurations that are within the stability region displayed in the left panel. Each point corresponds to a single configuration and is colour-coded according to the DM-fraction $N_\mathrm{b}/(N_\mathrm{b} + N_\mathrm{f})$. The solid black-white line shows the mass-radius curve for pure fermionic matter. In both cases, a vector field with a mass of $m=3.01 \e{-11}\,eV$ and no self-interactions was considered in addition to the FSG EOS \cite{Hempel:2009mc} for the fermionic part.}
    \label{fig:results:fermion-proca-stars:stability-and-MR-curve-FSG}
\end{figure}

We investigate the effect that different EOS have on FPS. In the first case in \autoref{fig:results:fermion-proca-stars:stability-and-MR-curve-APR}, we use the APR EOS \cite{Schneider:2019vdm} for the fermionic part and we chose a vector boson mass of $m=1.34 \e{-10}\,eV$ with no self-interaction for the bosonic part. When considering the left panel, we notice that the shape of the stability curve (black curve) is affected by the choice of the EOS. On the $\rho_c$-axis, it converges to a value of around $7.5\rho_\mathrm{sat}$, which is higher than the corresponding value of $\rho_c$ when the DD2 EOS is used (compare to \autoref{fig:results:fermion-proca-stars:stability-and-MR-curve-lamda0}). This is due to the fact that the APR EOS is softer than the DD2 EOS, meaning that the nuclear matter is easier to compress and higher central densities can be supported by the EOS. The easier compressibility also shows itself through smaller NS radii (see the right panel). In the limit of pure PS, the stability curve converges to the same value as it does when the DD2 EOS is used (compare to \autoref{fig:results:fermion-proca-stars:stability-and-MR-curve-lamda0}). The MR region shows similar qualitative behaviour as in the DD2 case. The high DM-fraction limit in particular shows a convergence to the solution to pure Proca stars. The APR EOS also evidently allows higher central amplitudes of the vector field $E_0$, compared to the DD2 EOS with equal boson mass and self-interaction strength. \autoref{fig:results:fermion-proca-stars:stability-and-MR-curve-FSG} shows different FPS configurations but this time, the FSG EOS \cite{Hempel:2009mc} was used for the fermionic part. For the bosonic part we used a boson mass of $m=3.01 \e{-11}\,eV$ and no self-interaction. Again, the FSG EOS is a soft EOS and thus reaches higher central densities $\rho_c$ for pure NS. For the case of pure NS, the FSG EOS is excluded as a possible EOS by current observational constraints (see \autoref{fig:results:fermion-boson-stars:comparison-to-measurements}), as it cannot reach the bound of $M=2.35^{+0.17}_{-0.17}\,M_\odot$ \cite{Romani:2022jhd}. However, adding dark matter to the pure NS can significantly increase the maximum gravitational mass of the combined system so that the FSG EOS is able to reach the observational bound on the maximum NS mass in the presence of DM. In fact, the MR curve of the pure DDS EOS is entirely contained within the stability region of the FPS with the FSG EOS. This again raises the point that some FPS solutions are degenerate with some NS solutions (see \autoref{fig:results:fermion-proca-stars:MR-curves-FBS-FPS-comparison}), when allowing for different DM-fraction and DM masses. To ascertain whether and which types of mixed DM-NS systems might exist, it will be crucial to perform sophisticated parameter searches of the system and obtain more measurements to constrain the DM and NS properties in future studies.

%% file: conclusions.tex
\markright{Conclusions and Outlook}
\section{Conclusions and Outlook} \label{sec:conclusions:conclusions-and-outlook}

In this work, we studied the impact that bosonic dark matter has on the radius and tidal deformability of neutron stars. The dark matter was modelled as either a massive, self-interacting complex scalar field or as a massive, self-interacting complex vector field. The DM was further assumed to only interact gravitationally with the fermionic neutron star matter. We derived the equations of motion describing static spherically symmetric fermion boson stars (with a scalar field) and fermion Proca stars (with a vector field) and computed their properties numerically. We derived the equations describing linear perturbations of fermion boson stars induced by an external gravitational tidal field and computed their tidal deformability numerically. For fermion Proca stars, we presented their equations of motion for the first time, found a scaling relation between the frequency, field and metric components and we derived an analytical upper bound on the vector field amplitude. \\

For FBS, we found that the presence of the scalar field can lead to DM core and cloud solutions. For the scalar field masses $m$ and the self-interactions $\lambda$, which result in core-like configurations, we found an increased compactness and smaller tidal deformability $\Lambda_\mathrm{int}$ compared to pure NS. This was found to be the case for DM masses of $m \gtrsim 1.34 \e{-10}\,eV$. However, large self-interactions $\Lambda_\mathrm{int}$ can also lead to cloud-like solutions in certain cases. For some FBS, observing only the fermionic radius and the total gravitational mass would appear to violate the Buchdahl limit. \\
The comparison of our results to available observations of NS masses and radii shows that their uncertainties are currently too large (apart from some pulsar mass measurements) to derive clear quantitative constraints on the DM properties (mass and self-interaction strength). The degeneracy between different FBS models and different EOS further complicates this. For some small boson masses $m \lesssim 1.34 \e{-11}\,eV$, the presence of DM can significantly increase the NS gravitational mass while leaving the fermionic radius approximately constant. This makes previously excluded EOS possible again, if they are used in a combined NS-DM system. Additionally, the unusually light neutron star HESS J1731\ensuremath{-}347 is difficult to explain using current EOS models of pure NS matter, but might be explained by a NS which also contains a significant fraction of DM. The observational constraint on the tidal deformability from the event GW170817 ($\Lambda_\mathrm{tidal} \leq 800$ at $M_\mathrm{tot} \approx 1.4 \,\mathrm{M}_\odot$) is currently not strong enough to significantly constrain the DM properties, even though the results seem to favour DM core-like configurations. With the ongoing joint run of the GW detectors LIGO, Virgo and KAGRA, we expect to obtain more observational data, which will enable us to derive quantitative constraints. We leave the quantitative analysis of constraints for future works. \\
In addition to solving the FBS self-consistently by integrating the scalar field, we also considered an effective EOS for the scalar field to reduce the complexity of the FBS system to a two-fluid model. The two-fluid model was recently used by \cite{Leung:2022wcf} to compute the tidal deformability of FBS. In this work, we compared the results obtained using the two-fluid system and by solving the full system self-consistently. We found that for scalar boson masses of $m = 6.7 \e{-11}\,eV$ and self-interaction strengths of $\Lambda_\mathrm{int} > 300 - 400$ with $\Lambda_\mathrm{int} = \lambda/8 \pi m^2$, the usage of the effective EOS and the two-fluid system is typically justified. We do not expect this conclusion to be dependent on $m$ but rather only on $\Lambda_\mathrm{int}$. The agreement between the full system and the two-fluid system increases with larger $\Lambda_\mathrm{int}$. Still, even for large $\Lambda_\mathrm{int}$, we find that a significant number of configurations show large relative errors of $> \mathcal{O}(10^2)$. \\
In the future, it would be interesting to study the impact that the scalar field has on the inspiral of binary compact objects. This was already studied for non-self-interacting scalar fields  \cite{Bezares:2019jcb}. It will however be necessary to extend these simulations to the self-interacting case, since the self-interaction can significantly modify the tidal properties of a NS, even at small DM-fractions, and thus modify the observed GW signal. \\

For FPS, we found an overall similar qualitative behaviour to FBS. We showed that the presence of the vector field can lead to core-like and to cloud-like solutions. Similarly to FBS, we found core-like solutions for $m \gtrsim 1.34 \e{-10}\,eV$ and small $\Lambda_\mathrm{int}$ and cloud-like solutions when $m \lesssim 1.34 \e{-11}\,eV$ and $\Lambda_\mathrm{int}$ is large. We computed radial profiles of FPS and found that the existence of a maximum possible vector field amplitude limits the effect of DM on the NS when the self-interaction $\Lambda_\mathrm{int}$ is large. The maximum amplitude implies a maximum possible amount of vector boson DM accretion and could thus be used to set bounds on the DM properties. \\
For stable FPS configurations, we found that many of the general qualitative trends that apply to FBS also apply to FPS, but vector DM leads to higher FPS masses and larger gravitational radii for equal $m$ and $\Lambda_\mathrm{int}$. This could imply a larger tidal deformability of FPS compared to FBS. Also, a measurement of the gravitational radius would favour larger vector boson masses compared to scalar boson masses. \\
For FPS configurations of constant DM-fraction, we found that the effect of vector DM on NS properties (total gravitational mass and fermionic radius) is larger compared to FBS with equal DM-fraction, mass $m$ and self-interaction strength $\Lambda_\mathrm{int}$. One therefore needs a larger amount of scalar DM to cause the same effect as vector DM. The degeneracies between different EOS and mixed systems are also present for FPS. For different boson masses and DM-fractions, FPS and FBS can both be degenerate with each other and also be degenerate with pure NS with a different EOS. We found an especially high degree of similarity between FBS solutions with no self-interaction and a boson mass of $m = 1.34 \e{-11}\,eV$ with FPS solutions where the vector boson mass is larger by a factor of $1.671$. We expect the similarity in the behaviour to hold also for different boson masses (and also for non-zero self-interactions), as long as the vector boson mass is scaled accordingly by the right factor. The prevalence of degenerate solutions highlights the importance of measuring additional observables, such as the tidal deformability, to break the degeneracies. \\
We confirmed the existence of higher modes that are stable under first-order radial perturbations. Also, we found that higher modes lead to higher total gravitational masses of the mixed FPS systems. Using FPS with different EOS for the fermionic part, we could explicitly confirm that for certain DM masses, previously excluded EOS are able to fulfil observational bounds if DM is present. Mixed systems of DM and NS matter can therefore be consistent with all current observational constraints, if suitable boson masses and self-interaction strengths are chosen. \\

Mixed systems of fermions and bosons are a fascinating field of research and can, as we have shown, be relevant to the search for dark matter within neutron stars. In the future, it will be crucial to obtain more observational data and constraints to narrow down the allowed ranges for the DM properties. For example as a next step, the tidal deformability of fermion Proca stars could be computed to gain an additional observable to constrain their properties. One could build upon the work of \cite{Herdeiro:2020kba}, where the tidal deformability of pure Proca stars was computed. More knowledge about the nuclear matter EOS is also needed to accurately constrain the DM properties and to reduce degeneracies between EOS effects and effects due to the presence of DM. In future works, measurements of the total gravitational mass, the fermionic radius and the tidal deformability can be used to systematically constrain the DM mass and self-interaction. But first, it would be useful to find an EOS-independent method of characterizing FBS/FPS to exclude EOS effects when probing DM properties. We leave this question for a future work.

%% file: Appendix.tex
%\markright{Appendix}
\section{Appendix}
\begin{appendices}

\section{List of Abbreviations} \label{sec:appendix:list-of-abbreviations}
\begin{itemize}
\item[] \textbf{ALP -- axion-like particle(s)}
\item[] \textbf{BH -- black hole(s)}
\item[] \textbf{BS -- boson star(s)}
\item[] \textbf{BSM -- (physics) beyond (the) standard model}
\item[] \textbf{CP -- charge (and) parity (symmetry/transformation)}
\item[] \textbf{DM -- dark matter}
\item[] \textbf{$\chi$EFT -- chiral effective field theory}
\item[] \textbf{EH -- Einstein-Hilbert (action)}
\item[] \textbf{EHKG -- Einstein-Hilbert-Klein-Gordon (action/equation(s))}
\item[] \textbf{EHT -- Event Horizon Telescope}
\item[] \textbf{EOS -- equation(s) of state}
\item[] \textbf{EP -- Einstein-Proca (action/equation(s))}
\item[] \textbf{FBS -- fermion boson star(s)}
\item[] \textbf{FPS -- fermion Proca star(s)}
\item[] \textbf{GR -- general relativity / general relativistic}
\item[] \textbf{GW -- gravitational wave(s)}
\item[] \textbf{KAGRA -- Kamioka Gravitational Wave Detector}
\item[] \textbf{$\Lambda$CDM -- $\Lambda$ (cosmological) cold dark matter / the standard model of cosmology}
\item[] \textbf{LIGO -- Laser Interferometer Gravitational-Wave Observatory}
\item[] \textbf{MACHOS -- massive astrophysical compact halo object(s)}
\item[] \textbf{MOND -- modified Newtonian dynamics}
\item[] \textbf{MR -- mass-radius (diagram)}
\item[] \textbf{NICER -- Neutron star Interior Composition Explorer}
\item[] \textbf{NS -- neutron star(s)}
\item[] \textbf{ODE -- ordinary differential equation(s)}
\item[] \textbf{PBH -- primordial black hole(s)}
\item[] \textbf{PN -- post-Newtonian (expansion/formalism)}
\item[] \textbf{PS -- Proca star(s)}
\item[] \textbf{QCD -- quantum chromodynamics}
\item[] \textbf{SIDM -- self-interacting dark matter}
\item[] \textbf{SM -- standard model (of particle physics)}
\item[] \textbf{SUSY -- Supersymmetry}
\item[] \textbf{TeVeS -- Tensor-Vector-Scalar (gravity/theory)}
\item[] \textbf{TOV -- Tolman-Oppenheimer-Volkoff (equations)}
\item[] \textbf{ULDM -- ultralight dark matter}
\item[] \textbf{WDM -- wave dark matter}
\item[] \textbf{WIMP -- weakly interacting massive particle(s)}

\end{itemize}
% \item[] \textbf{}

\section{Units} \label{sec:appendix:units}

In this work, we use units in which the gravitational constant, the speed of light and the solar mass are set to $G = c = M_\odot = 1$. As a direct consequence, distances are measured in units of $\approx 1.48\,km$, which corresponds to half the Schwarzschild radius of the Sun (also called the gravitational radius of the Sun). The Planck mass is $M_p = \sqrt{\hbar c / G} \approx 1.1 \times 10^{-38} M_\odot$. Since $G = c = M_\odot = 1$ it follows that $\hbar \approx 1.2 \times 10^{-76} \neq 1$. \\

Boson stars (with a scalar field) are described using the Klein-Gordon equation, which in SI units and flat spacetime reads $(\square - (mc/\hbar)^2)\varphi = 0$. The term $mc/ \hbar$ is the inverse of the reduced Compton wavelength $\lambda_c = \hbar /mc$, which sets the typical length scale for the system even in the self-gravitating case. We assume that the typical length scale of the boson is similar to the gravitational radius $GM_\odot/c^2$, which in the case of mass scales of $\sim 1\,M_\odot$ is approximately $1.48\,km$. With $m = \hbar / c \lambda_c$, this therefore leads to a mass scale of the bosonic particle of $1.336 \e{-10}\,eV$. Previous works such as e.g. \cite{DiGiovanni:2021ejn} thus specify the mass of the scalar particle in these units. A mass of $m=1$ in our numerical code \cite{Diedrichs-Becker-Jockel} then also corresponds to $1.336 \e{-10}\,eV$. This choice of the boson mass then automatically leads to boson stars with masses in the range of $\sim 1\,M_\odot$. The same reasoning can also be applied to the case where the boson is a vector boson. This is valid since all components of a vector field also fulfil the Klein-Gordon equations individually.

\section{Derivation of Fermion Proca Stars} \label{sec:appendix:derivation-of-fermion-proca-stars}

We here provide additional in-between algebra steps of the derivation of equation \eqref{eq:fermion-boson-stars:fermion-proca-stars:TOV-equations-B} in the main text. To obtain the equations of motion for the vector field component $B(r)$, one uses the $\nu=t$ component of the Proca equation \eqref{eq:fermion-boson-stars:fermion-proca-stars:proca-equations}. Given the spherical symmetric static metric ansatz \eqref{eq:fermion-boson-stars:fermion-proca-stars:metric-ansatz} and the ansatz for the vector field \eqref{eq:fermion-boson-stars:fermion-proca-stars:vector-field-harmonic-ansatz} chosen in this work, one obtains, after minor algebra, equation \eqref{eq:fermion-boson-stars:fermion-proca-stars:general-eq-FBS-B}:
\begin{align}
 (E'' - \omega B') - \left( \frac{a'}{a\,} + \frac{\alpha'}{\alpha\,} - \frac{2}{r} \right) (E' - \omega B) = V'(A_\rho \bar{A}^\rho ) \: a^2 E \: . \label{eq:appendix:derivation-of-FPS:general-eq-FBS-B}
\end{align}

This equation contains derivatives of the vector field components $E$ and $B$. The end goal is to find a first-order differential equation for the component $B(r)$. We therefore need to eliminate the second radial derivative $E''$ by taking the radial derivative of the equation of motion for $E'$ \eqref{eq:fermion-boson-stars:fermion-proca-stars:general-eq-FBS-E}:
\begin{align}
 E''  = \frac{dE'}{dr} = \frac{d}{dr} \left\{ - V'(A_\rho \bar{A}^\rho ) \frac{B \alpha^2}{\omega} + \omega B  \right\} \: . \label{eq:appendix:derivation-of-FPS:second-derivative-E}
\end{align}

We define the second derivative of the potential $V$ by using the chain rule
\begin{align}
	V''(A_\rho \bar{A}^\rho ) = \frac{d}{dr} V'(A_\rho \bar{A}^\rho ) = \frac{d V'(A_\rho \bar{A}^\rho )}{d (A_\rho \bar{A}^\rho ) } \: \frac{d (A_\rho \bar{A}^\rho )}{dr} \: .
\end{align}

The potential depends on the magnitude of the vector field. The radial derivative of the magnitude is given by

\begin{align}
	\frac{d (A_\rho \bar{A}^\rho )}{dr} = \frac{d}{dr} \left( \frac{B^2}{a^2} - \frac{E^2}{\alpha^2}  \right) =  \left( \frac{2 B B'}{a^2} - \frac{2 B^2 a'}{a^3} - \frac{2E E'}{\alpha^2} + \frac{2E^2 \alpha'}{\alpha^3} \right) \: .
\end{align}

Using these expressions, we can write the second derivative of the $E(r)$-component \eqref{eq:appendix:derivation-of-FPS:second-derivative-E} as
\begin{align}
 E'' = - V''(A_\rho \bar{A}^\rho ) \left( \frac{2 B B'}{a^2} - \frac{2 B^2 a'}{a^3} - \frac{2E E'}{\alpha^2} + \frac{2E^2 \alpha'}{\alpha^3} \right) \frac{B \alpha^2}{\omega}  + \omega B'  -  V'(A_\rho \bar{A}^\rho ) \left(  \frac{B' \alpha^2}{\omega}  + \frac{2 B \alpha \alpha'}{\omega}  \right) \: . \label{eq:appendix:derivation-of-FPS:second-derivative-E-rewritten}
\end{align}

Then, we re-arrange equation \eqref{eq:appendix:derivation-of-FPS:second-derivative-E-rewritten} to isolate terms with $B'$:
\begin{align}
\begin{split}
 E'' &= - B' \left( V''(A_\rho \bar{A}^\rho ) \frac{2}{\omega} \frac{B^2 \alpha^2}{a^2} + V'(A_\rho \bar{A}^\rho ) \frac{\alpha^2}{\omega} \right)   + V''(A_\rho \bar{A}^\rho ) \left( \frac{2 B^2 a'}{a^3} + \frac{2E E'}{\alpha^2} - \frac{2E^2 \alpha'}{\alpha^3} \right) \frac{B \alpha^2}{\omega}  \\
 &+ \omega B'  -  V'(A_\rho \bar{A}^\rho ) \: \frac{2 B \alpha \alpha'}{\omega} \: .
\end{split}
\end{align}

Now we can substitute this expression for $E''$ in the equation of motion for $B$ \eqref{eq:appendix:derivation-of-FPS:general-eq-FBS-B}. Note that the terms with $\omega B'$ cancel out. The final expression for $B'$ can now be obtained by solving the resulting expression for $B'$, leading to
\begin{align}
\begin{split} \label{eq:appendix:derivation-of-FPS:general-eq-FPS-B}
B' &= \left\{ V''(A_\rho \bar{A}^\rho ) \left( \frac{2B^2 a'}{a^3} + \frac{2E E'}{\alpha^2} - \frac{2E^2 \alpha'}{\alpha^3} \right) \frac{B \alpha^2}{\omega} - V'(A_\rho \bar{A}^\rho ) \left( a^2 E + \frac{2 B \alpha \alpha'}{\omega} \right) \right. \\
&- \left. \left( \frac{a'}{a\,} + \frac{\alpha'}{\alpha\,} - \frac{2}{r} \right) (E' - \omega B) \right\} \left( V''(A_\rho \bar{A}^\rho ) \frac{2}{\omega} \frac{B^2 \alpha^2}{a^2} + V'(A_\rho \bar{A}^\rho ) \frac{\alpha^2}{\omega} \right)^{-1} \: . 
\end{split}
\end{align}

\newpage
%\markright{\currsection \ \ $\cdot$ \ \  Declaration of Authorship}
\markright{Appendix D \ \  Declaration of Authorship}
\section*{Declaration of Authorship}

I hereby declare that I have composed the present thesis myself and without use of any other than the cited sources and aids. Sentences or parts of sentences quoted literally are marked as such; other references with regard to the statement and scope are indicated by full details of the publications concerned. This thesis was not yet, even in part, used in another examination or as a course performance.
\\
$\:$ \\
\\
Place, Date: ...................................................................... Signature: .............................................
\\
$\:$ \\
$\:$ \\
$\:$ \\
$\:$ \\
$\:$ \\
\\
Hiermit erkläre ich, dass ich die Arbeit selbstständig und ohne Benutzung anderer als der angegebenen Quellen und Hilfsmittel verfasst habe. Alle Stellen der Arbeit, die wörtlich oder sinngemäß aus Veröffentlichungen oder aus anderen fremden Texten entnommen wurden, sind von mir als solche kenntlich gemacht worden. Ferner erkläre ich, dass die Arbeit nicht -- auch nicht auszugsweise -- für eine andere Prüfung verwendet wurde.
\\
$\:$ \\
\\
Ort, Datum: ..................................................................... Unterschrift: ..........................................

\end{appendices}